\newif\ifShowKeys
\ShowKeystrue
\ShowKeysfalse

\documentclass[11pt,a4paper]{article} 
\usepackage[height=8.8in,width=6.45in]{geometry}

\usepackage[hidelinks]{hyperref}
\usepackage{xspace}

\usepackage{tocloft}

\ifShowKeys \usepackage{showkeys} \fi

\usepackage{amsmath, amssymb,amsthm}
\usepackage{amsthm}
\numberwithin{equation}{section}

\usepackage{bm,environ,mathrsfs,array,arydshln}
\usepackage{booktabs,float,slashed,hyperref}
\usepackage[mathcal]{euscript}
\usepackage{tensor} 						
\usepackage{mathabx}
\usepackage{simpler-wick}

\usepackage{aurical}
\usepackage[T1]{fontenc}

\usepackage[nodayofweek]{date time}
\usepackage{graphicx,epsfig,epic}
\usepackage{tikz} 
\usetikzlibrary{decorations.pathreplacing,decorations.markings,snakes}
\tikzset{middlearrow/.style={decoration={markings, mark= at position 0.5 with {\arrow{#1}} ,
}, postaction={decorate}}}

\usepackage{framed}						
\definecolor{shadecolor}{rgb}{0.95,0.95,0.97}

\usepackage{comment}

\allowdisplaybreaks

\newcommand{\red}[1]{\textcolor{red}{#1}}

\newcommand{\bs}{\begin{shaded}}
\newcommand{\es}{\end{shaded}}
\def\ba#1\ea{\begin{align}#1\end{align}}		
\newcommand{\be}{\begin{equation}}
\newcommand{\ee}{\end{equation}}
\newcommand{\mc}{\mathcal }

\newcommand{\la}{\label}

\newcommand{\lp}{\notag \\ & }

\DeclareMathOperator{\tr}{\text{tr}}

\newcommand{\wt}{\widetilde}

\newcommand{\cf}{\textit{cf.} }
\newcommand{\ie}{\textit{i.e.} }
\newcommand{\eg}{\textit{e.g.} }

\makeatletter
\DeclareFontFamily{OMX}{MnSymbolE}{}
\DeclareSymbolFont{MnLargeSymbols}{OMX}{MnSymbolE}{m}{n}
\SetSymbolFont{MnLargeSymbols}{bold}{OMX}{MnSymbolE}{b}{n}
\DeclareFontShape{OMX}{MnSymbolE}{m}{n}{
<-6>  MnSymbolE5
   <6-7>  MnSymbolE6
   <7-8>  MnSymbolE7
   <8-9>  MnSymbolE8
   <9-10> MnSymbolE9
  <10-12> MnSymbolE10
  <12->   MnSymbolE12
}{}
\DeclareFontShape{OMX}{MnSymbolE}{b}{n}{
<-6>  MnSymbolE-Bold5
   <6-7>  MnSymbolE-Bold6
   <7-8>  MnSymbolE-Bold7
   <8-9>  MnSymbolE-Bold8
   <9-10> MnSymbolE-Bold9
  <10-12> MnSymbolE-Bold10
  <12->   MnSymbolE-Bold12
}{}

\let\llangle\@undefined
\let\rrangle\@undefined
\DeclareMathDelimiter{\llangle}{\mathopen}%
 {MnLargeSymbols}{'164}{MnLargeSymbols}{'164}
\DeclareMathDelimiter{\rrangle}{\mathclose}%
 {MnLargeSymbols}{'171}{MnLargeSymbols}{'171}
\makeatother

\newmuskip\pFqmuskip

\newcommand*\pFq[6][8]{%
  \begingroup 
  \pFqmuskip=#1mu\relax
  \mathchardef\normalcomma=\mathcode`,
  \mathcode`\,=\string"8000
  \begingroup\lccode`\~=`\,
  \lowercase{\endgroup\let~}\pFqcomma
  {}_{#2}F_{#3}{\left[\genfrac..{0pt}{}{#4}{#5};#6\right]}%
  \endgroup
}
\newcommand{\pFqcomma}{{\normalcomma}\mskip\pFqmuskip}

\def\XXint#1#2#3{{\setbox0=\hbox{$#1{#2#3}{\int}$}
     \vcenter{\hbox{$#2#3$}}\kern-.5\wd0}}

\newcommand{\sql}{\sqrt\lambda}
\newcommand{\gs}{g_{\text{s}}}

\newcommand{\veva}[1]{\Big\langle  #1 \Big\rangle}
\newcommand{\vev}[1]{\langle  #1 \rangle}
\newcommand{\vevz}[1]{\langle  #1 \rangle}

\newcommand{\WWW}{\mathsf{W}}

\newcommand{\RRR}{\mathsf{R}}

\newcommand{\UUU}{\mathsf{U}}
\newcommand{\TTT}{\mathsf{T}}

\newcommand{\FFF}{\mathsf{F}}

\newcommand{\gym}{g_{_{\rm YM}}}
\newcommand{\model}{FA-orientifold\xspace}
\newcommand{\oldmodel}{SA-orientifold\xspace}

\def \no { \nonumber}
  
\def \foot{\footnote}\def \ci{cite}\def \l {\lambda}\def \iffa {\iffalse}
\def \RR {{R}} \def \ov {\over }\def \a  {\alpha} \def \ha {{1\ov 2}}
\def \ed {\end{document}}

\def \l {\lambda}
\def\foot{\footnote}
\def \adss {${\rm AdS}_5 \times S^5~$ }
\def \ov {\over}
\def \tr {{\rm tr\,}}
\def \ha {{1 \over 2}}
\def \td {\tilde}
\def \ci{\cite}

\def \N  {{\cal N}}
\def \aa  {{\rm a}}
\def \te {\textstyle} 
\def \Z {{\cal Z}}
\def \RR {{L}}
\def \aa {{\rm a}}   
\def \rr {{\rm r}}
\newcommand{\rf}[1]{(\ref{#1})}
\def \Z  {\mathbb{Z}}
\def \ads {AdS$_5\, $}
\def \const {{\rm const}}
\def \four {\frac{1}{4}}

\def \ori {{\rm ori}}
\def \orb {{\rm orb}}
\def \N {{\cal N}}

\def \ori {{\rm ori}}
\def \orb {{\rm orb}}
\def \N {{\cal N}}

\def \bF {\bar F}  \def \tF {\widetilde  F}
\def \gym  {g_{_{\rm YM}}}
\def \WW   {\vev{\mc W}}
\def \Z  {\mathbb{Z}}\def \l {\lambda}
 \def \nf {n_{_{\rm F}}}   \def \na {n_{_{\rm A}}}   \def \ns {n_{_{\rm S}}}   \def \nad {n_{_{\rm Adj}}}
\def \gym {g_{_{\rm YM}}}
\def \SA {SA-orientifold\ }  \def \FA {FA-orientifold\ } \def \SS {{\rm S}}
\def \nn   {{\rm n}}
 \def \half {\tfrac{1}{2}}
 \def \La {\Lambda} \def \Z {{\cal Z}}
\def \SS  {{\rm S}}
\def \gym {g_{_{\rm YM}}}
 \def \rr  {{\rm r}}
 \def \HH   {{\rm H}}
\def \ff {{\rm f}}
 \def \FC  {{\cal F}}

\newcommand{\aaa}{\hat a}

\begin{document}

\begin{titlepage}


\begin{tabbing}
\hspace*{11.5cm} \=  \kill 
\> Imperial-TP-AT-2021-03 \\
\end{tabbing}

\vspace*{15mm}
\begin{center}
{\Large\bf  Strong coupling expansion of free energy and BPS Wilson loop in  }\vskip 6pt
{\Large\bf    
$\N=2$  superconformal  models
  with fundamental  hypermultiplets }

\vspace*{10mm}

{\Large M. Beccaria${}^{\,a}$, G. V. Dunne${}^{\,b}$ and   A.A. Tseytlin${}^{\,c,}$\footnote{\ Also at the Institute of Theoretical and Mathematical Physics, MSU and Lebedev Institute, Moscow}
}

\vspace*{3mm}
	
${}^a$ Universit\`a del Salento, Dipartimento di Matematica e Fisica \textit{Ennio De Giorgi},\\ 
		and I.N.F.N. - sezione di Lecce, Via Arnesano, I-73100 Lecce, Italy
			\vskip 0.2cm
${}^b$ Department of Physics, University of Connecticut, Storrs, CT 06269-3046, USA
			\vskip 0.2cm
${}^c$ Blackett Laboratory, Imperial College London SW7 2AZ, U.K.
			\vskip 0.2cm
			
\vskip 0.1cm
	{\small
		E-mail:
		\texttt{matteo.beccaria@le.infn.it, \ gerald.dunne@uconn.edu,   \ tseytlin@imperial.ac.uk}
	}
\vspace*{0.8cm}
\end{center}

\begin{abstract}  
As a continuation of the study  (in arXiv:2102.07696   and arXiv:2104.12625) of  strong-coupling expansion  of 
non-planar corrections in $\N=2$   4d superconformal models 
we consider two special  theories  with gauge groups   $SU(N)$ and $Sp(2N)$. They   contain 
 $N$-independent  numbers  of hypermultiplets  in  rank 2  antisymmetric   and  fundamental representations
 and are planar-equivalent to  the corresponding $\N=4$  SYM theories. 
These $\N=2$  theories  can be realised  on  a system of $N$ D3-branes with  a finite number of D7-branes  and  O7-plane;  
 the  dual string theories   should  be  particular   orientifolds of  \adss   superstring. 
Starting with the localization matrix model representation for the $\N=2$  partition function on $S^4$ 
 we  find   exact differential relations  between 
 the  $1/N$  terms in the corresponding  free energy $F$ and  the  $\ha$-BPS  Wilson loop expectation value $\WW$
  and  also compute their   large 't Hooft coupling  ($\lambda \gg 1$) expansions. 
The structure of these  expansions  is  different from the previously  studied   models
without   fundamental  hypermultiplets. 
In the more tractable  $Sp(2N)$ case   we find  an  exact  resummed expression for the  
 leading  strong coupling terms at each order in the $1/N$  expansion. 
We also determine the   exponentially suppressed  at large $\lambda$ 
 contributions  to  the  non-planar corrections  to $F$ and $\WW$ and   comment on   their  resurgence properties. 
We  discuss   dual  string theory interpretation of these  strong coupling expansions. 
\end{abstract}
\vskip 0.5cm
	{
	}
\end{titlepage}
\tableofcontents
\vspace{1cm}


\setcounter{footnote}{0}
\section{Introduction  and summary}

An  important problem  in understanding detailed workings of  AdS/CFT duality 
is to  study    $1/N$ corrections  to  superconformal  gauge theory observables  and their  matching to     string loop corrections. BPS Wilson loop in $\N=4$  super Yang-Mills   theory provides a remarkable example 
when   its expectation value 
$\WW$ as    a function of  $N$ and $\l= \gym^2 N$ 
 can be found  exactly \cite{Drukker:2000rr}. 
Expanding first at large $N$ and then  at large $\l$  one finds in the $SU(N)$ theory 
 \ba
\vev{\mc W}^{\N=4}_{_{\rm SU(N)}} = e^{\frac{\lambda}{8N}(1-{1}/{N})}L^{(1)}_{N-1}\big(-\tfrac{\lambda}{4N}\big)= & N\,e^{\sqrt\lambda}\sum_{p=0}^{\infty}
 c_p \frac{\lambda^{\frac{6p-3}{4}}}{N^{2p}}\Big[1+{\te \mc O\big(\frac{1}{\sqrt\lambda}\big)}\Big]\no \\
 =& e^{2\pi\,T}\sum_{p=0}^{\infty}c'_{p}\,\Big(\frac{\gs}{\sqrt T}\Big)^{2p-1}\Big[1+\te \mc O\big(T^{-1}\big)\Big]   \ , \la{1}
\ea
where  we   expressed  the result in terms of the 
 string coupling and tension of  the dual AdS$_{5}\times S^{5}$  string  theory
\be
\la{2}
\gs =  \frac{\lambda}{4\pi N},\qquad\qquad  \qquad T = \frac{\RR^2}{2\pi \a'}=\frac{\sqrt\lambda}{2\pi}\ , \qquad \qquad
{ 1 \ov N} = {\gs\ov \pi T^2}  \ . 
\ee
As was argued in \cite{Giombi:2020mhz},  the particular structure \rf{1} of the  small $\gs$, large $T$   expansion  of $\WW$ is indeed  expected  on  the   string-theory  side
   and may apply also to other closely related   theories  with less supersymmetry. 
   
Indeed, the same   expansion  \rf{1} was   found recently 
  for  two special $\N=2$  4d superconformal models -- $SU(N) \times SU(N)$  "orbifold" 
\ci{Beccaria:2021ksw}  and $SU(N)$  "orientifold"  \ci{Beccaria:2021vuc}   that  are planar-equivalent  to $\N=4$ SYM theory. Here the localization approach  \cite{Pestun:2007rz,Pestun:2016zxk}  allows one  to 
 expresses  the expectation value 
  $\WW$  in terms of a non-trivial 
  matrix model integral.  One is  then able to extract 
the  large $\l$  behaviour of the  leading  non-planar $1/N$  correction, 
   finding that it scales as $\l^{3/2}$  relative to the planar (i.e. $\N=4 $  SYM)  term,   in agreement with \rf{1}. 

 The aim of the present  paper is  to consider     two  other  ($SU(N)$ and  $Sp(2N)$) 
  examples of $\N=2$  "orientifold"  superconformal  models  for which $\WW$
 can be  also computed using  the localization matrix model of  \cite{Pestun:2007rz} (see also \ci{Fiol:2014fla,Fiol:2015mrp,Fiol:2020bhf}).  
  These   models are   still  planar-equivalent to $\N=4$ SYM  but in contrast to the 
  "orientifold"
 model  studied  in  \ci{Beccaria:2021vuc} ($\N=2 $  vector multiplet 
   coupled to   hypermultiplets   in symmetric and  in antisymmetric $SU(N)$ representation) 
will contain a finite ($N$-independent)  number $\nf$ of hypermultiplets 
 in the fundamental representation. 
 The later are  effectively related to the presence of  (a finite number of) D7-branes in the dual string theory 
  description and thus  to a  different type of the 
orbifold/orientifold of \adss string theory    than in the  previous  case of $\nf=0$
\ci{Fayyazuddin:1998fb,Aharony:1998xz,Park:1998zh,Ennes:2000fu}.
We shall  find that   here   the structure of the large $N$, large $\l$ expansion of the  BPS 
Wilson loop  expectation   value $\WW$  
 will be   {\it different} from \rf{1},  raising   an interesting 
 question of how to explain this   on the dual string theory side. 
 
\subsection{Review of $\N=2$ models}

 \noindent  Let us   first  review  4d $\N=2$  superconformal   gauge theories   we are interested in. 
The condition of conformal invariance of an $SU(N)$  model  with a number of 
hypermultiplets  in
the adjoint, fundamental, rank-2 symmetric, and rank-2 antisymmetric representations  is \ci{Koh:1983ir,Howe:1983wj}
\be\la{3}
{SU(N)}: \qquad \qquad 
\beta_{1} = 2N-2N\,\nad-\nf-(N+2)\,\ns-(N-2)\,\na = 0 \ . 
\ee
The non-zero number of adjoints   can only be 
 $\nad=1$   when  we find the $\N=4$ SYM ($\nf=\na=\ns=0$). For $\nad=0$   we get $\N=2$ 
 superconformal models  with $\nf= 2N-(N+2)\,\ns-(N-2)\,\na$.
 To have planar  equivalence with $\N=4$ SYM (and thus  a relatively simple  AdS  dual) 
  the number $\nf$  should  not depend on $N$. 
 This implies that $\ns+\na=2$  and thus there are only two non-trivial solutions 
 that we shall  refer to as "SA" (symmetric+antisymmetric)   and "FA" (fundamental+antisymmetric)  models
 \be\la{4}
SU(N): \qquad \qquad  \ \text{SA}:\, \quad (\nf, \ns, \na) = (0,1,1)\ ,\qquad \qquad   \text{FA}:\,\quad  (\nf, \ns, \na) = (4,0,2)\ .
\ee
Both    $\N=2$   theories  are dual to certain orbifold/orientifold projections of \adss  superstring
\ci{Ennes:2000fu} and  for that reason we shall refer to them respectively 
 as   the "SA-orientifold"    and the  "FA-orientifold". 
It is  the SA-orientifold   model that was discussed in \cite{Beccaria:2021vuc}   and here we shall   study 
the FA-orientifold model. 

For completeness,  let us recall  that the 4d conformal anomaly  a  and c coefficients  of 
an  $\N=2$ superconformal model are determined by  the free-theory   values, i.e.  in terms of the  total number of the  vector  multiplets and hypermultiplets (counting also 
dimensions of  their representations): 
${\rm a} = \tfrac{5}{24}\,\nn_{\rm v}+\tfrac{1}{24}\,\nn_{\rm h}$,\  ${\rm c} = \tfrac{1}{6}\,\nn_{\rm v }+\tfrac{1}{12}\,\nn_{\rm h}$.  The resulting explicit values  are given   below
\iffa \ba
SU(N): \ \ \   \ \ \  & \N=4 {\rm SYM} \ \ \ \ \ \    {\rm a} = {\rm c} =  \frac{1}{4}N^{2}-\frac{1}{4} \ , \no \\
& \N=2 {\rm  SA} \ \ \ \ \ \    {\rm a} =  \frac{1}{4}N^{2}-\frac{5}{24} \ , \quad  
  {\rm c} =  \frac{1}{4}N^{2}-\frac{1}{6} \ , \la{222} \\
& \N=2 {\rm  FA} \ \ \ \ \ \    {\rm a} = \frac{1}{4}N^{2}+\frac{1}{8}N-\frac{5}{24} \ , \quad 
 {\rm c} = \frac{1}{4}N^{2}+\frac{1}{4}N-\frac{1}{6}  \ , \no \\
\ea
\fi
{\small 
\begin{table}[H]
\be\notag
\def\arraystretch{1.3}
\begin{array}{lccc}
\toprule
 {SU(N) } & &  {\rm a} & \ \ \ {\rm c}  \\
\midrule
\N=4 \ {\rm SYM}&  & \frac{1}{4}N^{2}-\frac{1}{4}&\ \ \  \frac{1}{4}N^{2}-\frac{1}{4}  \\
\N=2\ \text{SA} &  & \frac{1}{4}N^{2}-\frac{5}{24} &\ \ \  \frac{1}{4}N^{2}-\frac{1}{6}   \\
\N=2\ \text{FA} &   & \frac{1}{4}N^{2}+\frac{1}{8}N-\frac{5}{24}&\ \ \  \frac{1}{4}N^{2}+\frac{1}{4}N-\frac{1}{6}   \\
\bottomrule
\end{array}\notag
\ee
\end{table}  }
Similarly,   in the case of the   $Sp(2N)$  gauge group the condition of 
conformal invariance of the $\N=2$  model containing the  adjoint,  fundamental and antisymmetric  hypermultiplets
reads  \cite{Koh:1983ir} (cf. \rf{3})\foot{In 
this paper we shall   denote by $Sp(2N)$ the compact symplectic group 
$USp(2N)= U(2N) \cap  Sp(2N, C)$  (sometimes also denoted as $Sp(N)$)  so that $Sp(2) = SU(2)$. 
The dimensions of  its adjoint, fundamental and antisymmetric representations are, 
respectively,   $\dim {\rm Adj} =\dim [Sp(2N)] = N(2N+1)$,
 $\dim {\rm F} = 2N$, $\dim {\rm A} = N(2N-1)-1$. 
 Note   while the groups $Sp(2N)$ and $SO(2N)$   and their representations 
 are formally  related   by  $N \to - N$  \ci{Mkrtchian:1981bb}, the index  of a  representation 
   that enters the 1-loop beta-function  is always  positive (i.e. its  sign is changed 
   at the same time  with   taking $N\to - N$). 
      Thus the conformal invariance condition 
   is not invariant  and has different solutions for the two groups. 
   For example,  the   antisymmetric  representation  of $Sp(2N)$   is  mapped to the 
   symmetric traceless  representation of $SO(2N) $   with the  index $2N+2 $     which is larger than  the index 
    of the adjoint $SO(2N) $ representation $2N-2$.
    Thus  there are no $SO(2N) $  conformal  theories with  hypermultiplets  in  the  symmetric traceless   representation \cite{Koh:1983ir}.
    }
\be\la{5} Sp(2N):\qquad \ \ 
\beta_{1} = 2N+2-(2N+2)\,\nad-\nf-(2N-2)\,\na=0\ .
\ee
The $Sp(2N)$   $\N=4$ SYM  theory corresponds to $\nad=1$, $\nf=\na=0$. For  $\nad=0$   demanding
planar equivalence to $\N=4$ SYM implies that $\nf$ should be independent of $N$ and thus  the only  solution  is 
the FA-orientifold   model  with $\nf=4, \ \na=1$
\be\la{55}
Sp(2N): \qquad \qquad \text{FA} : \,\quad  (\nf, \na) = (4,1)\ .\qquad \qquad
\ee
The corresponding  conformal anomaly coefficients are given  below:   
\begin{table}[H]
\be\notag
\def\arraystretch{1.3}
\begin{array}{cccc}
\toprule
{Sp(2N)} &  & {\rm a} & \ \ \ \ {\rm c}  \\
\midrule
\N=4 \ \text{SYM}  &  & \frac{1}{2}N^{2}+\frac{1}{4}N &\ \ \ \  \frac{1}{2}N^{2}+\frac{1}{4}N  \\
\N=2 \  \text{FA}\  \ \ &  & \frac{1}{2}N^{2}+\frac{1}{2}N-\frac{1}{24}  &\ \ \ \  \frac{1}{2}N^{2}+\frac{3}{4}N-\frac{1}{12}\\
\bottomrule
\end{array}\notag
\ee
\end{table}

\subsection{Summary of the  results}

Let us now summarise the main results of this paper  starting with the $SU(N)$ case. 
As  in  the case of the \SA  \ci{Beccaria:2021vuc}   the  structure of the 
 localization matrix model implies  that  the  leading $1/N$ 
corrections  to 
the Wilson loop expectation  value can be expressed   in terms of the corresponding corrections
 to the gauge  theory  free energy  $F(\l, N)=-\log Z $ on 4-sphere. 
   For that reason the  main  effort 
 goes into  the study for   the large $N$ expansion of $F$. 
 
 To recall, in the  case of the  $SU(N)$ 
 $\N=4$   SYM theory  where  the partition function $Z$  is given by 
 the Gaussian matrix model \ci{Drukker:2000rr,Pestun:2007rz}
 one finds  (after subtracting the "trivial" UV divergence in a particular scheme, see
 Appendix \ref{A}) \ci{mehta,Marino:2012zq,Russo:2012ay,
 Rodriguez-Gomez:2016cem}
\be
\la{11}
SU(N): \qquad \qquad   F ^{\mc N=4}(\l)= - \te \ha  (N^2 -1)\log \l  +  C(N)\ , 
\ee
where  $C(N)$  (given by \rf{AA4} or   by log of Barnes function)  does
 not depend on $\l$.\foot{Since the large $N$ expansion of $C(N)$ contains 
$\log N$ term (cf. footnote \ref{a29}), its comparison  with string theory  would  require a non-perturbative  definition of the latter.
Given that 
$F$   is scheme-dependent, $C(N)$ may be in principle  eliminated  by  changing the scheme  (e.g.   by redefining  the 
matrix model measure). We shall ignore this $\l$-independent constant  in what follows. }
The  large $N$ expansion 
of the free energy of the  $\N=2$  FA-orientifold model  which is  planar-equivalent  to the  $\N=4$   SYM   may be represented as 
\be \la{12}
SU(N): \qquad  \qquad F(\l) = F^{\mc N=4}(\l)   +  N\, F_{1}(\l) + F_{2}(\l) +\mc O\te ( { 1 \ov N})\ .
\ee
The $F_1$ term   was absent in the case of the  \SA in \ci{Beccaria:2021vuc} (it  is related to the presence  of the fundamental hypermultiplets in the spectrum of this $\N=2$ model). 
 $F_1$   admits an explicit integral representation in terms of Bessel functions \rf{eq:F1}
 allowing to find its  strong coupling expansion  
\ba\la{13}
F_1 &\stackrel{\lambda\gg 1}{=}  f_1 \l + f_2 \log \l   + f_3 +f_{4}\,\lambda^{-1}+\mc O\big(
 e^{-\sqrt{\lambda}}\big)\ , \\
\la{14}
f_1 &= \tfrac{\log 2}{4\pi^{2}}\ ,\ \ \ \ \   f_2=-\tfrac{1}{4}\ , \  \ \ \ f_{3}=\tfrac{1}{2}\log\pi+\tfrac{7}{6}\log 2+\tfrac{3}{4}-6\log\mathsf{A},\ \ \ \  f_{4} = -\tfrac{\pi^{2}}{4} \ ,\ea
where $\mathsf{A}$ is the  Gleisher's constant.\foot{Note that $\log 2$  in $f_1$ originates from the Dirichlet $\eta$-function  value 
$\eta(1) = \sum_{k=1}^\infty {(-1)^{k-1}\ov k} =\log 2$ (see \rf{b8}).}
There is just a  finite number of  "polynomial" in  large $\l$   corrections  and an infinite number of exponential 
$e^{- (2n+1)  \sqrt{\lambda}}$  corrections  
reflecting the asymptotic nature of the strong coupling expansion (see  \rf{eq:F1exp}; here we omit the $\lambda^{-1/4}$ prefactor of  $e^{-\sqrt{\lambda}}$).

$F_2$   may be   written as  the sum of the two different contributions: 
a simpler one $\tF_2$   which is  related to $F_1$  by a differential relation    and  a more complicated one 
$\bF_2$  which  turns out to be   the same as  the leading $1/N^2$ correction to $F$ 
 in the SA-orientifold  case in  \ci{Beccaria:2021vuc}   
\be \la{15}
F_2(\l)= \tF_2(\l) + \bF_2 (\l)\ , \qquad \qquad\qquad  
 \tF'_2 
= -\tfrac{\lambda}{2}\big[(\lambda\, F_1)''\big]^{2} \ , 
\ee
where $(...)'= {d\ov d \l} (...)$.
As a result,\footnote{\la{foot4} {The analysis in \cite{Beccaria:2021vuc} showed that the leading large $\lambda$ term in 
 $\bF_{2}$ is definitely $\lambda^{1/2}$.  The derivation of its  coefficient $k_{1}=\frac{1}{2\pi}$
was  based on partially  heuristic analysis of the determinant of an infinite matrix, whose matrix elements admit
 an asymptotic expansion for large $\lambda$. 
 A comparison
with Pad\'e resummation of the determinant revealed that $ \frac{1}{2\pi} $  may  actually be  a lower estimate of the exact value of $k_1$. This issue  will not be relevant for the   large $\l$  expansion in the models considered
here where $\tF_{2}$ is dominant over $\bF_{2}$ at large coupling. 
}}
 \ba
\la{17}
  \tF_2  \stackrel{\lambda\gg 1}{=}  &\ 
   \ p_1\lambda^{2}+p_2\,\lambda  +p_{3}\log \lambda+p_4  +  { \mc O\big(
    e^{-\sqrt{\lambda}}\big)}\ , \\ 
 \la{18}    \bF_2   \stackrel{\lambda\gg 1}{=}   &\ \ k_1  \l^{1/2} + k_2 \log \l + k_3 +  \mc O(\lambda^{-1/2})   \ , \\
    p_1  = & - f_1^2     \ , \ \ \quad p_2 = - 2 f_1 f_2   \ ,\ \  \quad  \te  p_3=-\ha f_2^2 \  ,\ \ ... \la{19} \ , \ \ \ \ \ 
\te k_1 = {1\over 2 \pi} \ ,  ...\ .\ea
where  the values of $f_i$ were given in \rf{14}.  The  form of  the exponential corrections in 
$ \tF_2 $  follows from those in $F_1$  and  the relation  in \rf{15},  and similar  corrections are expected in 
$ \bF_2 $. 

The  large $N$ expansion of the circular   $\ha$-BPS   Wilson loop expectation   value  in this 
$\N=2$ theory 
can be written as 
\be
\la{20}
SU(N): \qquad  \qquad 
\vev{\mc W} = N\, W_0(\l) + W_1(\l) +\tfrac{1}{N}\big[W_{0,2}(\l)+W_2(\l) \big]+   \mc O\te ( { 1 \ov N^2}) \ , 
\ee
where $W_0$ and $ W_{0,2}$ are the leading     $\N=4$ SYM  contributions following from \rf{1} \cite{Erickson:2000af,Drukker:2000rr}
\ba
W_0 = \tfrac{2}{\sql}I_{1}(\sql)
\  , \ \ \ \qquad \qquad 
W_{0,2}= \tfrac{1}{48}\,\big[-12\sql\,I_{1}(\sql)+\lambda\,I_{2}(\sql)\big] \ ,
 \la{21} 
\ea
while $W_1$ and $W_2$  are the  genuine $\N=2$ corrections. 
As we will show,    they    can be  expressed in terms
of the $1/N$ corrections $F_1$ and $F_2$ to the free energy \rf{12}   by the following remarkable  differential relations 
(cf. \rf{15})
\be\la{22}
W'_1  = -\tfrac{\lambda}{4}\,W_0\, 
(\lambda\, F_1)''\ , \qquad \qquad \ \ 
W_2 = -\tfrac{\lambda^{2}}{4}\,W_0 \, 
F'_2\ .
\ee
Using   \rf{13}--\rf{19} in  \rf{22} and normalizing to the   leading planar value 
\be \la{555}
W_0\stackrel{\lambda\gg 1}{=}  \sqrt{\tfrac{2}{\pi}} \l^{-3/4} e^{\sql} \big[1+  \mc O{\te ( { 1 \ov \sql}})\big] \ , \ee 
   we  then find 
for the  strong coupling expansions of $W_1$ and $W_2$
\ba
\frac{W_1}{W_0}  \stackrel{\lambda\gg 1}{=}  &\ \ \te  -f_1 \,\lambda^{3/2}+ {3\ov 2} f_1\l   -  ( \tfrac{3}{8} f_1 + \ha f_2)  \l^{1/2} + 
\mc O\te ( { \l^0})  \ , \la{24}
\\
 \la{25}
\frac{W_2}{W_0}  \stackrel{\lambda\gg 1}{=} &\ \ \te  \ha f_1^2  \,\lambda^{3}  
  + \ha f_1 f_2\,  \l^2 - {1\ov 8} k_1 \l^{3/2}  + \mc O\te ( { \l}) \ . 
\ea
Like $F_1$ in \rf{12},  the $W_1$  term in \rf{20}  was  absent in the case of the \SA  in \ci{Beccaria:2021vuc}    
(where there were no odd powers in  $1/N$  series). 
Also,  in the \SA case   the expansion of  $W_2/W_0$      started   with the 
$k_1 \l^{3/2}$ term that  originated from  the $\bF_2$ term  in \rf{18}  in view of     \rf{22}. 
The expressions \rf{24},\rf{25}   also  contain exponential  corrections as follows from \rf{13},\rf{17} and \rf{22}.

Similar results   are found in the  case of the $Sp(2N)$  \FA model \rf{55}   which is 
 more tractable   as the corresponding localization $\N=2$ 
matrix model is   simpler than in the $SU(N)$ case. 
Here\foot{Here  we shall  use the   same definition  for $\l$ as in the $SU(N)$ case, i.e.  $\l=\gym^2 N$
(i.e. without extra factor of 2 as, e.g., in \ci{Giombi:2020kvo}).}  
\ba \la{26}
Sp(2N): \qquad 
&F = F^{\mc N=4}   +  N\, \FFF_{1}(\l) + \FFF_{2}(\l)  +   \te  { 1 \ov N}    \FFF_{3}(\l)  +  \te   { 1 \ov N^2}    \FFF_{4}(\l)   
  +\mc O\te ( { 1 \ov N^3})
\ ,   \\
 &F ^{\mc N=4}= - \half N ( 2 N +1)  \log \l\  . \la{662}
\ea
 It turns out that 
the structure of the corresponding matrix model   implies 
  that $\FFF_{1}$,  $\FFF_2$  and $\FFF_3$   can be expressed in terms of  the  function
$F_1$ in \rf{12}  (and its integral $\tF_2$ in \rf{15})   that appeared   in the $SU(N)$   case 
 \ba
& \FFF_{1}=2F_{1} \ , \qquad \ \ \ \ 
 \FFF_{2}  = \tfrac{1}{2}(\lambda F_{1})'+ 2\,\wt F_{2} \ , \qquad \ \ \ \ \  \tF'_2 
= -\tfrac{\l}{2}\big[(\lambda\, F_1)''\big]^{2} \ , \la{27}\\
 &\FFF_{3}  = \te \frac{\l^{2}}{24}\big( \l F_{1}\big)'''  - \frac{\l^2}{4}\big[ \big(\l F_{1}\big)''\big]^{2}  +\frac{\l^{3}}{3}\big[ \big(\l F_{1}\big)''\big]^{3}\ , \qquad \ \ \ \FFF_4=\te -\frac{2\lambda^{2}}{4!} \big( \l^3  \big[ (\l F_{1})'' \big]^{4} \big)' +...   \ , 
 \la{2700}
 \ea
 Similar  expressions   in terms of derivatives of $F_1$  appear to  exist also  for higher $\FFF_n$ 
 terms in  \rf{26}. 
 
 Computing the strong-coupling expansion  of $\FFF_n$ we find that (cf. \rf{12},\rf{13},\rf{17}) 
  \be\la{1779}
     F=  F^{\N=4} + \Delta F \stackrel{\lambda\gg 1}{=} \Delta F_{\rm pol}  - ( N^2 + N 
     + \tfrac{3}{16} ) \log \l   - \tfrac{\pi^2}{2} \tfrac{N}{\l }    +   \mc O (  e^{-\sql})   \ , \ee
     where $\Delta F_{\rm pol} $   stands for the polynomial in $\l$  part of the  strong coupling  expansion.
      Note   that $\log \l$  term in \rf{1779}   receives contributions only at orders $N^2, N$ and $N^0$   while the $\l^{-1}$ term appears 
 only at order $N$. 

      Remarkably,       the sum  of the leading large $\l$ terms in $\Delta F_{\rm pol} $   at each order in $1/N$ 
  appears to   have  a  closed $\log$ expression  ($f_1= {\log 2 \ov 4 \pi^2}$ as  in \rf{14})
    \ba    
     \Delta F_{\rm pol} =& 
     N  \big[  2 f_1\l   + \mc O(\l^{0} ) \big]   +    \big[ 2  f_1^2  \l^2 + \mc O(\l) \big]  
              + \tfrac{1}{N}   \big[ \tfrac{8}{3} f_1^3 \l^3 + \mc O(\l^{2} ) \big]  + \mc O ( \tfrac{1}{N^2} ) 
              \no \\
              =  &   N^2 \FC( \tfrac{\l}{ N}  ) + ...\ , \qquad \qquad  
      \FC( \tfrac{\l}{ N}  ) =   \log \big ( 1+ 2 f_1 \tfrac{\l}{N}\big) \ . \la{1.25}
\ea
 Combined  with the $N^2 \log \l$ term  in \rf{1779}  the leading  strong-coupling  expression for  $F$  is  then 
 \be \la{999}
 F\stackrel{\lambda\gg 1}{=} - N^2  \log \l + N^2 \FC( \tfrac{\l}{ N}  ) + ...= 
 N^2 \log \big(\l^{-1}  +   2 f_1  N^{-1}\big) + ...=  N^2 \log\big[ N^{-1}  \big(\gym^{-2}   +   2 f_1\big) \big] + ...  
 \ ,  \ee 
suggesting   possible role  of a finite redefinition  of  the inverse  coupling constant.

 The  large $N$ expansion of the Wilson loop expectation  value  here   can be written as (cf. \rf{20}) 
 \ba
\la{29} Sp(2N): \qquad  
\vev{\mc W} =  \vev{\mc W}^{\N=4}  + \Delta \vev{\mc W}  \ , \qquad \ \ \
\Delta  \vev{\mc W}  = \WWW_1 +\tfrac{1}{N}\WWW_2   +\tfrac{1}{N^2}\WWW_3  +  \mc O\te ( { 1 \ov N^3}) \ , 
\ea
where the $\N=4$  $Sp(2N)$  SYM   contribution is  
 \ci{Fiol:2014fla,Giombi:2020kvo} (cf. \rf{1}) 
\ba
\la{30}
&\vev{\mc W}^{\N=4} = 2\, e^{\frac{\lambda}{16N}} \sum_{k=0}^{N-1}L_{2k+1}\big(-\tfrac{\lambda}{8N}\big)
= N\, \WWW_0 + \WWW_{0,1} + \tfrac{1}{N} \WWW_{0,2} +   \mc O\te ( { 1 \ov N^2}) \ , \\
&\WWW_{0}  =  \tfrac{4}{\sql}I_{1}(\sql) = 2W_{0}\ , 
\qquad \WWW_{0,1}  = \tfrac{1}{2}I_{0}(\sql)-\half\ , \qquad 
\WWW_{0,2}  = \tfrac{\l}{96}\,I_{2}(\sql).\la{300}\ea
As in the $SU(N)$ case, one finds that the  $\N=2$ corrections $\WWW_1$ and $\WWW_2$  are expressed in terms of 
$\FFF_1=2 F_1 $  and $\FFF_2$ as in \rf{22} so that 
\be
\la{31}
\WWW'_{1} =  -\tfrac{\lambda}{4}\WWW_{0}\big(\lambda F_{1}\big)''\ , \qquad \qquad   
\WWW_{2} =  -\tfrac{\lambda^{2}}{8}\WWW_{0}\,  
\FFF'_{2}
= -\tfrac{\lambda^{2}}{8}\WWW_{0} \Big[  \tfrac{1}{2}    \big(\lambda F_{1}\big)'' -  \l \big[\big(\lambda F_1\big)''\big]^{2} \Big]
\ . 
\ee
Comparing   $\WWW_1$ and $\WWW_0$  with $W_1$   and $W_0$ 
in the $SU(N)$ case in \rf{21},\rf{22}  we conclude that their   ratio is the same for  any $\l$. 
The  analog of the  strong-coupling expansions in \rf{24},\rf{25}   is\foot{Note    that     the leading terms  in  \rf{34}  and \rf{25}   are the same  but subleading terms have different structure.} 
\ba
\frac{\WWW_{1}}{\WWW_{0}} &  \stackrel{\lambda\gg 1}{=}  \frac{W_{1}}{W_{0} }=  -f_{1}\,\lambda^{3/2}+\tfrac{3}{2}\,f_{1}\,\lambda- (\tfrac{3}{8} f_{1}+\tfrac{1}{2}f_{2})\l^{1/2} 
+\mc O ( {\l^{0}})  
\ ,\la{33}   \\
\frac{\WWW_{2}}{\WWW_{0}} &  \stackrel{\lambda\gg 1}{=}   \tfrac{1}{2} f_1^2\lambda^{3}-\tfrac{1}{8}f_1 (1 - 4  f_2 )\lambda^{2}-\tfrac{1}{16}f_{2}( 1 - 2 f_2) \lambda+ \mc O ( e^{-\sql} ) \ . \la{34}
\ea
Similar relations  between  higher order $1/N$ terms $\FFF_n$  in free energy \rf{26}   and $\WWW_n$ in \rf{29}  are expected 
also in general,  with the dominant  large $\l$ term in $\FFF_n$ determining  the  strong coupling asymptotics of 
$\WWW_n$. In particular, 
\be \la{1325}
\WWW_3 = - \tfrac{\l^{3/2}}{4!} \WWW_0 \big[ \l ( \l F_1)'']^3 +... \ ,  \qquad \ \ \ 
\frac{\WWW_{3}}{\WWW_{0}}  \stackrel{\lambda\gg 1}{=}  - \tfrac{1}{3!} f_1^3\lambda^{9/2} + \mc O ( {\l^{4}}) \ . \ee
Combining  the leading terms in \rf{33},\rf{34} and \rf{1325} suggests that the dominant  (at each order in $1/N$) 
 strong coupling terms in $\Delta  \vev{\mc W}$  in \rf{29} exponentiate as 
\be \la{132}
\vev{\mc W} =  ( N \WWW_0 +...) +    \Delta  \vev{\mc W}\stackrel{\lambda\gg 1}{=} N  \WWW_0  \exp\big[  - f_1 \tfrac{\l^{3/2}}{ N}  \big]
+ ...
\ .  \ee 
This may be compared with similar exponentiation of the leading large $\l$ terms in the $\N=4$ SYM  case:
as one finds from \rf{1}  in $SU(N)$ case  \ci{Drukker:2000rr}   and from  \rf{30} in the $Sp(2N)$ case (see Appendix \ref{C}) 
\ba \la{xxx}
SU(N): \ \ \ \ \   &\vev{\mc W}^{\N=4} \stackrel{\lambda\gg 1}{=}  N W_0 \exp\big[\, \tfrac{\l^{3/2}}{ 96 N^2} \big] +... \ , \\
Sp(2N): \ \ \ \ \  &\vev{\mc W}^{\N=4} \stackrel{\lambda\gg 1}{=}  
 2 N W_0 \,  ( 1 + \tfrac{\l^{1/2}}{ 8 N } ) \exp\big[\, \tfrac{\l^{3/2}}{ 96\, (2N)^{2}} \big]  + ...\ , \la{yyy} \ea
where $W_0$ is  given by \rf{555}.
Note that the $ ( 1 + \tfrac{\l^{1/2}}{ 8 N } )$ prefactor   that generates odd powers of $1/N$ in the expansion of  $\vev{\mc W}^{\N=4} $ in $Sp(2N)$ case  in  \rf{yyy}  can be absorbed into $e^{\sql}$  in $W_0$  by  shifting $N\to N + {1\ov 4}$ 
in the definition of $\l = \gym^2 N$ (assuming one keeps only the leading large $\l$ term at each order in $1/N$).\foot{We  thank S. Giombi   for this observation.} 

\subsection{Comments on  dual string theory interpretation} 

Let us now  discuss   string theory  interpretation of these  strong-coupling expansions
derived on the gauge theory side. 
The  $SU({N})$  FA-orientifold  (i.e. the  $\N=2$  $SU({N})$  superconformal model 
with $\nf=4$ and $\na=2$) may be  engineered  in flat-space  type IIB superstring 
  as a low-energy limit of the  worldvolume theory on a stack of 
coincident  $N$ D3-branes  in the
presence of  four  D7-branes and one O7-plane (see \cite{Ennes:2000fu}  and references  there).\foot{ 
This implies modding  out by the orientifold group
$
G_{\ori } = \Z_{2,\orb} \times \Z_{2,\ori},
$
where $\Z_{2,\orb} =\{1,\, I_{6789}\}$
and $\Z_{2,\ori}=\{1,\, I_{45}\, \Omega\,  (-1)^{F_L}\}$. The inversions $I_{n_1...n_r}$ 
act on the $\mathbb R^6$  (with directions 4,...,9) transverse to the D3-branes  as
 $
\Z_{2,\orb}:  \  x_{6,7,8,9} {\rightarrow}   -x_{6,7,8,9} \,
$  and $
\Z_{2,\ori}: \  x_{4,5} {\rightarrow}- x_{4,5}$.
The fixed-point set of $\Z_{2,\ori}$ is the hyperplane $x_{4,5}=0$,
which corresponds to the position of  the 
O7-plane and four  D7-branes,
while the fixed set of $\Z_{2,\orb}$ is the  hyperplane $x_{6,7,8,9}=0$.}
Taking the large-$N$  near-horizon  limit of the underlying brane configuration
one   concludes   that the dual  string  theory   should  be a 
  projection AdS$_5{\times}  S'^5, \ \ S'^5=S^5/G_\ori$,   of the  original \adss type IIB theory  \cite{Ennes:2000fu}.
Here 
$\Z_{2,\orb}$  of   $G_{\ori } = \Z_{2,\orb} \times \Z_{2,\ori}$ acts as $\varphi_1 \to \varphi_1 + \pi$, $\varphi_2 \to \varphi_2 + \pi$
and $\Z_{2,\ori}$ acts as $\varphi_3 \to \varphi_3 + \pi$ on the coordinates of $S^5$
with the metric
$     ds_5^2  =d\theta_1^2 + \cos^2\theta_1 \,(d\theta_2^2 + \cos^2\theta_2\, d\varphi_1^2+\sin^2\theta_2\,d\varphi_2^2)+
\sin^2\theta_1\, d\varphi_3^2.$

Similarly, the  dual string theory for the $Sp({2N})$  FA-orientifold  (i.e. the  $\N=2$  $Sp({2N})$  superconformal model 
with $\nf=4$ and $\na=1$)  corresponds  \ci{Fayyazuddin:1998fb,Aharony:1998xz}  
    to the near-horizon  limit of  $N$ D3-branes  with  8 D7-branes  stuck on one 
O7-plane, i.e. is the type IIB   superstring on AdS$_5{\times}  S'^5, \ \ S'^5=S^5/\Z_{2,\ori}$ 
(D7  is wrapped on AdS$_5{\times} S^3$  where $S^3$ is fixed-point locus  of $\Z_{2,\ori}$). 

In both  $SU(N)$  and $Sp(2N)$  cases,  the presence of   D7-branes 
  introduces the  new D3-D7 open string sector (with massless   modes  being related to the fundamental hypermultiplets in the corresponding gauge theory). 
 That means, in particular,   that  the dual string theory perturbation theory  will involve  both 
 closed-string  and open-string  world-sheet topologies, i.e. corrections 
 of both even and odd    powers in $\gs$, corresponding to  even  and odd powers of $1/N$ on the   gauge theory side. 
 
 While in the $SU(N)$   $\N=2$  model   one expects  contributions from only orientable surfaces
  (with  topologies of  2-sphere  with 
 holes and handles) in the $Sp(2N)$  case  there should be additional  contributions with non-orientable crosscups
  (as is also suggested by the  structure of the  $1/N$ expansion of 
  perturbative gauge theory diagrams, cf.   \ci{Fiol:2014fla}).
 In  the  $Sp(2N)$  $\N=4$ SYM case 
  all odd-power  $1/N$  contributions   should come from crosscups \cite{Witten:1998xy}, 
  while in the  $Sp(2N)$  $\N=2$ \FA   model 
  there should  be additional contributions  from world sheets  with  boundaries
     introduced  due to the presence of D7-branes   
     (and related to  the presence of  fundamental hypermultiplets on the gauge theory side), see also 
     \ci{Aharony:1999ti}.

 Accounting for  the open string  (type I, or disc) term in the  dual string theory effective action that here 
  may be  interpreted  as the  D7-brane world-volume action allowed  to give  \ci{Aharony:1999rz,Blau:1999vz}
  the  holographic  interpretation of the order $N$ term in  the  (super)conformal anomalies of the $Sp(2N)$  \FA (cf. table below eq. \rf{55}).

The AdS/CFT   duality  suggests that  the  conformal gauge theory free energy  $F$ on $S^4$   should  be  matched 
with  the  string  partition function $Z_{\rm str}$ 
 in AdS$_5 \times S'^5$. The  leading 2-sphere topology   contribution to  the (properly defined)  $Z_{\rm str}$  
  is  approximated  by the type IIB supergravity action (plus $\a'$-corrections).  
  In particular, in the maximally supersymmetric 
   $\N=4$   $SU(N)$ SYM case one  can  match the 
  leading  $N^2$ term  in the free   energy 
   $F= 4 \aa \log ( \La \, \rr)+ {\ff}_0, \  \aa= \four ( N^2 -1)$,  
  with the leading  term in the supergravity  action proportional  to the (IR divergent)   volume of \ads 
(reproducing, in particular,  the conformal anomaly  \ci{Liu:1998bu, Henningson:1998gx}).
Here $\La$ is a UV  cutoff, $\rr$ is  the radius of $S^4$   and $\ff_0$  is a  regularization scheme dependent constant
(cf. \rf{a1}).  In  the particular scheme
selected by the localization matrix model representation for  the gauge-theory partition function 
 $Z= e^{-F}$  
(with  the $\l$-independent measure) one finds that 
 $F^{\N=4}= - \frac{1}{2}\,(N^2-1) \log { \l}  $. Then       the  leading $N^2$  term  in $F^{\N=4}$  can be 
  matched  \cite{Russo:2012ay}  with  the on-shell value of the   supergravity 
   term in the string   effective action in 
 \adss (assuming  particular  IR  cutoff in the  ${\rm AdS}_5$ 
  volume).\foot{On the 
\ads side the IR cutoff  $\ell $  is  measured in units  of  the  \ads radius $L$    and is related to  the product of 
the radius  $\rr$ of $S^4$ 
and UV cutoff $\Lambda$ as  $ {\rm r } \Lambda = {L {\ell} \ov \a'} = \sql { {\ell} \ov L}$ \cite{Russo:2012ay}.
Then the   regularized  \ads  volume (with  power  $\ell^n$  divergences  dropped) 
scales as $\log { {\ell} \ov L} \to -  \log \sql + \log (\Lambda {\rr })$, 
suggesting  that  $F=  4 \aa \log ( \Lambda \rr) + ... \to - 2 \aa \log \lambda+ ...$.}
  The  subleading $ \frac{1}{2}\, \log { \l}$ term  should come from the  1-loop   (torus) 
    contribution  to $Z_{\rm str}$, which is again proportional to the regularized  \ads 
    volume   and   receives contributions only from short multiplets, i.e. 
    is the same as the 1-loop supergravity correction \ci{Beccaria:2014xda}. 

The  localization matrix model result 
for the large $N$,  large $\l$ expansion  of the free energy of the $SU(N) $ \FA  model 
in \rf{12}--\rf{19} may be   written as 
 \ba
\la{128}
F(\lambda; N) \stackrel{\lambda\gg 1}{=} & - \tfrac{1}{2}\,N^2 \log { \l}   +  N\big(f_1\, \l + f_2\, \log \l +  f_3 + ...\big) \no\\
&+ \big( p_1\,    \lambda^2 +  p_2 \l +  k_1 \l^{1/2} +  k'_2\, \log \l   + k'_3 + ...  \big)
+ \mc O \big(\tfrac{1}{ N}\big)    \ , \ea
where $k'_2=  k_2  + p_3, \ k'_3= k_3 + p_4$.  The  leading $1/N$ terms in the $Sp(2N)$  \FA    case are similar 
(see \rf{1779},\rf{1.25},\rf{999}). 

  Let us  note that  in the $SU(N)$ case 
  the  $-2 \log \l$  term in \rf{128}  has 
   the  coefficient $ \frac{1}{4}\,N^2  +  \frac{1}{8}\,N  - \ha k'_2 $.
  In the $Sp(2N)$   case  the analog of this coefficient in \rf{1779}  is 
    $ 
    \ha N^2 + \ha N  - { 3 \ov 16} $. 
   Thus in both   cases   not only  the $N^2$  term 
   (as expected  from the  planar equivalence)\foot{\la{nin}In the  case of the 
   $\N=4 $ SYM   theory with the group $Sp(2N)$ which may be viewed as  an orientifold projection of $U(2N)$ theory
   and which  is    dual to  type IIB string 
   on AdS$_5 \times \mathbb{RP}^5$  \cite{Witten:1998xy} 
   the presence of the O3-plane (carrying  RR charge of $1\ov 4$) 
    leads to the effective  shift   of $N$  by $1\ov 4$   and thus to  the expression  $L^4 = 4 \pi \gs ( 2N + \ha)\a'^2$
    for the AdS radius. 
    As a result, one reproduces  both 
       leading $N^2$ and $N$  terms in the conformal anomaly  from
         the on-shell   value of  the 10d supergravity action \cite{Blau:1999vz,Giombi:2020kvo}.
      For example, the $\N=4$   $Sp(2N)$ SYM  free energy in \rf{662}  may 
      be   written as  $F= - N^2  \log \l  - \ha  N \log \l $
       or  as  $F=- {1 \ov 4}  \big[ (2N+\ha )^2 -  {1 \ov 4}  \big] \log \l $.
        From the flat space perspective, the   shift  $N \to N + \four $ may  be  equivalently 
         attributed 
         to the crosscup contributions (cf. \cite{Blau:1999vz}).  
       One  may also  interpret   the odd-power $1/N$  terms   
       in the  Wilson loop  expectation  value  of  the $\N=4$  $Sp(2N)$ theory \ci{Fiol:2014fla} (see \rf{7.17},
      \rf{7.18}) 
       as  coming  from the  crosscup contributions,  but  they   can also  be  formally 
       generated (at least in the large $\l$ expansion)  by  shifting  $ N \to N + \four$  
       in the   semiclassical string tension prefactor $e^{2\pi T}$
        ($2\pi T= \sql = {L^2\ov \a'}$ with $\gym^2= 2 \times 4 \pi \gs$, \ $\l= \gym^2 N$) 
       of the  even-power $1/N$ terms in \rf{yyy}
       (we thank S. Giombi for a discussion of this issue).}
    but also  the order $N$ term  
  is the  same  as in the a-anomaly coefficients 
   of the two theories (see  the tables below  eq. \rf{4}). At the same time, 
  the order $N^0$   coefficient  of $\log \l$  in the $Sp(2N)$ case   does not match the  one in the 
  conformal anomaly. This is   not surprising:
 as  discussed in Appendix \ref{A} below, in contrast to  what happens in the $\N=4$  SYM  case, 
   in the $\N=2$  theory  cases 
  there is no a priori reason   why  the $\log \l$  term in the  strong-coupling limit of the  free energy
  derived from the localization matrix model  should have the conformal a-anomaly as its coefficient.
  
Rewriting \rf{128}   in terms of the  dual  string theory  coupling  and string tension  as defined in \rf{2}    we get 
  (renaming coefficients  to absorb  factors of $2$  and $\pi$)\foot{In contrast to the $\N=4$ SYM case, in the $\N=2$  $Sp(2N)$ case we shall  assume that $N$  is not shifted in the definition of \ads   radius and string  tension
  and will also  ignore possible  extra factor of 2  in the relation between $\gs$ and $\gym^2$.}
 \ba
\la{131}
F(T, \gs) \stackrel{T\gg 1}{=} & -   {\pi^2  T^4\ov \gs^2}   \log ({2\pi   T})  
 +  {\pi  T^2\ov \gs} \big(f'_1\, T^2 +  f'_2\, \log T +  f'_3 + ...\big) \no\\
& +\big( p'_1\,  T^4  +  p'_2 T^2 +  k'_1 T +  k''_2\, \log T  + k''_3 + ...  \big) 
+ \mc O (\gs)    \ .
\ea
The leading (2-sphere)  term in  the tree-level 
string theory effective action   $  {1\ov \gs^2\a'^4 }\int d^{10} x \sqrt g\, (R+ ...)  $   evaluated on the AdS$_5 \times S'^5$  background   is  expected to  match the   $1\ov \gs^2$   term in \rf{131} 
 (after   using, as in the $\N=4$ SYM case  \cite{Russo:2012ay},  the    IR cutoff  related to $T$   in the AdS  volume).  
 
 The  $1\ov \gs$ term  in \rf{131} should come from the disc    contribution, and,   
 in the $Sp(2N)$ case,     also    from  the  crosscup topology. 
 In particular,  one may expect  the  ${T^2\ov \gs}\log T$ term  to originate  from  the  curvature squared term 
 $  {1\ov \gs \a'^2 }\int d^{8} x \sqrt g\, R R $   in the  D7-brane  action (with  D7-brane   wrapping AdS$_5$ 
 and $S^3$ from $S'^5$). The background value of this term  is proportional to the \ads  volume  and thus
  after the same IR regularization  it  should give  the  ${T^2\ov \gs}\log T$  contribution. 
 In  \cite{Blau:1999vz} 
   the  $  {1\ov \gs \a'^2 }\int d^{8} x \sqrt g\, R R $    term  was  shown  to reproduce the order $N$ term in  the conformal 
  anomaly  of  the $Sp(2N)$ \FA model. 
   This is  consistent  with the above observation that the order $N$ term in 
   the coefficient  of the $\log \l$ in \rf{128} or $\log T$ in \rf{131}   is  the same as in the a-anomaly coefficient
   of the  corresponding $\N=2$   superconformal model. 

The  interpretation of the $T^4\ov \gs$ term in \rf{131}  is not immediately clear. Naively,   such term could come from the 
D7-brane tension, i.e. $  {1\ov \gs \a'^4 }\int d^{8} x \sqrt g\, $ but this term should cancel against the orientifold (crosscup) 
contribution (cf.  \cite{Schnitzer:2002rt}), so that  the leading  term in 
 the D7-brane action  should be  the above curvature-squared term. 
The order $\gs^0$ terms in \rf{131}  should come from the closed-string  (torus) and open-string (annulus or disc with crosscup)   1-loop    corrections.  Since the compact  $S'^5$  part 
  of the background is not smooth (orbifold  action has fixed points) 
they  may  originate   from  "localized"  contributions   (rather  than "extensive" contributions    proportional to  the  volume of 
AdS$^5 \times S'^5$   like terms in the local  part of the  string  effective action).

The  resummed  expression for  leading  strong coupling     
 terms in the  free energy of the $Sp(2N)$ theory \rf{1779},\rf{1.25} written in terms of the  string coupling and string tension 
 in \rf{2}  is (we use that $f_1= {\log 2 \ov 4 \pi^2}$)
  \be\la{91}
     F\stackrel{T\gg 1}{=}     {\pi^2  T^4\ov \gs^2} \Big[  \log \big ( 1 +\tfrac{2\log 2}{ \pi}  \gs  \big) + ... \Big] 
        -
         2 
         \Big({\pi^2 T^4 \ov \gs^2}   + {\pi T^2 \ov \gs}  
         + \frac{3}{16} \Big) \log ( 2 \pi T)  
          - \frac{\pi }{8\gs  }    +   \mc O (  e^{-2\pi T})   \ . \ee
        Remarkably, the  leading $\log$ term 
        (dots   stand for terms that are subleading in $1/T$ at each order in $\gs$)
         has non-trivial dependence only on the  string coupling.
        The special $ - \tfrac{\pi }{8\gs  }  $        term (that   also  depends only  on $\gs$)   should be  a  particular crosscup contribution. 
        The  exponential corrections   should  have a   world-sheet instanton   interpretation, i.e.    should be related to world sheets wrapping compact $S^2$ parts of $S'^5$   that are   non-contractable  and thus stable due to orbifolding
        (see  also         discussion in  section 6.3).

The   large $N$, large $\l$ 
 expansion of  the Wilson loop expectation values in the $SU(N)$  and $Sp(2N)$  \FA  models  
may be written as (see \rf{20},\rf{24},\rf{25},\rf{555}     and \rf{29},\rf{33},\rf{34})
\ba \label{126}
\WW\stackrel{\l\gg 1}{=}      &e^{\sqrt \l} \Big[ \te    N ( b_0  \l^{-3/4}  +  b_{01}  \l^{-1/4} +   ...)+ (b_1  \l^{3/4}  + b_{12} \l^{1/4} +  ...) \no \\ 
 &\qquad  +  \tfrac{1}{ N}\  ( b_2 \l^{9/4}  + b_{21}  \l^{5/4} + ...)  +  \mc O \big(\tfrac{1}{ N^2}\big)  \Big] \ . 
  \ea
  Expressed in terms of the string coupling and tension in \rf{2} the leading strong coupling terms in 
   \rf{126} become
    \be\la{127}
\WW \stackrel{T\gg 1}{=}      e^{2\pi T}  \Big(  b'_0  {T^{1/2}\over \gs}   + b'_1  T^{3/2}   +  b'_2  {\gs T^{5/2}}  + ...\Big)
=    {T^{1/2}\over \gs}  e^{2\pi T} \big( b'_0    + b'_1 \gs T   +  b'_2 \gs^2 T^2  + ...\big) \ . 
\ee
The computation of $\WW$  on the string side     should proceed   in a  similar way  as for the circular loop in 
 the \adss case \ci{Drukker:2000ep,Giombi:2020mhz}  (the minimal surface   ending  on a circle at the boundary of \ads is the same AdS$_2$ one). The  crucial difference  is the presence of  a  
new  open-string sector and thus   extra "disc with holes"  
 and also  (in the $Sp(2N)$ case)  "disc with crosscups"  diagrams,  in addition to the 
 "disc with handles" ones.  
 In the $SU(N)$ case  the  structure of subleading terms in \rf{126},\rf{127} 
 is different compared to the  $\N=4$ SYM case in \rf{1}. 
In particular, the    order $\gs^0$  term  in \rf{127}   should  correspond to   the annulus contribution  (with one boundary with Dirichlet  and one -- with Neumann  boundary conditions). 

The prediction \rf{132} for the  resummation of the leading  large $\l$ terms  in the $Sp(2N)$  theory 
is the following   specification  of  \rf{127} 
\be \la{1288}
\vev{\mc W} \stackrel{T \gg 1}{=}   
 \frac{T^{1/2}}{\pi\,   \gs} \, e^{ 2 \pi T} \,    e^{  -  8 \pi^2 f_1 \gs T  } + ...  
= 
\frac{T^{1/2} }{ \pi\, \gs}\,  \exp\big[ 2\pi T \big(1   - \tfrac{\log 2 }{\pi}  \,   \gs \big)\big]  + ...  
\ ,   \ee 
where we used \rf{555} and  $f_1= {\log 2 \ov 4 \pi^2}$ from \rf{14}. 
Note that the  structure in the exponent  that involves  a function of $1 + c\,   \gs$  is similar to the one  of the first log  term in 
the free energy in  \rf{91}. 
The expression \rf{1288}  may be compared with the  leading-order one 
in the case of, e.g.,  $SU(N)$   $\N=4$   SYM   theory \rf{xxx} (the $Sp(2N)$ result \rf{yyy} is similar, cf. footnote \ref{nin})
\be \la{129}
\vev{\mc W} \stackrel{T \gg 1}{=}  
\frac{T^{1/2}}{ 2\pi\,  \gs}  \exp\big[ {2\pi T+\tfrac{\pi}{12}\tfrac{\gs^{2}}{T}} \big] 
+...\ , \ee
that  should represent  the sum of    handle insertions on the disc \ci{Giombi:2020mhz}. 
Similarly, \rf{1288} should be  summing   up the   leading   crosscup insertions.

Finally, let us  note that the  exact in $\l$  differential relations  like \rf{22}, \rf{31}  between the $1/N$ corrections  to the  free energy and  the Wilson loop  expectation value  that we find from the  localization 
matrix model representation on the  gauge theory side 
appear  to be   very non-trivial on the dual string theory side  where $F$  and $\WW$   are computed
using quite   different  procedures. It would be interesting  to uncover their string theory interpretation. 

\

The rest of this paper is organized as follows. We shall first discuss the $SU(N)$ case. 
In Section 2  we shall review the structure of the matrix model representation for the  partition function 
of the   $\N=2$  superconformal   \FA theory. In Section 3  we shall  find the explicit representations for the leading non-planar corrections $F_1$ and $F_2$  to its free energy. 

In Section 4 we shall  discuss the matrix model representation for the  Wilson loop  expectation value $\WW$ 
and in Section 5 find the general relations   between the $1/N$ terms in $\WW$ and  the free   energy $F$. 
Section 6 will  contain the results  of the strong-coupling expansion of the $1/N$ terms in $\WW$ and $F$.
In particular, in Section 6.3  we shall discuss the structure of exponentially small $e^{- n \sql}$ corrections to the leading non-planar term in 
$F$, their resurgence properties and comment on their possible string theory interpretation. 

Section 7   will be devoted to a similar analysis  in  the $Sp(2N)$   \FA model: matrix model representation, 
structure of $1/N$ corrections to the  free  energy  and $\WW$  and strong-coupling expansions. 
This case turns out be much simpler than the $SU(N)$ one   and we are able to 
determine the  structure of the large $\l$  asymptotics of free energy in rather explicit way. 

In Appendix A we  will  review the  general structure  of the partition function  of $\N=2$ models  as described  by 
the localization  matrix model  and explain how  it encodes the information 
about the value of the conformal anomaly  a-coefficient of the $\N=2$ model. 
Appendix \ref{B} will contain  some details of derivation of the strong-coupling expansion of $F_1$ using 
Mellin transform.
In Appendix \ref{C} we  will discuss   the relation   between the $1/N$ 
  coefficients  in the Wilson loop and  in  the free   energy  in the case of the $Sp(2N)$ theory
  and their  large $\l$ asymptotics. 
  
  \paragraph{Note added in v3:}
  The exact values of the   several leading coefficients $k_{n}$ in (\ref{18}) were  recently found 
analytically  in \ci{Beccaria:2022ypy}  (and also using a refined numerical method in \ci{Bobev:2022grf}).
 In  particular,   $k_{1} = {1\ov 8} $. The estimate  of $k_1$  as $1\ov 2 \pi$   suggested  in  \cite{Beccaria:2021vuc}
 was based 
 on  an approximate analytic treatment of $\bar F_2$, i.e.  was not rigorous
 (see comments in footnote \ref{foot4}).


\section{Matrix model representation   for    $\N=2$    $SU(N)$  theory}

Using  supersymmetric localization, the   partition function  of an $\mc N=2$   gauge theory 
on a sphere $S^{4}$ of unit radius may be written as a matrix integral over the eigenvalues $\{m\}_{r=1}^{N}$
of a $N\times N$ hermitian traceless matrix $m$ \cite{Pestun:2007rz}  (see also Appendix \ref{A}) 
\ba
\la{2.3}
\hat Z &\equiv e^{-F} = \mathscr{N}\,\int \mc Dm\,e^{-S_{0}(m)- S_{\rm int}(m)}\ , \qquad S_{0}(m) = \frac{8\pi^{2}N}{\lambda}\,\tr m^{2}, \qquad \lambda = \gym^{2}\,N\ , \\
\mc Dm &\equiv \prod_{r=1}^{N}dm_{r}\,\delta\big(\sum^N_{s=1}m_{s}\big)\,\big[\Delta(m)\big]^2 \ , \qquad \qquad \Delta(m) = \prod_{1\le r<s\le N}(m_{s}-m_{s}).
\ea
The ``interacting action'' $S_{\rm int}(m)$  that vanishes in the 
$\mc N=4$ theory  is non-trivial for the $\mc N=2$ theories. 
We will neglect the instanton contribution since we are going to consider the $1/N$ expansion. 
In the case   of the  $\mc N=2$  model containing hypermultiplets  in the fundamental, symmetric and antisymmetric representations of $SU(N)$  (with numbers subject to the conformal invariance condition \rf{3})
 one finds  (see \eg \cite{Billo:2019fbi})
\ba
\la{2.5}
S_{\rm int}(m) = & \sum^N_{r=1}\Big[\nf\log H(m_{r})+\ns\log H(2m_{r})\Big]\no\\  &+\sum_{r<s=1}^N\Big[(\ns+\na)\,\log H(m_{r}+m_{s})-2\log H(m_{r}-m_{s})\Big],
\ea
where $H$ is given in terms of the Barnes G-function\footnote{Note that the exponential prefactor in 
the r.h.s.  of \rf{2.4} cancels in $S_{\rm int}$  in superconformal models  (with $\nf$  satisfying  \rf{3}).}
\ba
H(x) &= \prod_{n=1}^{\infty}\Big(1+\frac{x^{2}}{n^{2}}\Big)^{n}\, e^{-\frac{x^{2}}{n}} = e^{-(1+\gamma_{\rm E})\,x^{2}}
\,{\rm G}(1+ix)\,{\rm G}(1-ix) \ . \la{2.4}
\ea
We will  normalize the $\N=2$  partition function (\ref{2.3}) to its $\mc N=4$ SYM value. 
After scaling the matrix $m\to a$ according to 
\be
a = \sqrt\frac{8\pi^{2}N}{\lambda}\,m\ ,\la{2.44}
\ee
the normalized partition function of the \model  in \rf{4}  ($\nf=4,\, \ns=0,\, \na=2)$  may be written as 
\ba
\la{2.8}
Z  &=  \vev{e^{-S_{\rm int}(a)}} = \int Da\,e^{-\tr a^{2}}\ e^{-S_{\rm int}(a)}\ ,\qquad\qquad 
 \int Da\,  e^{-\tr a^{2}}=1\ ,
\\
\la{2.9}
S_{\rm int}(a)  &\equiv \SS_1 + \SS_2  =\sum_{i=1}^{\infty}B_{i}(\lambda)\ \tr\Big(\tfrac{a}{\sqrt N}\Big)^{2i+2} +  \sum_{i,j=1}^{\infty}C_{ij}(\lambda)\ \tr\Big(\tfrac{a}{\sqrt N}\Big)^{2i+1}\tr\Big(\tfrac{a}{\sqrt N}\Big)^{2j+1}\ , 
\\
 \la{2.11}
 B_{i}(\l) &= 4\,\big( \frac{\lambda}{8\pi^2} \big)^{i+1}\frac{(-1)^{i}}{i+1}\zeta_{2i+1}(1-2^{2i})\ ,  
  \\
 \la{2.10}
 C_{ij}(\l) &= 4\,  \big( \frac{\lambda}{8\pi^2} \big)^{i+j+1}\,(-1)^{i+j}\,\zeta_{2i+2j+1}\frac{\Gamma(2i+2j+2 )}{\Gamma(2i+2)\, \Gamma(2j+2)} \ ,  \ea
 where $\zeta_{2i+1}\equiv \zeta(2i+1)$ are the Riemann $\zeta$-function values.
 
$Z$ in \rf{2.8}  is related to  the free energy  as 
\be
\la{2.12}
Z = e^{-\Delta F},\qquad \qquad  \Delta F =  F^{\mc N=2} -F^{\mc N=4} \ , \qquad \qquad 
F^{\mc N=4}= - \ha  (N^2 -1)\log \l\ .
\ee
Expanding $\Delta F$ at large $N$ we find  that the leading $N^2$ term   cancels due to planar equivalence\foot{Note, in particular,  that at large $N$  the  number of  hypers in 2 antisymmetric representations  $ 2 \times {N  (N-1)\ov 2} \approx N^2$ 
is the same as in the adjoint representation $N^2-1\approx N^2$.}  so that 
\be
\la{2.14}
\Delta F(\l) = N\, F_{1}(\l)+ F_{2}(\l) +\mc O\te ( { 1 \ov N})\ .
\ee
The order $N$ term  was absent in the case of the SA-orientifold  \ci{Beccaria:2021vuc} where $\nf=0$. 

The  weak coupling expansions  of $F_1$ and $F_2$ 
 are readily computed   by doing the  matrix model integrals in \rf{2.8}
 (here we set  $\hat\lambda = \frac{\lambda}{8\pi^{2}}$)
\ba
\la{2.15}
  F_{1} =&\te 3 \zeta _3 \hat{\lambda }^2-\frac{25}{2} \zeta _5 \hat{\lambda 
}^3+\frac{441}{8} \zeta _7 \hat{\lambda }^4-\frac{1071}{4} \zeta _9 
\hat{\lambda }^5+\frac{11253}{8} \zeta _{11} \hat{\lambda 
}^6-\frac{250965}{32} \zeta _{13} \hat{\lambda 
}^7 \lp\te 
+\frac{11713845}{256} \zeta _{15} \hat{\lambda 
}^8-\frac{53105195}{192} \zeta _{17} \hat{\lambda 
}^9+\frac{1100738457}{640} \zeta _{19} \hat{\lambda }^{10}+\cdots, \\
  F_{2} =&\te 5 \zeta _5 \hat{\lambda }^3-  \big(\frac{81}{2} \zeta _3^2+\frac{105 }{2}\zeta 
_7\big) \hat{\lambda }^4+(540 \zeta _3 \zeta _5+441 \zeta _9) 
\hat{\lambda }^5-    \big(1900 \zeta _5^2+\frac{6615}{2} \zeta _3 \zeta _7+  3465 \zeta _{11}\big) \hat{\lambda }^6  \lp\te 
+(24150 \zeta _5 \zeta 
_7+20655 \zeta _3 \zeta _9+\frac{212355 }{8}\zeta _{13}) \hat{\lambda 
}^7- \big(\frac{5044305}{64} \zeta _7^2 +\frac{1238895}{8} \zeta _5 \zeta _9+\frac{2126817}{16}  \zeta _3 \zeta _{11}   \lp\te  
\quad +\frac{6441435}{32} \zeta _{15} \big) \hat{\lambda }^8 +\big(\frac{500}{3} \zeta _5^3 +\frac{4125555 
}{4}\zeta _7 \zeta _9+1016400 \zeta _5 \zeta _{11}+\frac{1756755 }{2}\zeta _3 \zeta _{13} +\frac{12167155}{8} \zeta _{17} \big) \hat{\lambda 
}^9   \lp\te
-\big(5250 \zeta _5^2 \zeta _7+  \frac{54846477}{16} \zeta _9^2 +\frac{110007513 }{16}\zeta _7 \zeta _{11} +\frac{13635765 
}{2} \zeta _5 \zeta _{13}   \lp\te \ \ \ \ \ 
+\frac{189764289 }{32}\zeta _3 \zeta _{15} +\frac{91869921}{8} \zeta _{19} \big) \hat{\lambda }^{10}+\cdots\ . \la{216}
\ea
We shall see  that as in the case of the \oldmodel in \cite{Beccaria:2021vuc}, 
the  large $N$ expansion of the  BPS Wilson loop expectation value can be expressed
in terms of $F$, so it is   important  to study the latter first. 

\def \del {\partial}

 \section{Explicit   representation  for free energy corrections $F_1$ and  $F_2 $}
 \la{sec:F1F2}

Following the same strategy as  in \cite{Beccaria:2021vuc} we can find the  explicit representations of the 
leading and next-to-leading terms in the $1/N$ expansion of the free energy \rf{2.14}. 
To this aim, let us  introduce the generating function
 \ba
 X(\eta, \chi) = & \int Da\, e^{-\tr a^{2}}\,e^{V(\eta, \chi, a)}\equiv \vev{e^{V}} \ , \la{133} \\
  V(\eta, \chi, a) = & \sum_{i=1}^{\infty}\eta_{i}\,\tr\big(\tfrac{a}{\sqrt N}\big)^{2i+1}+
 \sum_{i=1}^{\infty}\chi_{i}\,\tr\big(\tfrac{a}{\sqrt N}\big)^{2i+2} \ . \la{331}
 \ea
 Expanding in  powers of the  "sources"  $\eta_i, \chi_i$   and evaluating the integrals over $a$ gives
 \ba
 \log X(\eta, \chi) =&\te  N\,\big(\frac{1}{2}\chi_{1}+\frac{5}{8}\chi_{2}+\cdots\big) 
 +\big(\frac{3}{16}\eta_{1}^{2}+\frac{15}{16}\eta_{1}\eta_{2}+\frac{5}{4}\eta_{2}^{2}+\frac{63}{32}\eta_{1}\eta_{3}+\frac{175}{32}\eta_{2}\eta_{3}+\frac{1575}{256}\eta_{3}^{2}+\cdots\big)  \lp\te 
 +\big(\frac{9}{8}\chi_{1}^{2}+\frac{9}{2}\chi_{1}\chi_{2}+\frac{75}{16}\chi_{2}^{2}+\cdots \big)+\mc O({1\ov N}) \no \\
 = &N\,R_{i}\chi_{i}+Q_{ij}\eta_{i}\eta_{j}+\widetilde Q_{ij}\chi_{i}\chi_{j}+\mc O(\tfrac{1}{N})\ , \la{324}
 \ea
 where we  assume summation over $i,j=1, ..., \infty$. 
 The linear in $\chi$ terms in \rf{324}  have the following general form 
 \ba
R_{i}\chi_{i} &= N^{-1} \sum_{i=1}^{\infty}\chi_{i}\ \veva{\tr\big(\tfrac{a}{\sqrt N}\big)^{2i+2}} =
 \sum_{i=1}^{\infty}\chi_{i}\, \frac{1}{2^{i+1}(i+2)} \binom{2i+2}{i+1},
 \ea
where  the coefficient $R_i$  may be written  as 
\be
\la{3.4}
R_{i} = \frac{2^{i+1}\,\Gamma(i+\frac{3}{2})}{\sqrt\pi\,\Gamma(i+3)}\ .
\ee
The infinite-dimensional  matrices $Q$ and $\wt Q$ in \rf{324}  can be expressed in terms of the connected 
  correlators of $\tr   a^n$
 (see \eg \cite{Beccaria:2020hgy}; here $\vev{AB}_{c}\equiv \vev{AB}-\vev{A}\vev{B}$)
\ba
\la{3.5}
\vev{\tr a^{2k_{1}+1}\, \tr a^{2k_{2}+1}} &= N^{k_{1}+k_{2}+1}\frac{2^{k_{1}+k_{2}+1}\,k_{1}\,k_{2}\Gamma(k_{1}+\frac{3}{2})\Gamma(k_{2}+\frac{3}{2})}{\pi\,(k_{1}+k_{2}+1)\Gamma(k_{1}+2)\Gamma(k_{2}+2)},\\
\vev{\tr a^{2k_{1}}\, \tr a^{2k_{2}}}_{c} &= N^{k_{1}+k_{2}}\frac{2^{k_{1}+k_{2}}\Gamma(k_{1}+\frac{1}{2})\Gamma(k_{2}+\frac{1}{2})}{\pi\,(k_{1}+k_{2})\Gamma(k_{1})\Gamma(k_{2})}.
\ea
The matrix $Q_{ij}$ is same as the one that appeared in the case of the \SA in   \cite{Beccaria:2021vuc} 
\be\la{3.7}
Q_{ij} =\frac{1}{\pi} \frac{2^{i+j}\,i\,j\,\Gamma(i+\frac{3}{2})\,\Gamma(j+\frac{3}{2})}{(i+j+1)\,\Gamma(i+2)\,\Gamma(j+2)}\ , 
\ee 
while for $\widetilde Q_{ij}$ we find 
\be
\la{3.8}
\widetilde Q_{ij} = \frac{1}{\pi}\,\frac{2^{i+j+1}\Gamma(i+\frac{3}{2})\Gamma(j+\frac{3}{2})}{(i+j+2)\,\Gamma(i+1)\Gamma(j+1)} = \frac{2\,(i+1)(j+1)(i+j+1)}{i\,j\,(i+j+2)}\,Q_{ij}.
\ee
Using (\ref{2.9}), the leading terms in the large $N$ expansion of the free energy $\Delta F$ in (\ref{2.14}) may
then  be represented as 
\ba
e^{-N F_{1}- F_{2}} &= e^{-C_{ij}\frac{\partial}{\partial\eta_{i}}\frac{\partial}{\partial\eta_{j}}-B_{i}\frac{\partial}{\partial\chi_{i}}}\,X(\eta,\chi)\Big|_{\eta=\chi=0} \lp
= e^{-C_{ij}\frac{\partial}{\partial\eta_{i}}\frac{\partial}{\partial\eta_{j}}-B_{i}\frac{\partial}{\partial\chi_{i}}}\,e^{N\,R_{i}\chi_{i}+Q_{ij}\eta_{i}\eta_{j} +\widetilde Q_{ij}\chi_{i}\chi_{j}}\Big|_{\eta=\chi=0}\ , \la{399}
\ea
where $B_i(\l) $ and $C_{ij}(\l) $ were defined in \rf{2.11},\rf{2.10}. 
  To compute \rf{399}  we may use  that 
\be e^{-B_i \del_i }  f(\chi_i ) =  f(\chi_i-B_i) \ , \qquad \qquad 
e^{- C_{ij} \del_i \del_j } = 
\int dy\,  e^{-{1\ov 4}  C^{-1}_{ij}  y_iy_j  +  y_i \del_i} \ . \la{400}\ee
This leads to an explicit weak coupling expansion of the leading large $N$ correction to the free energy:
\ba
 F_{1} &= \sum^\infty_{i=1} R_{i}\, B_i = -\frac{1}{\sqrt{\pi}}\sum_{i=1}^{\infty} \frac{(-1)^{i}}{(i+1)}\frac{\Gamma(i+\frac{3}{2})}{\Gamma(i+3)}\,(1-2^{-2i}) \, \zeta_{2i+1}\,\Big(\frac{\lambda}{\pi^2}\Big)^{i+1}\ . 
\la{3.10}
\ea
This weak coupling expansion is clearly convergent, with radius of convergence $\pi^2$. It can be summed up   into an integral representation using the identity:
\ba
(1-2^{-2i})\,  \zeta_{2i+1}=\frac{1}{\Gamma(2i+1)} \int_0^\infty dt\, \frac{t^{2i}}{e^t+1}\ .
\la{eq:zeta}
\ea
This  leads  to  the compact expression
\ba
F_1(\l)=\frac{2}{\sqrt{\lambda}} \int_0^\infty dt\,  \frac{e^{2\pi t}}{(e^{2\pi t}+1)^2}\Big[ \frac{J_1(2t\,\sqrt{\lambda})-t\, {\sqrt \lambda} +\frac{1}{2} (t \sqrt\lambda)^3}{t^2}\Big] \ . 
\la{eq:F1}
\ea
It is straightforward to check that the expansion of the Bessel $J_1$  function, combined with the identity (\ref{eq:zeta}), leads to the weak coupling expansion in (\ref{2.15}) and  (\ref{3.10}). However, the integral representation (\ref{eq:F1}) can also be used to analyze the strong coupling expansion, which is an asymptotic expansion, in contrast to the convergent weak coupling expansion (\ref{3.10}). The strong coupling expansion is discussed below in Section \ref{sec:sun-large}.

The next subleading correction to the free energy, the $O(N^0)$ term $F_{2}$ in (\ref{2.14}), may be naturally  split as 
\be 
F_2(\lambda) = \bF_2(\lambda) + \tF_2(\lambda) \ , \la{3.11}\ee
where $\bF_2$   comes from  the   $Q_{ij}\eta_i\eta_j$  part of \rf{399}   (i.e.   depends on $C_{ij}$ and $Q_{ij}$).
This $\bF_2$ part is identical to  the one   for the \oldmodel 
found  in \ci{Beccaria:2021vuc}  and    can be written as  
\ba
\bar  F_{2} (\lambda)&=\frac{1}{2}\log \det (1+ 4 C Q)  = \frac{1}{2}\log \det (1+M)\ ,\la{3.12}\\
M_{ij} &= 8\,\sqrt{2i+1}\sqrt{2j+1}\,\sum_{k=0}^{\infty} (-1)^{k}\,c_{ijk}\,\zeta_{2i+2j+2k+1}\, \Big(\frac{\lambda}{16 \pi^2}\Big)^{i+j+k+1} \ ,\\
c_{ijk} &= \sum_{m=0}^{k}\frac{\Gamma(2i+2j+2k+2)}{\Gamma(m+1)\, \Gamma(2i+m+2)\, \Gamma(k-m+1)\, \Gamma(2j+k-m+2)}\ . 
\ea
The properties of the weak coupling and strong coupling expansions of $\bar F_{2}(\l)$ have been studied in detail in \ci{Beccaria:2021vuc}.

The second term in (\ref{3.11}), denoted $\tF_{2}(\lambda)$,  comes from the $\wt Q_{ij} \chi_i \chi_j $ part  of \rf{399} (cf.\rf{400})
$e^{-B_{i}\frac{\partial}{\partial\chi_{i}}}\,\,e^{\wt Q_{ij}\chi_{i}\chi_{j}}\big|_{\chi=0} = e^{\wt{Q}_{ij}B_{i} B_{j}} 
 $. It can therefore be written as a double sum:
\be
\la{3.16}
 \tF_{2}(\lambda) = 
 -\sum^\infty_{i,j=1} \wt{Q}_{ij}B_{i} B_{j} \ ,  
\ee
where  the function $B_i(\l) $ was defined in \rf{2.11}  and  the coefficients 
$\wt Q_{ij}$ in \rf{3.8}  and we explicitly indicated summation over $i,j$. Thus, the weak coupling  series 
representation for  $\tF_{2}(\lambda)$ is (cf. \rf{3.10})
\ba
\la{3.19}
  \tF_2 (\l) &={1\ov \pi}  \sum_{i,j=1}^{\infty} \frac{(-1)^{i+j+1}\,(1-2^{-2i})(1-2^{-2j})\Gamma(i+\frac{3}{2})\Gamma(j+\frac{3}{2})}
{(i+j+2)\Gamma(i+2)\Gamma(j+2)}\zeta_{2i+1}\zeta_{2j+1}\Big( {\lambda\ov \pi^2}\Big)^{i+j+2}\ . 
\ea
Note that $ \tF_{2}(\l)$   is simpler than $\bF_2(\l)$, being  only quadratic in the zeta factors $\zeta_{2k+1}$, while $\bF_2(\l)$ involves sums over products of zetas to all orders. 
The weak-coupling  expansion of  the total  $F_2(\l)$ \rf{3.11} of course   agrees 
with the direct  expansion of $ F_{2}(\l)$ at weak coupling in  (\ref{216}).

Remarkably, there is a  direct differential relation between  $  \tF_2(\l)$ and $F_1(\l)$. 
Indeed, differentiating  $ \tF_2(\l)$ in \rf{3.19}  with respect to $\lambda$  we observe that 
  the double sum factorizes  in terms of the second derivative of the product $\l\,F_1(\l)$ with respect to $\lambda$, implying that 
\ba\la{3.20}
\frac{d}{d\lambda}  \tF_2 = -\frac{\lambda}{2}\Big[\frac{d^{2}}{d\lambda^{2}}(\lambda\, F_1)\Big]^{2}  \ .  \ea
Thus  the  form of $\tF_2(\l)$ is determined by that of  $F_1(\l)$. 
Using \rf{eq:F1}  we then get also 
\ba \la{321}
\frac{d}{d\lambda}  \tF_2 =2 \Big( \int_0^\infty dt\,  \frac{e^{2\pi t}}{(e^{2\pi t}+1)^2}\Big[J_1(2t\,\sqrt{\lambda})-t\sqrt{\lambda} \Big]\Big)^2 \ . 
\ea
This integral representation also permits a direct access to the strong coupling expansion of $\tF_2(\l)$.

\section{Wilson loop expectation value}

The $\mc N=2$ vector multiplet of the $\mc N=2$ theories
 contains the gauge vector $A_{\mu}$, a complex scalar $\varphi$, and two Weyl fermions.
The $\frac{1}{2}$-BPS Wilson loop depends only  on the fields of  the  vector multiplet   and is defined  as 
\be
\la{4.1}
\mc W = \tr\mc P\,\exp 
\oint\Big[i\,A_{\mu}(x) dx^{\mu}+\tfrac{1}{\sqrt 2}\big(\varphi(x)+\varphi^{+}(x)\big)\,ds\Big]\ , 
\ee
where the contour $x^{\mu}(s)$  represents  a circle of unit radius  and 
the  trace is  taken in the fundamental representation.
The expectation value of $\mc W$ may be  computed in the matrix model 
as   (\cf (\ref{2.8}))
\be 
\la{4.2}
\vev{\mc W} = \vev{ \tr e^{2 \pi m} }= \veva{ \tr  e^{\sqrt{\tfrac{\lambda}{2N}}\, a} }.
\ee 
Its large $N$ expansion  may be written as 
\be
\vev{\mc W} = N\,W_0(\l)+ W_1(\l) +\tfrac{1}{N}\Big( W_{0,2}(\l) +W_2(\l) \Big)+  \mc O\te ( { 1 \ov N^2}) \ ,   \la{4.3}
\ee
where we separated the $\mc N=4$ SYM parts 
\ba
&W_0\equiv \vev{\mc W}^{\mc N=4}_{0} = \tfrac{2}{\sql}I_{1}(\sql), \ \ \  \qquad W_{0,2} \equiv 
 \vev{\mc W}^{\mc N=4}_{2} = \tfrac{1}{48}\,\big[-12\sql\,I_{1}(\sql)+\lambda\,I_{2}(\sql)\big].
 \la{4.5} 
\ea
The  leading terms   in the weak-coupling expansions of the $\mc N=2$ parts $W_1$ and $W_2$ are found to be 
\ba
&W_1\equiv \vev{\mc W}_{1}^{\mc N=2} \ , \qquad \qquad  W_2\equiv \vev{\mc W}_{2}^{\mc N=2} \ \la{4.6}
, \\
\la{4.7}
W_1  = &\te  -\zeta_{3}\,\frac{3\lambda^{3}}{2\,(8\pi^{2})^{2}}\big(1+\frac{3 \lambda }{32}+\frac{\lambda ^2}{320}+\frac{\lambda 
^3}{18432}+\frac{\lambda ^4}{1720320}+\frac{\lambda 
^5}{235929600}+\frac{\lambda ^6}{44590694400}+\cdots\big)\notag \\
&\te  +\zeta_{5}\,\frac{75\lambda^{4}}{8\,(8\pi^{2})^{3}}\big(1+\frac{\lambda }{10}+\frac{\lambda ^2}{288}+\frac{\lambda 
^3}{16128}+\frac{\lambda ^4}{1474560}+\frac{\lambda 
^5}{199065600}+\frac{\lambda ^6}{37158912000}+\cdots\big)\notag \\
& \te -\zeta_{7}\,\frac{441\lambda^{5}}{8\,(8\pi^{2})^{4}}\big(
1+\frac{5 \lambda }{48}+\frac{5 \lambda ^2}{1344}+\frac{5 \lambda \
^3}{73728}+\frac{\lambda ^4}{1327104}+\frac{\lambda ^5}{176947200}+\cdots\big) + \cdots, \\
\la{4.8}
{W_2\ov \pi^2 W_0} 
= &\te -30 \zeta _5 \hat{\lambda }^4+(324 \zeta _3^2+420 \zeta _7) 
\hat{\lambda }^5-(5400 \zeta _3 \zeta _5+4410 \zeta _9) \hat{\lambda 
}^6  \lp\te  
+(22800 \zeta _5^2+39690 \zeta _3 \zeta _7+41580 \zeta _{11}) 
\hat{\lambda }^7  
-\big(338100 \zeta _5 \zeta _7+289170 \zeta _3 \zeta _9+
\frac{1486485 }{4}\zeta _{13} \big) \hat{\lambda }^8    \lp   \te  +
\big(\frac{5044305}{4} \zeta _7^2 +2477790 \zeta _5 \zeta _9 
+2126817 \zeta _3 \zeta 
_{11}+\frac{6441435 }{2}\zeta _{15}\big) \hat{\lambda }^9+\cdots.
\ea
Let us find the   closed  form of the series for the simpler  $W_{1}$ term that  is linear in $\zeta_{2n+1}$.
$W_1$  gets  contributions  from the single-trace term 
in (\ref{2.9}) that were absent in the case  of the \SA  in 
 \cite{Beccaria:2021vuc}.
 If we   write $S_{\rm int}$  in \rf{2.9} as  $\SS_1 + \SS_2 $ where 
 $\SS_1= 
  \sum_{i=1}^{\infty}B_{i}(\lambda)\ \tr\big(\tfrac{a}{\sqrt N}\big)^{2i+2} $ and $\SS_2$ is the double-trace term, 
   then  expanding  \rf{4.2}   to linear order in $\SS_1$    we   get 
\ba \vev{\mc W}  =
\frac{\int Da\, e^{-\tr a^{2}}  \,  e^{-\SS_1 -\SS_2 }\, \tr e^{\sqrt\frac{\lambda}{2N}a}\,}{\int Da\,e^{-\tr a^{2}} \,
e^{-\SS_1 -\SS_2 }  }\ \ \to\ \ 
 \frac{\veva{ (1-\SS_1)\, \tr e^{\sqrt\frac{\lambda}{2N}a}\,}}{\vev{1-\SS_1}} \ . 
\ea
Picking up  the part linear in $\SS_1$ gives 
\ba
W_1 &= -\veva{\tr \SS_1 \,  e^{\sqrt\frac{\lambda}{2N}a}}+\vev{\SS_1}\, \veva{\tr e^{\sqrt\frac{\lambda}{2N}a}} = -\veva{\SS_1\, \tr e^{\sqrt\frac{\lambda}{2N}a}}_{c} \lp
= -\sum_{p=0}^{\infty}\frac{1}{(2p)!}\Big(\frac{\lambda}{2N}\Big)^{p}\vev{\tr a^{2p}\ \SS_1}_{c}  = 
-\sum_{p=0}^{\infty}\frac{1}{(2p)!}\Big(\frac{\lambda}{2N}\Big)^{p}\sum_{i=1}^{\infty}B_{i}\, \veva{\tr a^{2p}\ \tr\Big(\frac{a}{\sqrt N}\Big)^{2i+2}}_{c}  \lp
= -4\sum_{p=0}^{\infty}\frac{1}{(2p)!}\Big(\frac{\lambda}{2N}\Big)^{p}\,\sum_{n=1}^{\infty}\Big(\frac{\lambda}{8\pi^{2}N}\Big)^{n+1}\frac{(-1)^{n}}{n+1}\zeta_{2n+1} (1-2^{2n})\,\veva{\tr a^{2p}\ \tr a^{2n+2}}_{c}\ . 
\ea
Using  (\ref{3.5}), we  then find 
\ba
\la{4.11}
W_1 &= 
 \frac{4}{\pi}\sum_{n=1}^{\infty}\sum_{p=0}^{\infty}\frac{(4\pi^{2})^{p}}{(2p)!}\,\frac{(-1)^{n}}{n+1}\zeta_{2n+1} (2^{2n}-1)
 \frac{\Gamma(p+\frac{1}{2})\Gamma(n+\frac{3}{2})}{(p+n+1)\Gamma(p)\Gamma(n+1)}\, \Big(\frac{\lambda}{4\pi^{2}}\Big)^{n+p+1}\  ,
 \ea
 which agrees with (\ref{4.7}).\foot{Let us note  that doing the sum over $p$ for each $n$ we obtain the 
 exact  form of the coefficients of  all $\zeta_{2n+1}$ terms
$$
W_1= -\zeta_{3}\,\frac{6\lambda^{2}}{2\,(8\pi^{2})^{2}}[2I_{2}(\sql)+I_{4}(\sql)]
 +\zeta_{5}\,\frac{15\lambda^{3}}{(8\pi^{2})^{3}}\Big[
5I_{2}(\sql)+4I_{4}(\sql)+I_{6}(\sql)
\Big]+\cdots
$$
matching Eq.~(3.29) of \cite{Billo:2019fbi}. 
}

Using the identity (\ref{eq:zeta}) we can resum this double series expansion into an explicit integral representation
\ba
W_1(\l)=&
2 \sqrt{\lambda } I_2(\sqrt{\lambda })  \int_0^\infty dt\,  \frac{e^{2\pi t}}{(e^{2\pi t}+1)^2} \Big[\frac{ J_1(2 t \sqrt{\lambda })}{4 t^2+1}-
  t \sqrt{\lambda}  \Big] \nonumber\\
   &+4 \sqrt{\lambda }  I_1(\sqrt{\lambda })  \int_0^\infty dt\,  \frac{e^{2\pi t}}{(e^{2\pi t}+1)^2} \frac{t\, J_2(2 t \sqrt{\lambda
   })}{4 t^2+1}\ . 
   \la{eq:W1}
\ea
It is straightforward to verify that the expansion of the Bessel  functions, combined with the identity (\ref{eq:zeta}), leads to the weak coupling expansion in (\ref{4.7}).

A closed  expression  for
  $W_2(\l)$  in \rf{4.3} will be given in the next section
  after  relating it to the corresponding terms  in the  free energy.

\section{General relations   between  the $1/N$ terms in $\vev{\mc W}$  and $F$}
\la{sec:rel}

The   coefficients  $W_1$ and $W_2$  in the large $N$ expansion (\ref{4.3}) 
of the Wilson loop  expectation value  turn out to 
have  close relation with the  $F_1$ and $F_2$   in the  free energy expansion (\ref{2.14})
(see also Appendix \ref{C}). 

To relate $W_1$ to $F_1$     let us  first 
write  (\ref{4.11}) as 
\ba
\la{5.1}
W_1 &= 
 -\frac{1}{\pi}\sum_{p=0}^{\infty} \frac{\l^{p}}{(2p)!}\, \frac{\Gamma(p+\frac{1}{2})}{\Gamma(p)}\,Y_{p}(\lambda)  = 
 -\frac{1}{\sqrt \pi}\sum_{p=0}^{\infty}\,\frac{{\lambda}^{p}}{4^p\, \Gamma(p)\Gamma(p+1)}\, Y_{p}(\lambda)\ , 
\\ \la{5.2}
Y_{p} (\l) &= \sum_{i=1}^{\infty}\,\frac{(-1)^{i+1}}{i+1}(1-2^{-2i})
 \frac{\Gamma(i+\frac{3}{2})}{(p+i+1)\Gamma(i+1)} \zeta_{2i+1} \Big({\lambda\ov \pi^2}\Big) ^{i+1} \ .
 \ea
We  notice that  differentiating \rf{5.1}   over  $\lambda$  leads to the expression   where 
the double sum factorizes.   Using   the expression for $F_1$  in \rf{3.10}    we   then  obtain 
\ba\la{5.4}
\frac{d}{d\lambda}W_1 &= \Big[-\frac{1}{\sqrt \pi}\sum_{p=0}^{\infty} \frac{{\lambda}^{p}}{4^p\, \Gamma(p)\Gamma(p+1)}\Big]\times 
\sqrt\pi\,  \frac{d^{2}}{d\lambda^{2}}(\l F_1)\ . 
\ea
This relation  may be written as  \be\la{5.5}
\frac{d}{d\lambda}W_1 = -\frac{1}{2}\sql\,I_{1}(\sql)\,\frac{d^{2}}{d\lambda^{2}}(\l F_1)\ .
\ee
Using also the expression for $W_0$ in  \rf{4.5}  we  conclude that 
\ba\la{5.6}
\frac{d}{d\lambda}W_1 &= -\frac{\lambda}{4}\,W_0\,\frac{d^{2}}{d\lambda^{2}}(\lambda\, F_1)\ . 
\ea
The 
 term $W_2$ in \rf{4.3}  turns out  to be   related to $ F_{2}$ in (\ref{2.14}),\rf{3.11}  by
\ba 
W_2 &= -\frac{\lambda^{2}}{4}\,W_0\frac{d}{d\lambda} F_2\ .\la{5.7}
\ea
 This can be  proved in the same  way as  in \ci{Beccaria:2021vuc}\foot{In 
 \ci{Beccaria:2021vuc}  we  used the notation    $ {W_2\ov W_0} =\Delta q $   and $F_2=\Delta F$.}
 by   expanding the Wilson loop factor  to leading order, using the large $N$ factorization of correlators
  and observing that  the insertion of $\tr a^2$  is the same as  the insertion of the Gaussian "action"  
 which,  in turn,  can be obtained by differentiating the matrix model integral over $\l$. 
 
Using that  in $F_2= \tF_2 + \bF_2$    and  \rf{3.20}   we may  represent  \rf{5.7} as 
\ba
W_2 &= \frac{\lambda^{3}}{8}\,W_0  \Big[\frac{d^{2}}{d\lambda^{2}}(\lambda\, F_1)\Big]^{2} 
-\frac{\lambda^{2}}{4}\,W_0\frac{d}{d\lambda} \bF_2 \ .   \la{5.8} \ea
In view of  (\ref{5.6}) the first term here is thus  related to the square of $\frac{dW_{1}}{d\lambda}$.

\section{{Strong coupling expansions of the $\N=2$    $SU(N)$ free energy and Wilson Loop}}
\label{sec:sun-large}

In this section we present results  for the large $\lambda$ expansions of the terms $F_{1}(\l)$ and $F_{2}(\l)$ in the large $N$ expansion (\ref{2.14}) of the free energy. Using the relations   \rf{5.6},\rf{5.7} 
these  will  also determine the expansion of the  terms $W_{1}(\l)$ and $W_{2}(\l)$ in the large $N$ expansion (\ref{4.3}) of the Wilson loop.

\subsection{Large $\lambda$ expansion of $F_{1}$ and $F_{2}$}

 The large $\lambda$ expansion of the first subleading large $N$ correction $F_{1}(\l)$ in (\ref{2.14}) for the free energy can be derived in several different but complementary ways. The simplest way is to use the representation 
\be\la{b6}
(1-2^{-2i})\, \zeta_{2i+1} = -\sum_{k=1}^{\infty}\frac{(-1)^{k}}{k^{2i+1}} \equiv \eta({2i+1}) \ .
\ee
where $\eta({2i+1})$ is the value of the Dirichlet $\eta$-function.
Then the expansion (\ref{3.10}) for $F_1$  yields 
\be
F_{1}(\l) = \sum_{k=1}^{\infty}\frac{(-1)^k}{4k}\Big[-\frac{\lambda }{\pi ^2}+\frac{8 k^4 \pi^{2}}{\lambda}  \Big(\sqrt{
1+\frac{\lambda }{\pi^{2}k^2}}-1\Big)-4 k^2 +8 
k^2 \log \Big( \ha  +\ha \sqrt{1+\frac{\lambda }{\pi^{2}k^2}}\Big)\Big].\la{b7}
\ee
Expanding at large $\lambda$  gives an expansion that can be evaluated using $\zeta$-function regularization
\ba\la{b8}
F_{1} \stackrel{\lambda\gg 1}{=} \sum_{k=1}^{\infty}(-1)^{k+1}\Big( \frac{\lambda}{4\pi^{2}k }+
 k \big[1+2 \log(2\pi k)
-\log \lambda \big]-\frac{4   \pi k^2}{\sqrt{\lambda }}+\frac{2   \pi ^2k^3 }{\lambda }+\cdots\Big)\ . 
\ea
Using  the $\eta$-function values
\be \te 
\eta(1)=\log2 \ , \quad \eta(-1)=\frac{1}{4}\ , \qquad   \eta'(-1)=-\frac{1}{4} -\frac{\log(2)}{3} + 3 \log A \ , \quad 
\eta(-2)=0, \quad  \eta(-3)=-\frac{1}{8} \ , \ee
where $A$ is Glaisher's constant,  we  thus obtain the strong coupling expansion
\ba
\la{6.2}
& F_1(\l) \stackrel{\lambda\gg 1}{=}  f_1 \l + f_2 \log \l   + f_3 +f_{4}\,\lambda^{-1}+\mc O\big(\lambda^{-1/4}\, e^{-\sqrt{\lambda}}\big)\ , \\
\la{6.3}
&f_1 =\te  \frac{\log 2}{4\pi^{2}}\ ,\ \ \qquad   f_2=-\frac{1}{4}\ , \ \ \qquad f_{3}=\frac{3}{4} +\frac{7}{6}\log 2 + \frac{1}{2}\log\pi
-6\log\mathsf{A}\ ,\ \ \qquad  f_{4} = -\frac{\pi^{2}}{4}\ .
\ea
Here  we indicated  that there is 
 only a finite number of power-law corrections:  as  will   be discussed below 
   in Section  \ref{sec63}  and Appendix \ref{B}, 
   all  further corrections 
 turn out to be  exponentially small as $\lambda\to +\infty$. 
 An indication of this  is that  
  all higher order corrections in \rf{b8}  have  coefficients  that are  expressed in terms of 
   $\eta$-function values  that  vanish. 
 
The strong coupling expansion (\ref{6.2})-(\ref{6.3}) for $F_1(\l)$ can be also  obtained from the integral representation (\ref{eq:F1}) using the Mellin transform method (see 
 Appendix \ref{B}), or by expanding the $\frac{e^{2\pi t}}{(e^{2\pi t}+1)^2}=\sum_{n=1}^\infty (-1)^{n+1} n \, e^{-(2n-1)\pi t}$ factor in the integral representation (\ref{eq:F1}) and integrating.

The $\bF_{2}$ part \rf{3.12} of $F_2$  in (\ref{3.11}) is same as in the  \oldmodel  and thus 
\cite{Beccaria:2021vuc} 
\be 
 \bF_2   \stackrel{\lambda\gg 1}{=}   k_1  \l^{1/2} + k_2 \log \l + k_3 + O(\lambda^{-1/2}) \ , \qquad \qquad 
k_1 =\te  {1\over 2 \pi} \ ,\ \ \ ...\ .   \la{6.4}
\ee
The strong coupling expansion of $  \tF_2$ in \rf{3.19} 
may be derived   directly   from \rf{3.20}   using \rf{6.2}\foot{Note that 
the value of the constant term  $p_4$  can not  be deduced  from   the differential relation \rf{3.20}
and requires separate   derivation using the method of Appendix \ref{B}  that gives 
$p_4= \frac{1}{16}+\frac{\log 2}{12}+\frac{\log \pi}{16}-\frac{3}{4}\log\mathsf{A}$.}
\ba
\la{6.5}
  \tF_2  \stackrel{\lambda\gg 1}{=}  &\ \ p_1\lambda^{2}+p_2\,\lambda  +p_{3}\log \lambda+p_4 +  
{  \mc O\big( \lambda^{5/4}\, e^{-\sqrt{\lambda}}\big)} \ , \\ 
\la{6.6}
p_1 = &- f_1^2     \ , \quad\ \ \ \ \  p_2 = - 2 f_1 f_2   \ , \quad\ \ \   \ \   \te p_3=-\ha f_2^2 \  ,\ \ \   ... \ , \ea
where  $f_{i}$  have  the values listed in (\ref{6.3}). { Notice that, as for $F_1(\l)$ in (\ref{6.2}), there is only a finite number of power law corrections, followed by exponentially suppressed terms, whose origin is discussed below in Section \ref{sec63}.}

\subsection{Large $\lambda$ expansion of $W_{1}$ and $W_{2}$}

Using the   relations   (\ref{5.6}), (\ref{5.7}), (\ref{5.8})  allows  us   to  find the  strong coupling expansions of $W_1$ and $W_2$ from those of $F_1$ and $F_2$.
In particular, from \rf{6.2} and the
 expansion of $W_0$ in \rf{4.5} 
\be 
W_0 \stackrel{\lambda\gg 1}{=}  \sqrt {\te { 2 \ov \pi}} \l^{-3/4} e^{\sqrt \l}\,  \Big( \te 1 - {3\ov 8 \sqrt \l} -{15\ov 128  \l} +...\Big) - i \sqrt {\te { 2 \ov \pi}} \l^{-3/4} e^{-\sqrt \l}\,  \Big( \te 1 + {3\ov 8 \sqrt \l} -{15\ov 128  \l} +...\Big)
\ , \la{610}
\ee
we find  (dropping  exponentially suppressed parts, cf. \rf{6.2}) 
\be
\frac{W_1}{W_0} = \te -f_1 \,\lambda^{3/2}+ {3\ov 2} f_1\l   - {1\ov 8} ( 3 f_1 + 4 f_2)  \l^{1/2} + \mc O (\l^0)    \ . \la{6.9}
\ee
Comparing \rf{6.4} and   \rf{6.5} 
 we   observe that $\tF_2$  dominates   over $\bF_2$   at the first  two  leading  orders of expansion  in  $\l \gg 1$.  
As a result, the dominant   contribution to $W_2$ comes from the first term in \rf{5.8} 
\be 
\Big[\frac{W_2}{W_0} \Big]_1\equiv  \frac{\lambda^{3}}{8}\, \Big[\frac{d^{2}}{d\lambda^{2}}(\lambda\, F_1)\Big]^{2} 
  \stackrel{\lambda\gg 1}{=}\  
\te     \ha f_1^2 \l^3  + \ha f_1 f_2\l^2 + \mc O (\l) \ , \la{6.10}
\ee
where 
we used  \rf{6.2}. 
The contribution to \rf{6.10} coming from $\bF_2$ term in \rf{5.8} is 
\be \la{6.11} 
\Big[\frac{W_2}{W_0} \Big]_2 \equiv  -\frac{\lambda^{2}}{4}\,\frac{d}{d\lambda} \bF_2    \stackrel{\lambda\gg 1}{=}  
{\te  - {1\ov 8} k_1 \l^{3/2} -  {1 \ov 4}   k_2 \l }+ \mc O (\l^{1/2})  \ , 
 \ee
so that in total 
\be \la{6.12}
\frac{W_2}{W_0}  = \Big[\frac{W_2}{W_0} \Big]_1  + \Big[\frac{W_2}{W_0} \Big]_2    \stackrel{\lambda\gg 1}{=}\ {\te   \ha f_1^2  \,\lambda^{3}  
  + \ha f_1 f_2\  \l^2 - {1\ov 8} k_1 \l^{3/2} } + \mc O (\l)  \ , 
\ee
where the values of $f_1,f_2$ and $k_1$ are given in \rf{6.3},\rf{6.4}.

\subsection{Exponentially suppressed corrections at large $\l$  \la{sec63}}

The leading   large $N$  correction to the free energy  $F_1(\l)$ has, in addition to the 
"perturbative"  terms in \rf{6.2}, also
 exponentially suppressed corrections in the 
 large $\l$ limit. 
 These 
 can be computed directly from the integral representation (\ref{eq:F1}). 
It is actually slightly simpler to begin with the combination $\frac{d^{2}}{d\lambda^{2}}(\lambda F_{1})$ which appears in the relation to $W_1$ as in (\ref{5.5}). From the integral representation (\ref{eq:F1}) we deduce that 
\ba
\la{6.15}
\frac{d^{2}}{d\lambda^{2}}(\lambda F_{1}) &= -\frac{2}{\sql}\int_{0}^{\infty}dt\,\frac{e^{2\pi t}}{(e^{2\pi t}+1)^{2}}\big[J_{1}(2t\sql)-t\sql\, \big] \lp
= \frac{\log 2}{2\pi^{2}}-\frac{1}{4\lambda}+\frac{2}{\pi^{2}}\sum_{n=0}^{\infty}\Big[K_{0}\big((2n+1)\sql\big)+\frac{K_{1}\big((2n+1)\sql\big)}{(2n+1)\sql}\Big] \ . 
\ea
Both these expressions are exact, but the first expression in terms of Bessel $J$-functions is well suited to a small $\lambda$ expansion, while the second expression in terms of Bessel $K$-functions is well suited to a large $\lambda$ expansion.
As $\l\to +\infty$  each Bessel $K$-function in \rf{6.15}  is  
given   by the exponentially small factor  $e^{-(2n+1)\sqrt{\lambda}}$, multiplied by an asymptotic series in $1\ov \sql$. Thus we obtain an expansion in the form of an "instanton sum", with each exponential multiplied by a "fluctuation expansion" in inverse powers of $\sqrt{\l}$:
\ba
\la{6.15b}
\frac{d^{2}}{d\lambda^{2}}(\lambda F_{1}) 
\stackrel{\lambda\gg 1} {=}  \frac{\log 2}{2\pi^2}-\frac{1}{4\lambda}
+\frac{\sqrt{2}}{\pi^{5/2}}\sum_{n=0}^\infty \frac{e^{-(2n+1)\sqrt{\lambda}}}{\sqrt{(2n+1)\sqrt{ \lambda}}}
\sum_{k=0}^\infty \frac{(-1)^k \left(k^2+\frac{3}{4}\right) \Gamma\left(k+\frac{1}{2}\right)\Gamma\left(k-\frac{3}{2}\right)}{2^k\, \Gamma(k+1) \big[(2n+1)\sqrt{\lambda}\, \big]^k}
\ea
The reconstruction of $F_{1}(\l)$ from this expansion requires two integrations, and 
the integration constants are easily fixed by the comparison with \rf{6.2},\rf{6.3}. As a result,  we  find that $F_1$ in \rf{6.2}   may be represented as 
\be  F_1  \stackrel{\lambda\gg 1} {=} F_1^{\rm pol} + F_1^{\rm exp} \ , \qquad \qquad 
F_1^{\rm pol}  =  f_1 \l + f_2 \log \l   + f_3 +f_{4}\,\lambda^{-1} \ , \la{666}\ee
Here $F_1^{\rm pol}$  is the "polynomial"   in $\l\gg1$  part, with a finite number of nonzero coefficients $f_j$ as in (\ref{6.2})--(\ref{6.3}), and  $F^{\rm exp}_1$ is the 
   exponentially small contribution  given  by 
\ba 
F^{\rm exp}_1(\lambda) \stackrel{\lambda\gg 1} {=}  & -\frac{1}{\pi} \Big(\frac{2}{\pi}\Big)^{3/2} \sum_{n=0}^\infty \frac{1}{(2n+1)^2} \sum_{k=0}^\infty 
 \frac{(-1)^k \left(k^2+\frac{3}{4}\right) \Gamma\left(k+\frac{1}{2}\right)\Gamma\left(k-\frac{3}{2}\right)}{2^k\, \Gamma(k+1)}  \nonumber\\
 &\hskip 2cm \times \Big \{ \Gamma\Big(\tfrac{3}{2}-k, (2n+1)\sqrt{\lambda}\Big) -\frac{\Gamma\big(\tfrac{7}{2}-k, (2n+1)\sqrt{\lambda}\big) }{(2n+1)^2 \lambda} \Big\} \la{617}
 \ .
 \ea
 Here the   sum over $n$  looks like an "instanton" expansion:
 for  each $n$ and $k$
 the incomplete  $\Gamma$-function terms in  \rf{617}  are proportional   to 
  $e^{-(2n+1)\sqrt{\lambda}}$  when  $\lambda\to +\infty$.
 Using the expansions of these   $\Gamma$-functions   we find explicitly that 
 \ba
F_1^{\rm exp}(\lambda) \stackrel{\lambda\gg 1} {=}  2\Big(\frac{2}{\pi}\Big)^{3/2} \lambda^{-1/4} \sum_{n=0}^\infty \frac{e^{-(2n+1)\sqrt{\lambda}} }{(2n+1)^{5/2}} 
\sum_{l=0}^\infty 
 \frac{(-1)^l \big[ 4l(l+4)+3 \big] \, \Gamma\left(l+\frac{1}{2}\right)\Gamma\left(l-\frac{3}{2}\right)}{\pi\, 2^{l+2}\, \Gamma(l+1) \big[(2n+1)\sqrt{\lambda}\big]^l} \ . 
 \la{eq:F1exp}
\ea
 For each $n$, the fluctuation series is factorially divergent, but it is resurgent in the sense that the large $l$ behaviour is
encoded in the low $l$ terms. To see this explicitly, let us define the "fluctuation" coefficients from (\ref{eq:F1exp}):
\ba
c_l=
 \frac{(-1)^l \big[ 4l(l+4)+3 \big] \, \Gamma\left(l+\frac{1}{2}\right)\Gamma\left(l-\frac{3}{2}\right)}{\pi\, 2^{l+2}\, \Gamma(l+1)} \ . 
 \la{eq:cl}
 \ea
The first few  \underline{low-order}  values of $c_l$   are  given by 
 \ba
 c_l=
 \left\{1,\frac{23}{8},\frac{153}{128},-\frac{435}{1024},\frac{13755}{32768},-\frac{172935}{262144},\frac{5
   893965}{4194304},-\frac{126080955}{33554432}, 
\dots\right\}
   \la{eq:cls}
   \ea
 At large order, $l\to\infty$, these coefficients are alternating in sign and factorially divergent, and including the subleading corrections the large order behaviour can be written as:
    \begin{eqnarray}
   c_l\stackrel{l\to\infty}{=} \frac{(-1)^l}{\pi} \frac{\Gamma(l)}{2^l}\Big[1+\frac{2\cdot \frac{23}{8}}{(l-1)}+\frac{2^2\cdot \frac{153}{128}}{(l-1)(l-2)}+\frac{2^3\cdot (-\frac{435}{1024})}{(l-1)(l-2)(l-3)}+ \dots \Big]\ .
   \label{eq:large}
   \end{eqnarray}
      Notice that the numerators of the subleading corrections correspond precisely to the low order coefficients in (\ref{eq:cls}). The powers of 2 correspond to the difference between the two Bessel function saddles ($e^{-x}$  vs.  $e^{+x}$)  whose ratio is $e^{-2x}$. 
   Thus  we see that the subleading corrections to the \underline{large-order} growth of the fluctuation coefficients are 
   directly encoded in the \underline{low-order} fluctuation coefficients.  
 
  This  behaviour in (\ref{eq:large})  is the typical low-order/large-order resurgence relation  \ci{berry-howls,dunne-unsal,aniceto-basar-schiappa}.
      These resurgence properties are inherited from the large argument expansion of the Bessel function term in square brackets in the r.h.s. of (\ref{6.15}).
Furthermore, this resurgent behaviour of $F_1(\lambda)$ is inherited by the exponentially small  corrections to the Wilson loop  ratio $W_1(\lambda)/W_0(\lambda)$ in (\ref{6.9}), 
 due to the expression  \rf{5.6} relating $W_1(\lambda)$ to $F_{1}(\l)$.
Similar  exponential terms will appear in  the strong coupling expansion of $F_2$   and $W_2$    and also 
 in the  corresponding terms in  the $Sp(2N)$ theory case  discussed in the next section.

\

The  exponential $e^{- c \sql}$ corrections found here in  the $1/N$   term in  $\N=2$ free energy 
are  generally expected in observables  in   conformal  gauge  theory  with  an AdS string dual. 
The perturbative expansion (in inverse string tension)  in 2d string sigma model is expected  to be   asymptotic 
and such corrections may have a world-sheet theory origin (which may be different in different observables).
Similar   terms   appear, e.g., 
in the $\N=4$ SYM  theory in the large $\l$ expansion of the 
cusp anomalous dimension  (see \ci{Alday:2007mf,Basso:2009gh} and also 
 \ci{Aniceto:2015rua,Dorigoni:2015dha} for their  relation to  resurgence). 
 
 One may conjecture  that the $e^{-  (2k+1)  \sql}$ terms   in $F_1$  have a  string   instanton 
   interpretation in terms of world sheets
      wrapping  part of the compact internal space $S'^5$ 
   that has fixed points  under the orientifold/orbifold action on $S^5$ (see discussion in the  Introduction). 
   
 It is   useful to compare this  with what happens in the  case of the Wilson loop expectation 
 in $\N=4$  SYM theory (see \rf{1},\rf{4.3},\rf{4.5}).  The   large $\l$  expansion of the  Bessel  $I_1$ 
 function in $W_0$ in 
 \rf{4.5} leads to just two exponential terms in \rf{610}, with the subleading one  being imaginary
 (the same  pattern is found  also for higher $1/N$  terms in  $\WW$ in  \rf{1}).
 While  the leading $e^{\sql}$  term
 in \rf{610} represents  the expansion  near the minimal AdS$_2$  surface embedded in AdS$_5$,
  the second term may be interpreted\foot{An instanton interpretation  of this  second term was originally conjectured 
  in \ci{Drukker:2000rr}.} 
 \ci{Drukker:2006ga,Zarembo:2016bbk}
 as the contribution of an unstable  surface   wrapping $S^2$ of $S^5$.\foot{This   may be viewed as a limit of 
 the result found in the case of $1\ov 4$-BPS "latitude"  Wilson loop 
 where there are two solutions of disc topology  covering (in addition to AdS$_2$)  the  smaller  or bigger part of $S^2$ in $ S^5$.}
 Note that   higher order terms  $\sim  e^{-n\sql}$  do not appear,  as multiple  wrappings   would correspond 
 to multiply wrapped   Wilson loop. 
 
 In contrast, in the case of $F_1(\l)$ in the $\N=2$   theory we get  an infinite series of exponential terms 
 as here multiple wrappings  should be allowed\foot{To recall,  the $1/N$ correction  $F_1$   should  be given   by      string path integral over surfaces  of disc topology with free boundary.}   and they have  real coefficients
as the   corresponding  world-sheet  solutions  should be stable  due to orbifolding of $S^5$. 

Note that  the  appearance of 
  the imaginary term in  the formal large $\l$ expansion  of 
  $W_0$  is  related to fact  that the asymptotic expansion of the Bessel $I_1$ 
   function  about the dominant  $e^{\sqrt{\lambda}}$  term is  non Borel summable:  the coefficients 
   of the expansion about $e^{\sql}$ are factorially divergent and non-alternating in sign  and then  the  naive Borel summation integral  has an imaginary contribution, and this must be cancelled   against the $i e^{-\sql}$  term 
   as  total   $W_0$   should be real. 
 At the same time, the  exponentially small factors   $e^{-(2k+1)\sqrt{\lambda}}$   in $F_1$  are multiplied by asymptotic series that are Borel summable (note that the $c_l$ coefficients in \rf{eq:cl} 
 are factorially divergent but  alternate in sign)  and 
 therefore,  one finds only  real  exponentially suppressed  contributions. 

 In view of the relation \rf{5.6}   between $W_1$ and $F_1$    and the expansion of $W_0$ in \rf{610}
 the resulting  expression for the $1/N$   correction $W_1$  to the  Wilson loop  in the $\N=2$  
  theory will thus contain  two different   sources  of the subleading exponential   corrections since 
  \be\la{623}  {d\ov d\lambda} W_1 = -{1\ov 4} \l  W_0 { d^2 \ov d\l ^2}  ( \l F_1)  \sim    \big[w(\sql)\, e^{\sql}  + i w(-\sql) \, e^{-\sql}  \big] \sum^\infty_{k=0}  u_k(\sql)\, e^{-(2k+1) \sql}  \ . \ee
Thus, $ {d\ov d\lambda} W_1$   has a trans-series 
expansion involving an overall  $e^{\sqrt{\lambda}} $ factor, multiplied by 
even powers of $e^{-\sqrt{\lambda}}$. 
These alternate between being real and imaginary,\foot{From the string theory point of view, $W_1$ comes 
from  contributions of world sheets with annulus topology (with  one boundary  being  fixed by the Wilson loop 
 circle  and the other being free). Then the argument   about stability of   all wrappings of   subspace 
 in $S'^5$  (given above  for $F_1$   case)   should no longer  apply.}
 in such a way that the full trans-series is well-defined and real (as $W_1$ should be when $\lambda$ is real and positive). 
The same  structure also survives the  $\lambda$- integration that gives $W_1$. 
The resurgence properties of this final trans-series for $W_1$  would be interesting to study in more detail.\foot{An 
alternative approach is to start  directly with  the integral  representation for $W_1$ in \rf{eq:W1}
and perform the  large $\l$ expansion, getting both  perturbative and non-perturbative  contributions.}


\section{$\N=2$ superconformal $Sp(2N)$  theory}
Let us now  repeat similar analysis 
 in the case of the \FA  model \rf{55}  with the gauge group $Sp(2N)$. 

\subsection{Matrix model formulation}

The structure of the matrix model here is the same as in (\ref{2.3}). For the model 
 with  $\nad, \na$ and $\nf$   expressed in terms of them   using  the  finiteness condition \rf{5} 
the  interacting action in (\ref{2.3}) reads \cite{Fiol:2020bhf} (cf.  \rf{2.9}  and also Appendix \ref{A}) 
\ba
S_{\rm int}(a) = &\sum_{i=1}^{\infty}\Big(\frac{\lambda}{8\pi^{2}}\Big)^{i+1}\,\frac{(-1)^{i}}{i+1}\, \zeta_{2i+1}\, \Big\{ 2\,(2^{2i}-1)(\nad-\na-1)\,\tr \Big(\frac{a}{\sqrt N}\Big)^{2i+2} \no\\ & \qquad \ \qquad \qquad 
+\tfrac{1}{2}\,(\nad+\na-1)\sum_{k=1}^{i}{ \binom{2i+2}{2k}}\ \tr \Big(\frac{a}{\sqrt N}\Big)^{2i-2k+2} \ \tr \Big(\frac{a}{\sqrt N}\Big)^{2k}
\Big\},\la{71}
\ea
where the matrix $a$ is in the $2N$-dimensional fundamental representation of $Sp(2N)$. The expression \rf{71} 
 greatly simplifies for the \model 
where $\nad=0$, $\na=1$   (and $\nf=4$): only the single-trace term survives  so that (\cf (\ref{2.9}))\foot{There is no similar 
simplification  with  no double-trace terms in $S_{\rm int}$
in the $SU(N)$  case   \rf{2.5} (apart from "trivial"  $\N=4$ SYM case   where $S_{\rm int}=0$).}
\be\la{7222}
S_{\rm int}(a) =  \sum_{i=1}^{\infty} B_{i}(\lambda)\,\tr\Big(\frac{a}{\sqrt N}\Big)^{2i+2}\ , 
\ee
where $B_{i}$ is same as in (\ref{2.11}).

 Perturbative calculations are most efficiently performed by   the same methods as in \cite{Billo:2017glv} in  the $SU(N)$ case.
The matrix model variable is written in a basis 
of $\mathfrak{sp}(2N)$ generators in the fundamental representation with the following normalization
\be
a = \sum_{r=1}^{N(2N+1)}a^{r}\, T_{F}^{r}\ ,\qquad\qquad  \tr \big(T_{F}^{r}T_{F}^{s}\big) =\tfrac{1}{2}\delta^{rs}\ .
\ee
Then the   matrix model measure is simply
\be
Da = \mc N\prod_{r=1}^{N(2N+1)}da^{r} \ . \la{7237}
\ee
Integration is done with respect to  the Gaussian weight $e^{-\tr a^{2}}$  (\cf  (\ref{2.8})),  i.e.  it  reduces to repeated Wick contractions using $\vev{a^{r}a^{s}}=\delta^{rs}$
and the $Sp(2N)$ fusion/fission relations \cite{Cvitanovic:1976am,Huang:2016iqf}  
\ba
\la{7.7}
\tr (T^{a}\,M_{1}\,T^{a}\,M_{2}) &= \frac{1}{4}\tr M_{1}\ \tr M_{2}+\frac{1}{4}(-1)^{n_{2}}\tr (M_{1}\overline{M}_{2}), \\
\la{7.8}
\tr (T^{a}\,M_{1})\ \tr (T^{a}\,M_{2}) &= \frac{1}{4}\tr (M_{1}\,M_{2}) -\frac{1}{4}(-1)^{n_{2}}\tr (M_{1}\overline{M}_{2}),
\ea
where $M_{1}$ and $M_{2}$ are products of generators, $n_{2}$ is the number of factors in $M_{2}$, and $\overline{M}_{2}$ is the product in reverse order.
In particular, one finds   the following  useful correlators\footnote{Note that 
$\vev{ABC}_{c} = 
\vev{ABC}-\vev{A}\, \vev{BC}-\vev{B}\, \vev{AC}-\vev{C}\, \vev{AB}+2\vev{A}\, 
\vev{B}\, \vev{C}$, {\em etc.}}
\ba
\la{7.9}
& \vev{\tr a^{2n}} = \te N^{n+1}\,\frac{2^{1+n} \Gamma (\frac{1}{2}+n)}{\sqrt{\pi }\,\Gamma (2+n)}
\Big[1+\frac{n+1}{4N}+\frac{n(n^{2}-1)}{48N^{2}}+\frac{n(n^{2}-1)(n-2)}{192N^{3}}+\cdots\Big], \\
\no 
& \vev{\tr a^{2n}\ \tr a^{2m}}_{c} = \te N^{n+m}\,\frac{2^{n+m+1}\Gamma(n+\frac{1}{2})\Gamma(m+\frac{1}{2})}{\pi\,(n+m)\,\Gamma(n)\Gamma(m)} 
\,\Big[1+\frac{n+m}{4N}
 +\frac{(n+m)(1-2n-2m+n^{2}+nm+m^{2})}{48N^{2}}+\cdots\Big], \\
\no 
& \vev{\tr a^{2n}\ \tr a^{2m}\ \tr a^{2k}}_{c} = \te 
N^{n+m+k-1}\,\frac{2^{n+m+k+1}\Gamma(n+\frac{1}{2})\Gamma(m+\frac{1}{2})\Gamma(k+\frac{1}{2})}
{\pi^{3/2}\,\Gamma(n)\Gamma(m)\Gamma(k)}\Big(
1+\frac{n+m+k-1}{4N}+\cdots\Big)~, \\
& \vev{\tr a^{2n}\ \tr a^{2m}\ \tr a^{2k}\ \tr a^{2\ell}}_{c} = \te 
N^{n+m+k+\ell-2}\,\frac{2^{n+m+k+\ell+1}  \Gamma(n+\frac{1}{2})\Gamma(m+\frac{1}{2})\Gamma(k+\frac{1}{2})\Gamma(\ell+\frac{1}{2})}
{\pi^{2}\,\Gamma(n)\Gamma(m)\Gamma(k)\Gamma(\ell)}\no \\
& \qquad \qquad   \qquad \qquad \qquad\ \ \  \qquad \times (n+m+k+\ell-1) +\cdots~, \la{783} \\
&\vev{\tr a^{2n}\ \tr a^{2m}\ \tr a^{2k}\ \tr a^{2\ell} \ \tr a^{2s } }_{c}= \te 
N^{n+m+k+\ell + s -3}\,\frac{2^{n+m+k+\ell+ s +  1}  \Gamma(n+\frac{1}{2})\Gamma(m+\frac{1}{2})\Gamma(k+\frac{1}{2})\Gamma(\ell+\frac{1}{2}) \Gamma(s+\frac{1}{2}) }
{\pi^{2}\,\Gamma(n)\Gamma(m)\Gamma(k)\Gamma(\ell)   \Gamma(s )}\no \\
& \qquad \qquad  \qquad \qquad \qquad\ \ \  \qquad\times (n+m+k+\ell + s -1)  (n+m+k+\ell + s -2)  +\cdots~
\la{784}
\ea

\subsection{Free energy}

The free energy of the $Sp(2N)$ \model  has the same  structure of the 
$1/N$ expansion as in (\ref{2.14}), i.e.   after the subtraction of the $\N=4$ SYM  free   energy we have  (see \rf{662}) 
\be
\Delta F(\l) = N\,\FFF_{1}(\l)+\FFF_{2}(\l)+\tfrac{1}{N}\FFF_{3}(\l)+\tfrac{1}{N^2}\FFF_{4}(\l) +\tfrac{1}{N^3}\FFF_{5}(\l) 
+ \mc O(\tfrac{1}{N^{4}}) \ ,\la{79}
\ee
where we included two  more terms, compared to (\ref{2.14}).
To get the explicit expressions for the terms $\FFF_{1}(\l), \FFF_{2}(\l)$, and $\FFF_{3}(\l)$ 
we repeat the analysis in Section \ref{sec:F1F2}  (the  computation of $\FFF_4$  follows similar  steps). 
In this case we need   to  consider the analog of the generating function \rf{331}  containing only $\chi$-part 
 \be
 X(\chi) = \int Da\, e^{-\tr a^{2}}\,e^{V(\chi, a)}\ , \qquad 
 \qquad V(\chi, a) =  \sum_{i=1}^{\infty}\chi_{i}\,\tr\big(\tfrac{a}{\sqrt N}\big)^{2i+2}\ .\la{710}
 \ee
 Evaluating the integrals gives
 \ba
 \log X(\chi) = &
 \sum_{i=1}^{\infty} \veva{ \tr \big(\tfrac{a}{\sqrt N}\big)^{2i+2}}\,\chi_{i}
 +\tfrac{1}{2} \sum_{i,j=1}^{\infty}\veva{ \tr \big(\tfrac{a}{\sqrt N}\big)^{2i+2}\,\tr \big(\tfrac{a}{\sqrt N}\big)^{2j+2}}_{c}\,\chi_{i}\chi_{j}\lp
 +\tfrac{1}{3!} \sum_{i,j,k=1}^{\infty}\veva{ \tr \big(\tfrac{a}{\sqrt N}\big)^{2i+2}\,\tr \big(\tfrac{a}{\sqrt N}\big)^{2j+2}\,
 \big(\tfrac{a}{\sqrt N}\big)^{2k+2}}_{c}\,\chi_{i}\chi_{j}\chi_{k}+\cdots~.\la{712}
 \ea
 Using  (\ref{7.9})--(\ref{784}), this may be written as 
 \ba
 \log X(\chi) &= \RRR_{i}\chi_{i}+\UUU_{ij}\chi_{i}\chi_{j} +\TTT_{ijk}\chi_{i}\chi_{j}\chi_{k}+ \te O({1\ov N^{2}})  \ , \la{711}
 \ea
 where
 \ba
 \RRR_{i} = N\,\RRR_{i}^{(0)}+\RRR_{i}^{(1)}+\tfrac{1}{N}\RRR_{i}^{(2)}+\mc O(\tfrac{1}{N^{2}})\ , &\quad \qquad 
 \UUU_{ij} = \UUU_{ij}^{(0)}+\tfrac{1}{N}\UUU_{ij}^{(1)}+\mc O(\tfrac{1}{N^{2}})\ , \no \\
 \TTT_{ijk} = \tfrac{1}{N}\,\TTT_{ijk}^{(0)}&+\mc O(\tfrac{1}{N^{2}}), \la{714}
 \ea
 and
 \ba
 \RRR_{i}^{(0)} &= \frac{2^{i+2}\Gamma(i+\frac{3}{2})}{\sqrt\pi\,\Gamma(i+3)} = 2\,R_{i}, \qquad\qquad 
 \RRR_{i}^{(1)} = \frac{i+2}{2}\,R_{i},\qquad\ \ \
 \RRR_{i}^{(2)} = \frac{i(i+1)(i+2)}{24}\,R_{i}, \notag \\
 \UUU_{ij}^{(0)} &= \frac{2^{i+j+2}\,\Gamma(i+\frac{3}{2})\,\Gamma(j+\frac{3}{2})}{\pi (i+j+2)\,\Gamma(i+1)\,\Gamma(j+1)} = 2\,\wt Q_{ij},
 \qquad \qquad 
 \UUU_{ij}^{(1)} = \frac{i+j+2}{2}\,\wt Q_{ij}, \la{7.14} \\
 \TTT_{ijk}^{(0)} &= \frac{2^{i+j+k+4}\Gamma(i+\frac{3}{2})\Gamma(j+\frac{3}{2})\Gamma(k+\frac{3}{2})}
{6\pi^{3/2}\,\Gamma(i+1)\Gamma(j+1)\Gamma(k+1)} = \tfrac{1}{3}(i+1)(i+2)(j+1)(j+2)(k+1)(k+2)\,R_{i}R_{j}R_{k},\no 
 \ea
 with  $R_{i}$  and $\wt Q_{ij}$  being  the same as in (\ref{3.4}) and (\ref{3.8}). 
 
 The free energy $\Delta F$ in \rf{79} is then obtained  by acting on $-\log X$
 with the operator $\exp(-B_{i}\frac{\partial}{\partial\chi_{i}})$ and setting $\chi_{i}\to 0$.\foot{Equivalently, 
 we just start with $\exp \big[-  \sum_{i=1}^{\infty} B_{i}(\lambda)\,\tr\big(\frac{a}{\sqrt N}\big)^{2i+2} \big]$ (cf. \rf{7222}), 
   compute its expectation value   expanding in powers of $B_i$ terms using the connected correlators in \rf{7.9}--\rf{784}
  and then rewrite the result as $e^{-\Delta F}$.  }
  This replaces $\chi_{i}\to -B_{i}$ (cf. \rf{400})  and thus
\ba\la{716}
 \Delta F(\l) = \sum_{i=1}^{\infty} \,\RRR_{i}B_i -\sum_{i,j=1}^{\infty}\UUU_{ij}\, B_{i}B_{j} +\sum_{i,j,k=1}^{\infty} \TTT_{ijk}\, 
B_{i}B_{j}B_{k}\,+ \te O({1\ov N^{2}}) .
\ea
The $\FFF_{1}$ term  in \rf{79}  is then  simply 
\be\la{713}
\FFF_{1} (\l)= \sum_{i=1}^\infty    \RRR_{i}^{(0)} B_i  = 2\sum_{i=1}^\infty  \,R_{i}\, B_i = 2 F_{1}(\l),
\ee
where $F_{1}(\l)$ is the corresponding $SU(N)$ term in (\ref{3.10}).
Thus, $\FFF_1(\l)$ for the $Sp(2N)$ model also has an exact integral representation of the form in (\ref{eq:F1})  multiplied by  factor of $2$.

For the $\FFF_{2}$ term we obtain
 \ba
 \la{7.16}
 \FFF_{2}(\l) &= \sum^\infty_{i=1} \RRR_{i}^{(1)} B_{i}\,  -\sum^\infty_{i,j=1} \UUU_{ij}^{(0)} B_{i} B_{j} = \frac{1}{2}\sum^\infty_{i=1}(i+2)R_{i}\, B_i-2\,\sum_{i,j=1}^\infty \wt Q_{ij}B_{i} B_{j} \lp
 = \frac{1}{2}\frac{d}{d\lambda}\big[\lambda F_{1}(\l)\big]  +2\,\wt F_{2}(\l)\ , \qquad\qquad  \qquad \frac{d}{d\lambda}  \tF_2 = -\frac{\lambda}{2}\Big[\frac{d^{2}}{d\lambda^{2}}(\lambda\, F_1)\Big]^{2} \ , 
 \ea
 where $\tF_{2}(\l)$ is the same as in (\ref{3.16}), (\ref{3.19}),\rf{3.20}.
 
 We conclude that in this $Sp(2N)$ model  the $\FFF_{2}$ term is much simpler than in  the $SU(N)$ case in \rf{3.11} -- it  does 
   not contain the analog of the $\bar F_2$ term \rf{3.12}.
 In (\ref{7.16}), the first term is linear in the $\zeta_{2n+1}$-values, while the second is quadratic. The presence of this  first term 
 is related to the different structure of the  large $N$ expansion in  (\ref{7.9}) that contains the 
  $1/N$ term which was absent in the $SU(N)$ case.\footnote{Note,  for  example, that 
\be
\notag
\vev{\tr a^{6}} = \begin{cases}
\frac{5}{8 N^2} (N^{2}-1) (3-3 N^2+N^4) = \frac{5 N^4}{8}+0\times N^{3}-\frac{5 N^2}{2}+\cdots, & \ \ SU(N) \\
\frac{5}{32} N (1+2 N) (1+2 N+4 N^2) = \frac{5 N^4}{4}+\frac{5 N^3}{4}+\frac{5 N^2}{8}+\cdots, &\ \  Sp(2N)
\end{cases}\ .
\ee
}

Furthermore, since $F_1(\l)$ has a simple integral representation (\ref{eq:F1}), and $\tF_2(\l)$ is directly related to $F_1(\l)$ as in (\ref{3.20}), we see from (\ref{713}) and (\ref{7.16}) that in the $Sp(2N)$ model both $\FFF_{1}(\l)$ and $\FFF_{2}(\l)$ have explicit integral representations that permit precise analysis of both the convergent weak coupling expansion and the asymptotic  strong coupling expansion. This carries over to the Wilson loop corrections, as discussed  in the next  subsections.

Finally, from \rf{716} we conclude that the $1/N$ term $\FFF_{3}(\l)$ in \rf{79}  is given by 
\ba
&\FFF_{3}(\l) = \sum_{i=1}^{\infty}\RRR_{i}^{(2)} B_{i}\ -\sum_{i,j=1}^{\infty}\UUU_{ij}^{(1)} B_{i}B_{j} \ +\sum_{i,j,k=1}^{\infty}
\TTT_{ijk}^{(0)}\, B_{i}B_{j}B_{k} \no \\
&\quad =\tfrac{1}{24} \sum_{i=1}^{\infty} {i(i+1)(i+2)}\, R_{i}\, B_{i}-  \tfrac{1}{2} \sum_{i,j=1}^{\infty}({i+j+2})\, \wt Q_{ij}B_{i}B_{j}
+\tfrac{1}{3}\Big[\sum_{i=1}^{\infty}(i+1)(i+2)\, R_{i}\, B_{i}\Big]^{3}.\la{719}
\ea
Using that according to \rf{2.11}   we have  $B_{i}\sim \lambda^{i+1}$ and also the relation in \rf{3.20}, the expression  for $\FFF_3$  may be written  as (cf. \rf{7.16})  
\ba
\FFF_{3}(\lambda) = &\te  \frac{\l^{2}}{24}\big[ \l F_{1}(\l)\big]'''  +  \frac{\lambda}{2}\,\tF_{2}'(\l)+\frac{\l^{3}}{3}\Big( \big[\l F_{1}(\l)\big]''\Big)^{3} \ \no\\
= & \te \frac{\l^{2}}{24}\big( \l F_{1}\big)'''  - \frac{\l^2}{4}\big[ \big(\l F_{1}\big)''\big]^{2}  +\frac{2\l^{3}}{3!}\big[ \big(\l F_{1}\big)''\big]^{3}\ , 
 \la{720}
\ea
where  $f'(\l)\equiv {d\ov d \l} f(\l)$.

It is possible to generalize the above computation of $\FFF_3$   to the case of the next terms $\FFF_4$  and 
$\FFF_5$  in \rf{79}. 
The analog  of the last term  in \rf{720}    with highest number of powers of  derivatives over $\l$ or of 
 highest power in  $(\l F_1)''$     turns out to be (cf. \rf{719},\rf{783},\rf{784},\rf{7.14}) 
\ba
\FFF_{4} (\l) =  & -\tfrac{1}{4!}\sum_{i,j,k,\ell=1}^\infty   c_{ijk\ell} 
R_{i}R_{j}R_{k}R_{\ell}\, B_{i}B_{j}B_{k}B_{\ell}  + ...
=\te -\frac{2\lambda^{2}}{4!} \Big( \l^3  \big[ (\l F_{1})'' \big]^{4} \Big)' +... \ , \la{722} \\
\FFF_{5} (\l) =  & \tfrac{1}{5!}\sum_{i,j,k,\ell,s=1}^\infty   c_{ijk\ell s} 
R_{i}R_{j}R_{k}R_{\ell} R_s \, B_{i}B_{j}B_{k}B_{\ell} B_s   + ...
=\te \frac{2 \lambda^{2}}{5!}  \Big[  \l^2 \Big( \l^3  \big[(\l F_{1})''\big]^{5} \Big)'\Big]' +... \ , \la{723} \ea
where we used that,   as follows from \rf{783},\rf{784}, 
{\small 
\ba
c_{ijk\ell} \equiv  & 2(i+j+k+\ell+3) (i+1)(i+2)(j+1)(j+2)(k+1)(k+2)(\ell+1)(\ell+2)\  , \la{1722}\\
c_{ijk\ell s } \equiv  & \te 2(i+j+k+\ell+3) (i+j+k+\ell+s + 4)  (i+1)(i+2)(j+1)(j+2)(k+1)(k+2)(\ell+1)(\ell+2)\ \no  . 
\ea  }
These  terms  provide  the dominant contributions  in   $\FFF_4$  and   $\FFF_5$ 
at strong coupling:  $\FFF_4 \sim \l^{4}, \  \FFF_5 \sim \l^{5} $ (see   below). 
Comparing  the last term in \rf{720}   with \rf{722} and \rf{723}   we  observe a definite pattern  for generalization
of these leading terms
\be
\FFF_{k+2} (\l)  =  {\te {2 \ov (k+2)!}  \big(-\l^2 {d \ov d \l }\big)^{k-1}  }\Big( \l^3  [( \l  F_1)'']^{k+2} \Big) + ... \ , \qquad 
\ \ \ k=0,1,2,... \ . \la{7755}
\ee
Here the  $k=0$ case  represents the  second ($2 \td F_2$) term in  \rf{7.16} given by  the integral over $\l$, 
$\FFF_{2} (\l)  = \ha (\l F_1)'    - \int d \l   \, \l \, [( \l  F_1)'']^{2}  $. 

It  is natural to expect  that   the full   expressions for  higher order $1/N$ corrections $\FFF_n$  in  the free  energy in \rf{79} 
will    be expressed  in terms of  derivatives of $F_1(\l)$. 
The integral representation for $F_1$ \rf{eq:F1}  will then  imply  a similar representation 
not only for  $\FFF_2$ (cf.  \rf{7.16},\rf{321}) and $\FFF_{3}$  \rf{720} but also  for all $\FFF_n$.

\subsection{Strong coupling expansion of free energy}
 
 Given  the relations \rf{713},\rf{7.16}   and \rf{720} 
 the strong coupling expansions of the free energy
 terms $\FFF_1$, $\FFF_2$  and  $\FFF_3$  in \rf{79} follow  from the
  $SU(N)$ results for $F_1$   and $\wt F_2$  in \rf{6.2},\rf{6.3}  and  \rf{6.5} 
  and the leading  terms in $\FFF_4$  and $\FFF_5$   from \rf{722},\rf{723} 
  \ba
 \FFF_1 
 =
  &\ 2 f_1 \l +2 f_2 \log \l   +2 f_3 +2 f_4\lambda^{-1}  +  \mc O (  e^{-\sql})  \no\\ & = 
{  \tfrac{\log 2}{2\pi^{2}}\,\lambda}-\tfrac{1}{2}\,\log\lambda+ \const    - \tfrac{\pi^2}{2\l }   +  \mc O (  e^{-\sql})\ , \la{721}\\
 \FFF_{2}  
 =&\te-  2 f_1^2 \l^2 +  f_1 ( 1 - 4f_2)  \l + { \frac{1}{2}}  f_2 (  1 - 2 f_2 ) \log \l + \ha ( f_2+ f_3 + 4 p_4)  +   \mc O (  e^{-\sql}) 
 \ ,  \la{288}\\
  \FFF_3=&\  \tfrac{8}{3} f_1^3 \l^3 
     - f_1^2 (1-4 {f_2}) \l^2- {f_1} {f_2} (1-2 {f_2}) \l 
  -\tfrac{1}{24} {f_2} (1 + 6 f_2 - 8 {f^2_2})+    \mc O (  e^{-\sql})  \ , 
    \la{777}\\
    \FFF_4=&- 4 f_1^4 \l^4 + \mc O(\l^3)  
     \ ,  \la{773}\\ 
       \FFF_5= &\ \tfrac{32}{5}  f_1^5 \l^5 + \mc O(\l^4)  
         \ ,  \la{771}
 \ea
 Here  $ \mc O (  e^{-\sql})  $   stands for the corresponding  exponentially suppressed  corrections 
  $\sim \l^{-k/4} e^{-n \sql}$    that follow from the ones in $F_1$ in  \rf{666},\rf{eq:F1exp}.\foot{While 
  $F_1$ has exponentials that are odd powers of $e^{-\sql}$,  $\FFF_2$  (that 
  contains  squares of  derivatives of $F_1$   and cross-terms, cf. \rf{7.16})     has 
   both even and odd powers of  $e^{-\sql}$. Similarly, for $\FFF_3$ in \rf{720} one also  finds 
    both odd and even powers of $e^{-\sql}$. }

 We observe that the leading large $\l$ asymptotics of $\FFF_n$ appears to be $\l^n$.
Note  also  that $\FFF_3$  has no    $\log \l$ term  while  the order $\l^{-1}$   term  appears  only in  $\FFF_1$. 
Assuming  that all  higher 
  $\FFF_n$  terms  are expressed in  in terms of derivatives of $\l F_1$ as in \rf{7.16},\rf{720},\rf{722},\rf{723} 
 the  only $\log \l$ corrections will come from $\FFF_1$ and $\FFF_2$, i.e. 
  the coefficient of the   $\log \l$  term in $F$ receives  contributions  only from the $N^2, N$ and $N^0$
  orders in the $1/N$ expansion  while 
  the  $\l^{-1} $ term   in $F$   is exactly captured by \rf{721}.

   Including also the $\N=4$ SYM contribution in \rf{662} the full expression for the free energy expanded at large $\l$   may 
     be written  as  
     \ba\la{177}
     F= & F^{\N=4} + \Delta F \stackrel{\lambda\gg 1}{=} \Delta F_{\rm pol}  - ( N^2 + N 
     + \tfrac{3}{16} ) \log \l   - \tfrac{\pi^2}{2} \tfrac{N}{\l }    +   \mc O (  e^{-\sql})   \ , \\
     \Delta F_{\rm pol} =& 
     N \l \big[  2 f_1  + \mc O(\l^{-1} ) \big]   +    \l^2 \big[ 2  f_1^2  + \mc O(\l^{-1} ) \big]  
              + \tfrac{1}{N} \l^3  \big[ \tfrac{8}{3} f_1^3  + \mc O(\l^{-1} ) \big]  + \mc O ( \tfrac{1}{N^2} )\no \\
               =  &
      N^2 \FC( \tfrac{\l}{ N}  ) + ...\ , \\
      \FC(\tfrac{\l}{ N})  =  & 2 f_1 \tfrac{\l}{ N} +  2 f_1^2  \big( \tfrac{\l}{ N}  \big)^2 +  \tfrac{8}{3} f_1^3   \big( \tfrac{\l}{ N}  \big)^3
      - 4 f_1^4   \big( \tfrac{\l}{ N}  \big)^4 +  \tfrac{32}{5} f_1^5   \big( \tfrac{\l}{ N}  \big)^5 +    ...
     \ , \la{178}
     \ea
     where   $ \Delta F_{\rm pol}$ represents  the   polynomial  in $\l\gg 1 $ contributions with 
     $\FC( \tfrac{\l}{ N}  )$  being the sum of the leading   $\l^n$ terms at each order in $1/N$. 
     
  Remarkably, the coefficients  in \rf{178}  suggest that  $\FC$  has  the following exact form 
  \be \la{888}
 \FC( \tfrac{\l}{ N}  ) =  \log \big ( 1+ 2 f_1 \tfrac{\l}{N}\big)  \ .   \ee 
Using that according to   \rf{2}   we have ${\l\ov N} =   4 \pi \gs$       we conclude 
that    this  leading order term   expressed in terms of string  parameters  non-trivially 
 depends  just on  string coupling  ($ 8 \pi  f_1 = {2\ov \pi} \log 2 $) 
 \be \la{99}
 F=  N^2 \FC( \tfrac{\l}{ N}  )  + ...  = \frac{\pi^2 T^4}{\gs^2}  \log \big ( 1+ 8 \pi  f_1 \gs \big)  + ...     \ . \ee 
 This term should be summing the   leading   large string tension  contributions from each order 
 in  string topological expansion 
  
 The term  $ - \tfrac{\pi^2}{2} \tfrac{N}{\l }  = - \tfrac{\pi}{8} \frac{1}{\gs } $  in \rf{177} should also 
  have  a special origin on the string side, coming from a particular   crosscup or disc contribution 
 not involving (in contrast to  the $ \frac{1}{\gs }  $ term in \rf{99}) extra powers of string tension
 (and thus subleading compared to \rf{99} at large $T$).

 \subsection{Wilson loop}
 
 The $\frac{1}{2}$-BPS Wilson loop is again defined as in (\ref{4.1}). In the $Sp(2N)$ $\N=4$  SYM theory 
 its expectation value 
 (exact in $N$ and $\l$ defined still as  $\lambda=N\gym^{2}$)  is  given by the sum of the Laguerre polynomials 
 \cite{Fiol:2014fla}  (cf. \rf{1} \footnote{
 The Laguerre polynomials in (\ref{7.17}) are the basic ones, while in the $SU(N)$ case in \rf{1} we have the associated Laguerre polynomial
 arising from the sum in (\ref{7.17}) without parity restriction on the index, {\em i.e.} from the identity $L_{N}^{(1)}(x) = \sum_{k=0}^{N}L_{k}(x)$.
 })
\be
\la{7.17}
\vev{\mc W}^{\N=4} = 2\, e^{\frac{\lambda}{16N}} \sum_{k=0}^{N-1}L_{2k+1}\big(-\tfrac{\lambda}{8N}\big)\, .
\ee
The resulting  $1/N$ expansion is 
\be
\la{7.18}
\vev{\mc W}^{\N=4} = N\,\tfrac{4}{\sql}I_{1}(\sql)+\tfrac{1}{2}\,\big[I_{0}(\sql)-1\big]+\tfrac{1}{N}\,\tfrac{\l}{96}\,I_{2}(\sql)+ \mc O(\tfrac{1}{N^2}) \ .
\ee
Then  the $\N=2$   expectation value may be written as in \rf{29}\foot{To recall,   we define $\vev{\mc W}$  so that 
$\vev{1}=1$, i.e.  we divide over the   matrix model partition function  $Z= e^{-F}$.}
\be
\la{7.19}
\vev{\mc W} = N\,\WWW_{0}(\l)+\WWW_{0,1}(\l) +\WWW_1(\l)
+\tfrac{1}{N}\big[\WWW_{0,2}(\l)+\WWW_2(\l)\big]+\mc O(\tfrac{1}{N^2})\ , 
\ee
where the $\N=4$ parts $\WWW_{0,n}$  are given by  (\ref{7.18})
\ba
\la{7.20}
\WWW_{0} &\equiv \vev{\mc W}_{0}^{\N=4} = \tfrac{4}{\sql}I_{1}(\sql) = 2W_{0}, \qquad \qquad \WWW_{0,1} \equiv \vev{\mc W}_{1}^{\N=4} = \tfrac{1}{2}\big[I_{0}(\sql)-1\big]\ , \\
\la{7.21}
\WWW_{0,2} &\equiv \vev{\mc W}_{2}^{\N=4} = \tfrac{\l}{96}\,I_{2}(\sql) \ .
\ea
The relation between the genuine $\N=2$ parts $\WWW_{1}$ and  $\WWW_{2}$ in \rf{7.19}
 and the free energy terms in \rf{79} 
 is the same (up to factor of 1/2) as in $SU(N)$ case in \rf{5.6},\rf{5.7}  (see Appendix \ref{C})
 \ba
\la{7.22}
\WWW'_{1} = -\tfrac{\lambda}{8}\WWW_{0}\, 
(\lambda \FFF_{1})''\ , \qquad \qquad 
\WWW_{2} = -\tfrac{\lambda^{2}}{8}\WWW_{0}\,  \FFF'_{2}\ . 
\ea
 We  thus find  using  \rf{5.6} and \rf{7.16}   
 (cf. \rf{5.8}) 
 \ba \la{7233}
 \WWW_1(\l)= 2 W_1(\l) \ , \qquad\qquad 
 \WWW_2(\l)= 
-\tfrac{\lambda^{2}}{8}\WWW_{0} \Big( \half (\lambda F_{1})''    -  {\l}\big[(\lambda\, F_1)''\big]^{2} \Big)
\ . 
\ea
Like for  the free energy in \rf{720}--\rf{723}, these relations  can be extended  also  to  higher  $1/N$ orders.
 
 Using \rf{7.22},\rf{7233}   we  find  for the  strong-coupling expansion of the     coefficients in \rf{7.19}
\ba
\frac{\WWW_{1}}{\WWW_{0}} &=\frac{W_{1}}{W_{0} }=  -f_{1}\,\lambda^{3/2}+\tfrac{3}{2}\,f_{1}\,\lambda- (\tfrac{3}{8} f_{1}+\tfrac{1}{2}f_{2})\l^{1/2} 
+\mc O ( {\l^{0}})   =\te  -\frac{\log 2}{4\pi^{2}}\lambda^{3/2}+\mc O(\l)  \ ,\la{363}   \\
\frac{\WWW_{2}}{\WWW_{0}} &=    \tfrac{1}{2} f_1^2\lambda^{3}-\tfrac{1}{8}f_1 (1 - 4  f_2 )\lambda^{2}-\tfrac{1}{16}f_{2}( 1 - 2 f_2) \lambda+ \mc O ( e^{-\sql} )=\te   -\frac{\log^{2}2}{32\pi^{4}}\,\lambda^{3} + \mc O (\l^2)  \ . \la{364}
\ea
 Note  that like  $\FFF_n$ in free energy   the Wilson loop  coefficients 
  $\WWW_n$  have additional exponentially suppressed corrections $\sim e^{-\sqrt{\lambda}}$ at strong coupling, which are resurgent, and which follow directly from the exponentially suppressed corrections to $F_1(\l)$ derived in Section \ref{sec63}.

  Similar relations  between  higher order $1/N$ terms $\FFF_n$  in free energy \rf{26}   and $\WWW_n$ in \rf{29}  are expected 
also in general,  with the dominant  large $\l$ term in $\FFF_n$ determining  the  strong coupling asymptotics of 
$\WWW_n$ (see Appendix \ref{C}). In particular, 
\be 
\WWW_3 = - \tfrac{\l^{3/2}}{4!} \WWW_0 \big[ \l ( \l F_1)'']^3 +... \ ,  \qquad \ \ \ 
\frac{\WWW_{3}}{\WWW_{0}}  \stackrel{\lambda\gg 1}{=}  - \tfrac{1}{6} f_1^3\lambda^{9/2} + \mc O ( {\l^{4}}) \ . \ee
Comparing  to \rf{363},\rf{364}  thus suggests that the leading (at each order in $1/N$) 
 strong coupling terms in $\Delta  \vev{\mc W}$  in \rf{29} exponentiate as 
\be \la{1324}
\vev{\mc W} =  ( N \WWW_0 +...) +    \Delta  \vev{\mc W}\stackrel{\lambda\gg 1}{=} N  \WWW_0 \, \exp\big[  - f_1 \tfrac{\l^{3/2}}{N} \big] + ...
\ .  \ee 
This may be compared with similar exponentiation \ci{Drukker:2000rr} of the leading large $\l$ terms in the $\N=4$ SYM  case 
in \rf{xxx},\rf{yyy} that  on string side may be interpreted as representing sum of  separated handle insertions  into the 
disc diagram 
\ci{Giombi:2020mhz}. Similarly, \rf{1324}   may be interpreted as a sum of  crosscup insertions into the disc.


  \iffa
   that means<W> = ( N W0+ ...)_SYM  + W0 [ W1+ 1/N W2+ ...]= N W0 [ 1+ 1/N W1/W0+ 1/N^2 W2/W0+ ...] + ( ...)-- > N W0 exp [ - f1 la^3/2/N ]or N W0 exp [ - f1 /pi T g_s ]while in SYM case we had exp factor asexp[ sql + (la^3/2)/(96 N^2) ] = exp[ 2 pi T + pi/12 g_s^2/T ]
exp [ - f1 /pi T g_s ] should be exponentiation of disc or more likely crosscup insertion rather than
handle insertion
\fi
  
\subsection*{Acknowledgements}
We would like to thank M. Bill\`o, S. Giombi, M. L. Frau, A. Lerda and  A. Pini  for related discussions. 
MB was supported by the INFN grant GSS (Gauge Theories, Strings and Supergravity). 
GD was supported by the U.S. Department of Energy, Office of Science,
Office of High Energy Physics under Award Number DE-SC0010339.
AAT was supported by the STFC grant ST/T000791/1.

\newpage
 \appendix
 
 \section{Partition function  of  $\N=2$   matrix model  and conformal  anomaly
\la{A}}
Let us first  recall that the  conformal anomaly  coefficients  a  and c in $\mc N=2$  superconformal  models
are  not renormalized, \ie  are given  just by their free-theory  values found  by summing up contributions   of particular  fields  (see, e.g., \ci{Duff:1977ay}).  
In a model   with $\nn_{\rm v}$   vector multiplets and $\nn_{\rm h}$   hypermultiplets
one finds 
\be  \la{1a}
{\rm a} = \frac{5}{24}\nn_{\rm v}+\frac{1}{24}\nn_{\rm h} \ . \ee
In particular, in the $\N=4$   SYM theory  ($\nn_{\rm v}=\nn_{\rm h}$)  with group $G$  we get 
${\rm a} = {1\ov 4} \dim G$.
The free  energy of a   massless superconformal model 
  on  $S^4$ of radius $\rr$  may be written   as  
\be\la{a1}
  \hat F=-\log \hat Z=  4 \aa \log (\La\, \rr)   +  F_{\rm fin} ( \l, N)  \ ,  \ee
where $\La$ is a  UV cutoff,  i.e.  the $\rr$   dependence is  controlled  by the $\aa$-coefficient. 
The  free  energy   thus depends on a subtraction scheme   and   below we shall denote by  $F$    its regularized  value. 

 The  localization  matrix model  expression for the  partition function $Z$ of $\N=2$ gauge theory on $S^4$  is \cite{Pestun:2007rz}  
 \be
Z=e^{-F} = \int Da\,  e^{-\frac{8\pi^{2}N\,\rr^{2}}{\l}\,\tr a^{2}}  \,  \Z_\text{1-loop}(a)
\ , \qquad 
\qquad \Z_\text{1-loop}(a) = e^{-S_{\rm int}(a)} \ . \la{a2}
\ee
In the  $\N=4$ SYM case    
 $\Z_\text{1-loop}(a)=1$  and doing the Gaussian  integral  we get \ci{mehta,Marino:2012zq} 
  (for  $G=SU(N)$)
\ba 
& Z^{\N=4}= C_0(N) \,   \Big({8 \pi \rr^2 \ov \l}\Big)^{-\ha\dim G} \ , \ \ \ \ \ \qquad
C_0= (2 \pi)^{N/2}\,(2\pi N)^{-\frac{1}{2}N^{2}}\,{\rm G}(N+2)\ ,  \la{A4}\\
& F^{\N=4} = 
4{\rm a}\log \rr-2{\rm a}\log \lambda+ C(N)\ ,\qquad \aa =\te {1\ov 4} \dim G= {1\ov 4} (N^2-1) \  ,\la{a3} \\  &  C(N) =  {1\ov 2} (N^2-1) \log( 8 \pi)  - \log  C_0(N)  \ , \la{AA4}
\ea
where ${\rm G}(N+2)=\prod_{k=1}^N \Gamma(1+ k) $ is the Barnes G-function.\foot{\la{a29}
Note that the large $N$ expansion of  $C(N)$   may be written as

$C(N) = \frac{1}{4} N^2 \big[3+4 \log (4 \pi )\big]-N \big[\log N+\log(2\pi)-1\big]-\frac{5}{12}\log N+\mc O(N^{0})$. 
}

%
%
%
%
%
%
Setting $\rr=1$   we  conclude that in the  subtraction scheme assumed in the 
localization approach  $F^{\N=4} = -2{\rm a}\log \lambda$ (up to a  $\l$-independent 
constant). 
This was  noted   in 
  \cite{Russo:2012ay}  and an 
 AdS/CFT  interpretation of this  result  was  suggested.
 
One may wonder  what happens in  other  $\N=2$    superconformal models, in particular, 
 if  the   conformal anomaly   a-coefficient  is also encoded the $\log \l$ term of the 
  large $\l$  expansion of the free  energy $F$ on $S^4$. For the models   that are 
  planar-equivalent to $\N=4$ SYM this is certainly the case  at the leading $N^2$ order 
  but  as we shall see below  this does not need to be be true at subleading orders in $1/N$. 

For an $\N=2$  model with a collection of hypermultiplets in representation $R=\oplus R_{i}$ 
of a group $G$ with algebra $\mathfrak{g}$   one finds  \cite{Pestun:2007rz}\foot{We ignore  the instanton  contribution 
  since it is exponentially suppressed in the $1/N$  expansion we are interested in here.}
\be
\la{B.5}
\hat \Z_\text{1-loop}(a, \rr) = \prod_{n=1}^{\infty}\left(
\frac{\prod_{\alpha\in \text{roots}(\mathfrak{g})}\big[\rr^{-2}n^{2} +  (\alpha\cdot a)^{2}\big]}{\prod_{w\in \text{weights}(R)}\big[ \rr^{-2}n^{2}  +    (w\cdot a)^{2}\big]}
\right)^{n}\ .
\ee
$\hat \Z_\text{1-loop}$  coming from the ratio of 1-loop determinants on $S^4$ in a constant  scalar $a$   background 
 does not depend on $\l$  but  does  depend on $\rr$. Note that 
 the  product  over  roots   here includes  also the "massless"  contributions  of the zero roots 
 corresponding 
 to Cartan directions for which $\alpha\cdot a=0$ 
 (same also  applies  to the product over   weights in the case of the  adjoint representation).

The regularized  value of  $\hat \Z$ in \rf{B.5}  used in  \cite{Pestun:2007rz} was 
\be
\la{B.6}
\Z_\text{1-loop} (a\,\rr) = \frac{\prod_{\alpha\in \text{roots}(\mathfrak g)} \HH(i\,\alpha\cdot a\, \rr)}{\prod_{w\in\text{weights}(R)}\HH(i\, w\cdot a\, \rr)}\ ,
\ee
where $\HH(x) \equiv  {\rm G}(1+x)\, {\rm G}(1-x)$ is the  product of the Barnes G-functions. 
Notice that  here the contribution of the  "massless"  terms present in \rf{B.5}  is trivial as 
$H(0)=1$. As a  result, the contribution of \rf{B.6}  to the $\log \rr$   term in $F$  or to the conformal anomaly  is trivial -- the $\rr$ dependence can be absorbed into the rescaling of  the integration variable $a$    in \rf{a2}   and 
this the resulting $Z$  will depend on $\rr$  in the same way \rf{a3}
 as in the $\N=4$ SYM case. 

To properly account  for the conformal anomaly of the $\N=2$  model 
 we need to go back to the original unregularized 
 expression  \rf{B.5}  and compute  its dependence on the radius $\rr$. 
 Rearranging \rf{B.5}  using  that  
\be
\la{B.7}
\prod_{n=1}^\infty \big( \rr^{-2} n^2 + \mu^2 \big)^n 
= \prod_{n=1}^\infty \rr^{-2n}  \  \prod_{n=1}^\infty \big(  n^2 + \rr^2 \mu^2 \big)^n \ , 
\ee
where $\mu$ stands for $\alpha\cdot a$ or $w\cdot a$, 
we conclude that  the non-trivial  dependence on $\rr$  (that cannot be  absorbed into $a$)  is captured  by the  infinite product  factor that can be  defined using the standard Riemann 
 $\zeta$-function  regularization  as 
\be 
\prod_{n=1}^\infty \rr^{-2n}  = e^{- 2  \zeta(-1)  \log \rr\,   } = e^{{1\ov 6} \log \rr}  \ . \la{b77}\ee
As a  result, we find  from \rf{B.5}\foot{Here  we use that  the total  number of roots  
counting also the trivial Cartan ones is  the same as $\dim G$.}
\be 
\hat \Z_\text{1-loop} (a,\rr) \ \  \to \ \  e^{\frac{1}{6} (\dim G - \dim R)\log \rr}\,  \Z_\text{1-loop} (a\, \rr )\ . 
\ee
Redefining $\rr a  \to a $  to  account  for 
 the dependence on $\rr$ in  the free action in \rf{a2}   and in 
 $ \Z_\text{1-loop} (a\, \rr ) $  we need  also to include the contribution of the Gaussian measure or the $\N=4$ term in \rf{a3}, so that the total $\rr$ dependence of the $\N=2$ free energy is (cf. \rf{a1})
\be 
F =\big[{\rm dim}\, G  - \tfrac{1}{6} (\dim G - \dim R)\big] \log \rr  + ... 
= 4 \aa  \log \rr   + ..., \ \ \ \ \  \ \ \aa=\te {5\ov 24} \dim G + {1\ov 24} \dim R \,, \la{4a}\ee
in agreement  with the general expression for the a-anomaly in \rf{1a}. 

We  have thus shown  that it is  the  "bare" expression  for the matrix model integral 
\rf{a2}  using \rf{B.5} that  correctly  includes  the conformal  a-anomaly term in free energy. 
It is  clear  that the direct correlation  between the dependence on 
$\rr$ and  on $\l$  is   a  feature of  only the Gaussian  part of the integral in \rf{a2}. In particular,  
  the dependence of the $\N=2$ free energy on $\log \l$    beyond the leading planar limit 
  need not be controlled 
 by the  a-anomaly   coefficient   as that happened in the $\N=4$ SYM case in \rf{a3}. 
 
Nevertheless,  we have found (see discussion   below \rf{128}) 
 that  not only  the order $N^2$   but also 
the order $N$   coefficient of the  $\log \l$ term in the 
large $\l$  limit of the free energies   of the $SU(N)$ and $Sp(2N)$  \FA theories 
computed  in this paper   do  agree  with the corresponding terms in the  conformal a-anomalies.
We suspect that  the matching  of the order $N$ term   should be  also related to the 
fact that these  models are planar-equivalent to $\N=4$ SYM  theory.

 \section{Derivation of large $\lambda$ expansion of $F_1$    using  Mellin transform
  \la{B} }
 
 In the main text, we computed the large $\lambda$ expansion of $F_{1}$  using 
 the approach described in  (\ref{b6})-(\ref{b8}).
Here  we shall  compute the large $\lambda$ expansion of $F_{1}$ 
  given by the integral representation \rf{eq:F1} by 
applying the Mellin transform method (see \eg \cite{zagier,flajolet}).
The first step   is  to rewrite \rf{eq:F1}  in the form of a Mellin convolution 
 \be
h(x) \equiv (f\star g) (x)= \int_0^\infty dt \,  f(t\,x)\,g(t)\ , \qquad \qquad x=\sql \ . 
\ee
The  Mellin transform  is $\widetilde{h}(s) = \mc M[h](s) = \int_0^\infty dx \,  x^{s-1}\,h(x) = \widetilde{f}(s)\,\widetilde{g}(1-s)$.
If $\alpha< s< \beta$ is the fundamental strip of analyticity of $\widetilde{h}(s) $,
the asymptotic expansion of $h(x)$ for $x\to\infty$ is obtained from the poles of its Mellin transform in the region $s\ge \beta$. In particular, 
the pole $\frac{1}{(s-s_0)^n}$ gives a term $\frac{(-1)^n}{(n-1)!}\,\frac{1}{x^{s_0}}\,\log^{n-1} x$ in the asymptotic expansion of $h(x)$.
 
 Explicitly,  let us  first put   (\ref{eq:F1}) in the equivalent form\footnote{
 For an odd function $\hat f(t)$, we have the identity 
$ \int_{0}^{\infty} dt\,\frac{e^{2\pi t}}{(e^{2\pi t}+1)^{2}} \hat f(t) = \int_{0}^{\infty} dt\,\frac{e^{2\pi t}}{(e^{2\pi t}-1)^{2}} f(t)$ with $f(t) =  \hat f(t)-2\hat f(\tfrac{t}{2})$
and the inversion relation  $\hat f(t) = \sum_{k=0}^{\infty}2^{k} f(2^{-k}t)$.
 }
 \ba
F_1(\lambda) &= \frac{2}{\sql}\int_{0}^{\infty}dt\,\frac{e^{2\pi t}}{(e^{2\pi t}-1)^{2}}\,\frac{3t\,\sql-8J_{1}(t\sql)+J_{1}(2t\sql)}{t^{2}} \lp
= 2\sql\,\int_{0}^{\infty}dt\, f(t\sql)\, g(t) = 2\sql\ (f\star g)(\sql)\ ,
 \ea
 where
 \be
 f(t) = \frac{3t-8J_{1}(t)+J_{1}(2t)}{t^{2}}\ , \qquad \qquad g(t) = \frac{e^{2\pi t}}{(e^{2\pi t}-1)^{2}}\ . 
 \ee
 The Mellin transform of $g(t)$ is 
 \ba
\mc M\big[\frac{e^{2\pi t}}{(e^{2\pi t}-1)^{2}}\big] (s)  &= -\frac{1}{2\pi}\mc M\big[\frac{d}{dt}\frac{1}{e^{2\pi t}-1}\big] (s) = \frac{1}{2\pi}(s-1) \mc M\big[\frac{1}{e^{2\pi t}-1}\big]({s-1})\lp
=(2\pi)^{-s}\, \Gamma(s)\, \zeta(s-1)\ .
\ea
Computing the Mellin transform of $f$, then using $\widetilde{f\star g} = \widetilde f(s)\, \widetilde g(1-s)$, and  finally evaluating  the residues gives
\be\la{A.4} 
F_1 \stackrel{\lambda\gg 1}{=}\ \te    \frac{\log 2}{4}\,{\lambda\ov \pi^2}-\frac{1}{4}\log{ \lambda\ov \pi^2}\  +
(\frac{7}{6}\log 2+\frac{3}{4}-6\log\mathsf{A})\ -\frac{\pi^{2}}{4}\big({\l\ov \pi^2}\big)^{-1}+\dots\ ,
\ee
where $\mathsf{A}$ is Glaisher's  constant. There are no additional pole contributions beyond those giving (\ref{A.4}). This implies that dots in (\ref{A.4}) 
stand for the exponentially suppressed corrections (discussed in Section 6.3).

 \section{Strong coupling expansion of Wilson loop  in $Sp(2N)$ theory
  \la{C} }

Let  us first  consider the   expectation value of  the BPS Wilson loop (defined in fundamental representation) 
 in the $\N=4$  $Sp(2N)$ SYM theory 
 \cite{Fiol:2014fla} (see also \ci{Giombi:2020kvo})
 \be\la{c1}
\vev{\mc W}^{\N=4} = 2\,e^{\frac{\lambda}{16N}}\,\sum_{i=0}^{N-1}L_{2i+1}\big(-\tfrac{\lambda}{8N}\big)\ .
\ee
Using    
the integral representation of Laguerre polynomials
$L_{n}(x) = \frac{1}{2\pi i}\oint\frac{dt}{t^{n+1}}(t+x)^{n}e^{-t} $,
we can write
\ba
\la{F.3} 
\vev{\mc W}^{\N=4} &= \frac{1}{2\pi i}\oint dt\,\frac{8\,N\,e^{-t+\frac{\lambda}{16N}}\,(1-\frac{\lambda}{8tN})}{ \lambda\,(1-\frac{\lambda}{16 t N})}\,\Big[1-\Big(1-\frac{\lambda}{8Nt}\Big)^{2N}\Big] .
\ea
Expanding at large $N$ and observing that 
\be
\frac{1}{2\pi i}\int \frac{du}{u^{n}}\,e^{-x(u+u^{-1})} = (-1)^{n-1}\,I_{n-1}(2x)\ ,
\ee
we obtain  for the leading terms \ci{Giombi:2020kvo}
\ba
\la{F.5}
\vev{\mc W}^{\N=4} &= \te
4N\,\frac{I_1(\sql) }{\sql}+\frac{1}{2} \big[I_{0}(\sql)-1\big] 
+ \frac{\lambda I_{2}(\sql) }{96N} 
+\frac{1}{N^{2}}\big[-\frac{\lambda I_{0}(\sql) }{192}+\frac{\sql(\lambda+8)I_{1}(\sql)}{768}\big] +\cdots\ .
\ea
Let us denote the leading large $N$ term here as  
$\vev{\mc W}_{0} = N \WWW_0= 4N\,\frac{I_1(\sql) }{\sql}$  (cf. \rf{30},\rf{300}). 
Expanding  at large $\l$ and keeping only the  dominant  term at each order in $1/N$ 
 we find
 {\small
\ba
\frac{\vev{\mc W}^{\N=4}}{\vev{\mc W}_{0}} \stackrel{\l \gg 1 }{=} \te 
 1+\frac{\lambda^{1/2}}{8N}+\frac{\lambda^{3/2}}{384N^{2}} +\frac{\lambda^{2}}{3072N^{3}}+
\frac{\lambda^{3}}{294912N^{4}}+\frac{\lambda^{7/2}}{2359296N^{5}} 
 \te   +\frac{\lambda^{9/2}}{339738624N^6} +\frac{\lambda^{5}}{2717908992N^7}
 +\cdots\ .\la{c5}
\ea
}
A natural guess for the  sum of  this expansion is  
\be
\frac{\vev{\mc W}^{\N=4}}{\vev{\mc W}_{0}} \stackrel{\l \gg 1 }{=}  \big(1+\tfrac{\lambda^{1/2}}{8N}\big)\,\exp\big(\tfrac{\lambda^{3/2}}{384\,N^{2}}\big)\ .\la{c6}
\ee
This  expression  can be 
 proved rigorously starting from the exact relations between ${\vev{\mc W}^{\N=4}}$  in $U(N)$ and $Sp(2N)$ theories 
 given   in  \cite{Fiol:2014fla}
 \be 
 \vev{\mc W}^{\N=4}_{_{\rm Sp(2N)}} (\l) =  \vev{\mc W}^{\N=4}_{_{\rm U(2N)}} (\l) + \tfrac{1}{16 N} \int^\l_0 d\l'
 \, \vev{\mc W}^{\N=4}_{_{\rm U(2N)}} (\l') \ , \la{fiol}
 \ee
  and taking the large $\l$ limit.\footnote{The Wilson loop in the $\N=4$ $U(N)$ theory is 
  given by   $   \vev{\mc W}^{\N=4}_{_{\rm U(N)}}  (\l) =
  e^{\frac{\lambda}{8N}}L^{(1)}_{N-1}\big(-\tfrac{\lambda}{4N}\big)$  (cf. \rf{1}).}
 \iffa \ba
\notag 
\vev{\mc W}^{\mc N=4}_{\rm U(N)}(\lambda) &= 
N\,\frac{2 I_1}{\sqrt{\lambda }} +\frac{\lambda  I_2}{48 
N}+\frac{1}{N^{3}}\Big(\frac{\lambda ^{5/2} I_3}{9216}-\frac{\lambda ^2 
I_4}{11520}\Big)+\frac{1}{N^{5}}\Big(\frac{\lambda ^4 I_4}{2654208}-\frac{\lambda 
^{7/2} I_5}{1105920}+\frac{\lambda ^3 
I_6}{1935360}\Big)\lp
+\frac{1}{N^{7}}\Big(\frac{\lambda ^{11/2} 
I_5}{1019215872}-\frac{\lambda ^5 I_6}{212336640}+\frac{\lambda 
^{9/2} I_7}{137625600}-\frac{\lambda ^4 
I_8}{309657600}\Big)+\cdots.
\ea
  }\fi 
 
\iffa 
String variables for $Sp(2N)$ are \red{[we keep $\gamma_{1,2}$ to understand conventions...]}
\be
g^{2}_{\rm YM} = 4\gamma_{1}\pi\, \gs,\qquad L^{4}=8\gamma_{2}\pi\, \gs\,\alpha'^{2}\,(N+\tfrac{1}{4}),\qquad T = \frac{L^{2}}{2\pi \alpha'},
\ee
that we write as 
\be
\gs = \frac{\lambda}{4\pi \gamma_{1} N}, \qquad T = \sqrt\frac{\gamma_{2}}{\gamma_{1}}\frac{\sql}{\sqrt 2 \pi}\sqrt{1+\frac{1}{4N}}.
\ee
Thus, taking $\gamma_{2}=1$ and $\gamma_{1}=2$ shows that 
\be
\vev{\mc W} \stackrel{\rm LT}{=} 2N\sqrt{\frac{2}{\pi}}\lambda^{-3/4}e^{\sql}\left(1+\frac{\lambda^{1/2}}{8N}\right)\,\exp\left(\frac{\lambda^{3/2}}{384\,N^{2}}\right),
\ee
may be written
\be
\vev{\mc W} \stackrel{\rm LT}{=} \frac{1}{2\pi}\frac{T}{\gs}e^{2\pi T+\frac{\pi}{12}\frac{\gs^{2}}{T}},
\ee
where the 'orientifold' factor $1+\frac{\sql}{8N}$ comes from odd part of $e^{2\pi T}$.
Keeping only leading terms   in large $\l$ at each order in  $1/N$   we may also reinterpret this as 
$$
e^{\sql} \left(1+\frac{\sql}{8N}\right) = e^{\sql (1 + {1\ov 4N})^{1/2} }   $$

shifts of $\l$ in other places are subleading.
Thus   at least  in this approximation   all terms  with odd  powers of $1/N$ 
can be generates   by redefinition  $\l= \gym^2 N \to \gym^2 ( N + \four) = \l ( 1 + {1\ov 4 N}) $
\fi

Let us   now turn to the Wilson loop expectation value in the $\N=2$   $Sp(2N)$ theory 
given by the matrix model expectation value  as in \rf{2.8},\rf{4.2}   with the  single-trace interaction action in \rf{7222}
\be \la{cc2}
S_{\rm int} = B_{i} (\l)\,  \tr\aaa^{2i+2} \ , \qquad \qquad \aaa\equiv \frac{a}{\sqrt N} \ ,  \ee 
where here and below  we assume summation over $i=1, ..., \infty$  and $B_i(\l)$ is given by \rf{2.11}. 
Denoting   as in \rf{2.8}  by $\vev{...}$   the normalized expectation value in the Gaussian theory  (i.e. in $\N=4$ SYM case)   then 
\ba
\vev{\mc W} &= \frac{\vevz{\tr e^{\sqrt\frac{\l}{2}\aaa}\  e^{-B_{i}\tr\aaa^{2i+2}}}}{\vevz{e^{-B_{i}\tr\aaa^{2i+2}}}} = \sum_{k=0}^{\infty}\tfrac{1}{(2k)!}\left(\tfrac{\l}{2}\right)^{k}
\frac{\vevz{\tr \aaa^{2k}\  e^{-B_{i}\tr\aaa^{2i+2}}}}{\vevz{e^{-B_{i}\tr\aaa^{2i+2}}}} \lp
= 2N+\tfrac{\l}{4}\frac{\vevz{\tr \aaa^{2}\  e^{-B_{i}\tr\aaa^{2i+2}}}}{\vevz{e^{-B_{i}\tr\aaa^{2i+2}}}}+\sum_{k=1}^{\infty}\tfrac{1}{(2k+2)!}\left(\tfrac{\l}{2}\right)^{k+1}
\frac{\vevz{\tr \aaa^{2k+2}\  e^{-B_{i}\tr\aaa^{2i+2}}}}{\vevz{e^{-B_{i}\tr\aaa^{2i+2}}}} \lp
= 2N- \tfrac{1}{4N}{\partial_{\lambda^{-1} }  } \log \hat Z+\sum_{k=1}^{\infty}\tfrac{1}{(2k+2)!}\left(\tfrac{\l}{2}\right)^{k+1}
\frac{\vevz{\tr \aaa^{2k+2}\  e^{-B_{i}\tr\aaa^{2i+2}}}}{\vevz{e^{-B_{i}\tr\aaa^{2i+2}}}} \lp
= 2N-\tfrac{\l^{2}}{4N}\partial_{\lambda}\Delta F+\lambda\tfrac{N(2N+1)}{8N}+\sum_{k=1}^{\infty}\tfrac{1}{(2k+2)!}\left(\tfrac{\l}{2}\right)^{k+1}\partial_{B_{k}}\Delta F \ . \la{c8}
\ea
Here  $\hat Z = e^{-F} = \int Da' \  e^{  -\frac{1}{\lambda}\tr a'^{2} - S_{\rm int} (a') }$  is the total  partition function  as   in \rf{2.3} 
before rescaling  of integration variable by $\l^{1/2}$ in \rf{2.44}
     and  the total  free   energy $F= F^{\N=4} + \Delta F$ as in \rf{2.12}   with $F^{\N=4} $   given by \rf{662}.
     We  used that differentiating $\hat Z$ over $\l$     puts down the factor $\sim \tr a^2$. 
    The third term  in \rf{c8} comes from 
\be\la{c9}
\log \hat Z = -\Delta F + \log \int Da'\,  e^{-\frac{1}{\lambda}\tr a'^{2}} = -\Delta F+\tfrac{1}{2}N(2N+1) \log\lambda+\text{const}\ . 
\ee
 We also used the formal notation $ \partial_{B_{k}}\Delta F$ for the normalized $   \partial_{B_{k}} \vev{e^{-B_{i}\tr\aaa^{2i+2}}}
     = \vevz{\tr \aaa^{2k+2}\  e^{-B_{i}\tr\aaa^{2i+2}}} $.
Here (see  \rf{713},\rf{7.16})
\ba
\Delta F = &N \FFF_{1}+\FFF_{2}+\tfrac{1}{N}\FFF_{3}+\mc O ( \tfrac{1}{N^2} )  \ , \\
\FFF_1 = &2 \sum_i R_i  B_i \ , \qquad \qquad 
\FFF_{2} =  \tfrac{1}{2}\sum^\infty_{i=1}(i+2)\, R_i B_{i}  -2\sum^\infty _{i,j=1} \wt Q_{ij}B_{i} B_{j}\ .
\ea
where  numerical $R_i$ and $\wt Q$ are given by \rf{3.4},\rf{3.8} and $\l$-dependence is contained in $B_i$. 
 Defining $\WWW_n$   corrections to the $\N=4$   SYM value
 $\vev{\mc W}^{\N=4} = \vevz{\tr e^{\sqrt\frac{\l}{2}\aaa}}$   as  in \rf{7.19}, i.e. \be 
 \vev{\mc W} = \vev{\mc W}^{\N=4} +  
 \WWW_1 + \tfrac{1}{ N } \WWW_2 +  \tfrac{1}{ N^2 } \WWW_3 +\mc O ( \tfrac{1}{N^3} )    \ , \la{cc10} \ee
   we  see that  derivatives of  both $\FFF_n$   and $\FFF_{n+1}$ terms  in $\Delta F $ in \rf{c8} contribute to $\WWW_n$.
   In particular, $\partial_{B_{k}}\FFF_1= 2 R_k$   contributes to the  order $N$ (planar) part of $ \vev{\mc W} $  while 
 for $\WWW_1$   we  find 
 \ba\la{c10}
 \WWW_{1} &= -\tfrac{\lambda^{2}}{4}   \FFF'_{1}+\sum_{k=1}^{\infty}\tfrac{1}{(2k+2)!}\left(\tfrac{\l}{2}\right)^{k+1}\partial_{B_{k}}\FFF_{2}(B) 
 = -\tfrac{\lambda^{2}}{4} \FFF'_{1}-4\sum_{j,k=1}^{\infty}\tfrac{1}{(2k+2)!}\left(\tfrac{\l}{2}\right)^{k+1}\wt Q_{kj}B_{j}\ , 
 \ea
 where  $(...)' \equiv \del_\l (...)$.
 Since $B_{j}\sim \lambda^{j+1}$,   differentiating  $\WWW_1$ over $\l$  gives 
\ba
\WWW'_{1} 
 &= -\tfrac{1}{4} (\l^2 \FFF_1')'-\sum_{j,k=1}^{\infty}\tfrac{2(j+k+2)}{(2k+2)!}\left(\tfrac{\l}{2}\right)^{k}\wt Q_{kj}B_{j} 
 = -\tfrac{1}{4} (\l^2 \FFF_1')'-\sum_{j,k=1}^{\infty}\tfrac{2}{(2k+2)!}\left(\tfrac{\l}{2}\right)^{k}\tfrac{2^{j+k+1}\Gamma(j+\frac{3}{2})\Gamma(k+\frac{3}{2})}
 {\pi \Gamma(j+1)\Gamma(k+1)}B_{j} \lp
 = -\tfrac{1}{4} (\l^2 \FFF_1')'+\tfrac{\sqrt\pi}{2\sql}(\sql-2I_{1}(\sql))\sum_{j=1}^{\infty}\tfrac{1}{\pi}\tfrac{2^{j+1}\Gamma(j+\frac{3}{2})}
 {\Gamma(j+1)}B_{j} \lp
 =-\tfrac{1}{4} (\l^2 \FFF_1')'+\tfrac{1}{2\sql}(\sql-2I_{1}(\sql))\sum_{j=1}^{\infty}(j+1)(j+2)R_{j}B_{j}
\lp 
 = -\tfrac{1}{4}\ \WWW_{0}\sum_{j=1}^{\infty}(j+1)(j+2)R_{j}B_{j} = -\tfrac{\l}{8} \WWW_{0} (\l \FFF_{1})'' \ , 
 \ea
 where $ \WWW_{0} = {4\ov \sql} I_1(\sql)$ as in \rf{7.20}. This demonstrates  the  relation in \rf{7.22}. 
 Similarly one can show  also that 
 $\WWW_2 = -\tfrac{\l^2}{8} \WWW_{0}  \FFF_2'$. 
 
 The example of $\WWW_2$ suggests that    the dominant at large  $\l$ term in $\WWW_n$   comes from the dominant term  in the corresponding $\FFF_n$. Indeed,  from \rf{c8}  and  the expression for the  dominant  term in $\FFF_3$ in 
 \rf{719}  we get   for the leading order large $\l$ contribution
 \ba
 \WWW_{2} &      \stackrel{\l \gg 1 }{=}    \sum_{k=1}^{\infty}\tfrac{1}{(2k+2)!}\left(\tfrac{\l}{2}\right)^{k+1}\partial_{B_{k}}\tfrac{1}{3}\big[\sum_{i=1}^{\infty}(i+1)(i+2)R_{i}B_{i}\big]^{3} + ... \lp
 = \sum_{k=1}^{\infty}\tfrac{1}{(2k+2)!}\left(\tfrac{\l}{2}\right)^{k+1}(k+1)(k+2)R_{k}\big[\sum_{i=1}^{\infty}(i+1)(i+2)R_{i}B_{i}\big]^{2} + ... \lp
 = -\tfrac{1}{16}\l  \big[1 -\tfrac{2}{ \sql} I_{1}(\sql) \big] \, \big[\l(\l\FFF_{1})''\big]^{2} + ... 
  = \WWW_{0}\, \tfrac{\lambda}{32}[\l(\l \FFF_{1})'']^{2}+...  \ . \la{c14}
 \ea
 This is indeed the leading  at large $\l$ term  in the exact expression for  $\WWW_2$  in terms of  $\FFF_1=2F_1$  in 
 \rf{7233}. 
 
Applying the same   logic to  find the   large $\l$ contribution in  $\WWW_{3}$ we  use  the expression for the dominant
term in $\FFF_4$ in \rf{722}
 \ba
 \WWW_{3} &  \stackrel{\l \gg 1 }{=}     \sum_{m=1}^{\infty}\tfrac{1}{(2m+2)!}\left(\tfrac{\l}{2}\right)^{m+1}\partial_{B_{m}}\Big(-\tfrac{1}{4!}\sum^\infty_{i,j,k,\ell=1}c_{ijk\ell}R_{i}R_jR_kR_{\ell}B_{i}B_jB_k B_{\ell}\Big) + ... \lp
 = -\tfrac{1}{3!}\sum_{m=1}^{\infty}\tfrac{1}{(2m+2)!}\left(\tfrac{\l}{2}\right)^{m+1}R_{m}\sum_{i,j,k=1}^{\infty}c_{ijkm}R_{i}R_{j}R_{k}B_{i}B_{j}B_{k}  + ...\ , \la{c16}
 \ea
 where $c_{ijkm}$  is given in \rf{1722}. 
Summing  over $m$ and keeping only  leading  $e^{\sql}$ terms  
(i.e. terms proportional to  $\WWW_{0}=2 \sqrt{2 \ov 2} \l^{-3/4}  e^{\sql} +... $)  we get 
\ba
 \WWW_{3}  & \stackrel{\l \gg 1 }{=}   -\tfrac{1}{3!}\tfrac{\lambda^{3/2}}{8}\WWW_{0}\sum_{ijk}^{\infty}(i+1)(i+2)(j+1)(j+2)(k+1)(k+2)R_{i}R_{j}R_{k}B_{i}B_{j}B_{k} +...\lp
 = -\tfrac{1}{3!}\tfrac{\lambda^{3/2}}{8}\WWW_{0}\big[\sum_{i=1}^{\infty}(i+1)(i+2)R_{i}B_{i}\big]^{3} +...= -\tfrac{1}{3!}\tfrac{\lambda^{3/2}}{64}\WWW_{0}  \big[\l(\l\FFF_{1})''\big]^{3}+...\ .
 \ea
 Then    $\FFF_{1}   \stackrel{\l \gg 1 }{=}  2 f_1 \l + ...$  (see \rf{721})   gives 
 \be\la{c18}
 \tfrac{\WWW_{3}}{\WWW_{0}}   \stackrel{\l \gg 1 }{=}  -\tfrac{1}{6}f_{1}^{3}\lambda^{9/2} + ...\  .
 \ee

\newpage 
\bibliography{BT-Biblio}

\providecommand{\href}[2]{#2}\begingroup\raggedright\begin{thebibliography}{10}

\bibitem{Drukker:2000rr}
N.~Drukker and D.~J. Gross, \emph{{An Exact prediction of N=4 SUSYM theory for
  string theory}}, \href{http://dx.doi.org/10.1063/1.1372177}{\emph{J. Math.
  Phys.} {\bf 42} (2001) 2896--2914},
  [\href{http://arxiv.org/abs/hep-th/0010274}{{\tt hep-th/0010274}}].

\bibitem{Giombi:2020mhz}
S.~Giombi and A.~A. Tseytlin, \emph{{Strong coupling expansion of circular
  Wilson loops and string theories in AdS$_5 \times {\rm S}^5$ and AdS$_4
  \times {\rm CP}^3$}},
  \href{http://dx.doi.org/10.1007/JHEP10(2020)130}{\emph{JHEP} {\bf 10} (2020)
  130}, [\href{http://arxiv.org/abs/2007.08512}{{\tt 2007.08512}}].

\bibitem{Beccaria:2021ksw}
M.~Beccaria and A.~A. Tseytlin, \emph{{$1/N$ expansion of circular Wilson loop
  in $\mathcal N=2$ superconformal $SU(N)\times SU(N)$ quiver}},
  \href{http://dx.doi.org/10.1007/JHEP04(2021)265}{\emph{JHEP} {\bf 04} (2021)
  265}, [\href{http://arxiv.org/abs/2102.07696}{{\tt 2102.07696}}].

\bibitem{Beccaria:2021vuc}
M.~Beccaria, G.~V. Dunne and A.~A. Tseytlin, \emph{{BPS Wilson loop in $
  \mathcal{N} $ = 2 superconformal SU(N)
  \textquotedblleft{}orientifold\textquotedblright{} gauge theory and
  weak-strong coupling interpolation}},
  \href{http://dx.doi.org/10.1007/JHEP07(2021)085}{\emph{JHEP} {\bf 07} (2021)
  085}, [\href{http://arxiv.org/abs/2104.12625}{{\tt 2104.12625}}].

\bibitem{Pestun:2007rz}
V.~Pestun, \emph{{Localization of gauge theory on a four-sphere and
  supersymmetric Wilson loops}},
  \href{http://dx.doi.org/10.1007/s00220-012-1485-0}{\emph{Commun. Math. Phys.}
  {\bf 313} (2012) 71--129}, [\href{http://arxiv.org/abs/0712.2824}{{\tt
  0712.2824}}].

\bibitem{Pestun:2016zxk}
V.~Pestun et~al., \emph{{Localization techniques in quantum field theories}},
  \href{http://dx.doi.org/10.1088/1751-8121/aa63c1}{\emph{J. Phys.} {\bf A50}
  (2017) 440301}, [\href{http://arxiv.org/abs/1608.02952}{{\tt 1608.02952}}].

\bibitem{Fiol:2014fla}
B.~Fiol, B.~Garolera and G.~Torrents, \emph{{Exact probes of orientifolds}},
  \href{http://dx.doi.org/10.1007/JHEP09(2014)169}{\emph{JHEP} {\bf 09} (2014)
  169}, [\href{http://arxiv.org/abs/1406.5129}{{\tt 1406.5129}}].

\bibitem{Fiol:2015mrp}
B.~Fiol, B.~Garolera and G.~Torrents, \emph{{Probing $ \mathcal{N}=2 $
  superconformal field theories with localization}},
  \href{http://dx.doi.org/10.1007/JHEP01(2016)168}{\emph{JHEP} {\bf 01} (2016)
  168}, [\href{http://arxiv.org/abs/1511.00616}{{\tt 1511.00616}}].

\bibitem{Fiol:2020bhf}
B.~Fiol, J.~Mart\'\i{}nez-Montoya and A.~Rios~Fukelman, \emph{{The planar limit
  of $\mathcal{N}=2$ superconformal field theories}},
  \href{http://dx.doi.org/10.1007/JHEP05(2020)136}{\emph{JHEP} {\bf 05} (2020)
  136}, [\href{http://arxiv.org/abs/2003.02879}{{\tt 2003.02879}}].

\bibitem{Fayyazuddin:1998fb}
A.~Fayyazuddin and M.~Spalinski, \emph{{Large N superconformal gauge theories
  and supergravity orientifolds}},
  \href{http://dx.doi.org/10.1016/S0550-3213(98)00545-8}{\emph{Nucl. Phys.}
  {\bf B535} (1998) 219--232}, [\href{http://arxiv.org/abs/hep-th/9805096}{{\tt
  hep-th/9805096}}].

\bibitem{Aharony:1998xz}
O.~Aharony, A.~Fayyazuddin and J.~M. Maldacena, \emph{{The Large N limit of
  N=2, N=1 field theories from three-branes in F theory}},
  \href{http://dx.doi.org/10.1088/1126-6708/1998/07/013}{\emph{JHEP} {\bf 07}
  (1998) 013}, [\href{http://arxiv.org/abs/hep-th/9806159}{{\tt
  hep-th/9806159}}].

\bibitem{Park:1998zh}
J.~Park and A.~M. Uranga, \emph{{A Note on Superconformal ${\mathcal{N}}\!=2$
  Theories and Orientifolds}},
  \href{http://dx.doi.org/10.1016/S0550-3213(98)00814-1}{\emph{Nucl. Phys. B}
  {\bf 542} (1999) 139--156}, [\href{http://arxiv.org/abs/hep-th/9808161}{{\tt
  hep-th/9808161}}].

\bibitem{Ennes:2000fu}
I.~P. Ennes, C.~Lozano, S.~G. Naculich and H.~J. Schnitzer, \emph{{Elliptic
  Models, Type IIB Orientifolds and the AdS/CFT Correspondence}},
  \href{http://dx.doi.org/10.1016/S0550-3213(00)00580-0}{\emph{Nucl. Phys.}
  {\bf B591} (2000) 195--226}, [\href{http://arxiv.org/abs/hep-th/0006140}{{\tt
  hep-th/0006140}}].

\bibitem{Koh:1983ir}
I.~G. Koh and S.~Rajpoot, \emph{{Finite ${\mathcal{N}}\!=2$ Extended
  Supersymmetric Field Theories}},
  \href{http://dx.doi.org/10.1016/0370-2693(84)90302-2}{\emph{Phys. Lett.} {\bf
  135B} (1984) 397--401}.

\bibitem{Howe:1983wj}
P.~S. Howe, K.~S. Stelle and P.~C. West, \emph{{A Class of Finite
  Four-Dimensional Supersymmetric Field Theories}},
  \href{http://dx.doi.org/10.1016/0370-2693(83)91402-8}{\emph{Phys. Lett.} {\bf
  124B} (1983) 55--58}.

\bibitem{Mkrtchian:1981bb}
R.~L. Mkrtchian, \emph{{The Equivalence of Sp(2N) and SO(-2N) Gauge Theories}},
  \href{http://dx.doi.org/10.1016/0370-2693(81)91015-7}{\emph{Phys. Lett.} {\bf
  105B} (1981) 174--176}.

\bibitem{mehta}
M.~L. Mehta, \emph{{Random matrices}}.
\newblock Elsevier, 2004.

\bibitem{Marino:2012zq}
M.~Mari\~no, \emph{{Lectures on non-perturbative effects in large $N$ gauge
  theories, matrix models and strings}},
  \href{http://dx.doi.org/10.1002/prop.201400005}{\emph{Fortsch. Phys.} {\bf
  62} (2014) 455--540}, [\href{http://arxiv.org/abs/1206.6272}{{\tt
  1206.6272}}].

\bibitem{Russo:2012ay}
J.~G. Russo and K.~Zarembo, \emph{{Large $N$ Limit of ${\mathcal{N}}\!=2$
  $SU(N)$ Gauge Theories from Localization}},
  \href{http://dx.doi.org/10.1007/JHEP10(2012)082}{\emph{JHEP} {\bf 10} (2012)
  082}, [\href{http://arxiv.org/abs/1207.3806}{{\tt 1207.3806}}].

\bibitem{Rodriguez-Gomez:2016cem}
D.~Rodriguez-Gomez and J.~G. Russo, \emph{{Operator mixing in large $N$
  superconformal field theories on S$^{4}$ and correlators with Wilson loops}},
  \href{http://dx.doi.org/10.1007/JHEP12(2016)120}{\emph{JHEP} {\bf 12} (2016)
  120}, [\href{http://arxiv.org/abs/1607.07878}{{\tt 1607.07878}}].

\bibitem{Erickson:2000af}
J.~K. Erickson, G.~W. Semenoff and K.~Zarembo, \emph{{Wilson loops in N=4
  supersymmetric Yang-Mills theory}},
  \href{http://dx.doi.org/10.1016/S0550-3213(00)00300-X}{\emph{Nucl. Phys.}
  {\bf B582} (2000) 155--175}, [\href{http://arxiv.org/abs/hep-th/0003055}{{\tt
  hep-th/0003055}}].

\bibitem{Giombi:2020kvo}
S.~Giombi and B.~Offertaler, \emph{{Wilson loops in $ \mathcal{N} $ = 4 SO(N)
  SYM and D-branes in AdS$_{5}$ \texttimes{}
  \ensuremath{\mathbb{R}}\ensuremath{\mathbb{P}}$^{5}$}},
  \href{http://dx.doi.org/10.1007/JHEP10(2021)016}{\emph{JHEP} {\bf 10} (2021)
  016}, [\href{http://arxiv.org/abs/2006.10852}{{\tt 2006.10852}}].

\bibitem{Witten:1998xy}
E.~Witten, \emph{{Baryons and branes in anti-de Sitter space}},
  \href{http://dx.doi.org/10.1088/1126-6708/1998/07/006}{\emph{JHEP} {\bf 07}
  (1998) 006}, [\href{http://arxiv.org/abs/hep-th/9805112}{{\tt
  hep-th/9805112}}].

\bibitem{Aharony:1999ti}
O.~Aharony, S.~S. Gubser, J.~M. Maldacena, H.~Ooguri and Y.~Oz, \emph{{Large N
  field theories, string theory and gravity}},
  \href{http://dx.doi.org/10.1016/S0370-1573(99)00083-6}{\emph{Phys.Rept.} {\bf
  323} (2000) 183--386}, [\href{http://arxiv.org/abs/hep-th/9905111}{{\tt
  hep-th/9905111}}].

\bibitem{Aharony:1999rz}
O.~Aharony, J.~Pawelczyk, S.~Theisen and S.~Yankielowicz, \emph{{A Note on
  Anomalies in the AdS/CFT Correspondence}},
  \href{http://dx.doi.org/10.1103/PhysRevD.60.066001}{\emph{Phys. Rev. D} {\bf
  60} (1999) 066001}, [\href{http://arxiv.org/abs/hep-th/9901134}{{\tt
  hep-th/9901134}}].

\bibitem{Blau:1999vz}
M.~Blau, K.~S. Narain and E.~Gava, \emph{{On Subleading Contributions to the
  AdS / CFT Trace Anomaly}},
  \href{http://dx.doi.org/10.1088/1126-6708/1999/09/018}{\emph{JHEP} {\bf 09}
  (1999) 018}, [\href{http://arxiv.org/abs/hep-th/9904179}{{\tt
  hep-th/9904179}}].

\bibitem{Liu:1998bu}
H.~Liu and A.~A. Tseytlin, \emph{{D = 4 super Yang-Mills, D = 5 gauged
  supergravity, and D = 4 conformal supergravity}},
  \href{http://dx.doi.org/10.1016/S0550-3213(98)00443-X}{\emph{Nucl.Phys.} {\bf
  B533} (1998) 88--108}, [\href{http://arxiv.org/abs/hep-th/9804083}{{\tt
  hep-th/9804083}}].

\bibitem{Henningson:1998gx}
M.~Henningson and K.~Skenderis, \emph{{The Holographic Weyl anomaly}},
  \href{http://dx.doi.org/10.1088/1126-6708/1998/07/023}{\emph{JHEP} {\bf 9807}
  (1998) 023}, [\href{http://arxiv.org/abs/hep-th/9806087}{{\tt
  hep-th/9806087}}].

\bibitem{Beccaria:2014xda}
M.~Beccaria and A.~A. Tseytlin, \emph{{Higher spins in AdS$_{5}$ at one loop:
  vacuum energy, boundary conformal anomalies and AdS/CFT}},
  \href{http://dx.doi.org/10.1007/JHEP11(2014)114}{\emph{JHEP} {\bf 1411}
  (2014) 114}, [\href{http://arxiv.org/abs/1410.3273}{{\tt 1410.3273}}].

\bibitem{Schnitzer:2002rt}
H.~J. Schnitzer and N.~Wyllard, \emph{{An Orientifold of AdS(5) $\times$ T11
  with D7-branes, the associated $\alpha'^2$ corrections and their role in the
  dual N=1 Sp(2N + 2M) $\times$ Sp(2N) gauge theory}},
  \href{http://dx.doi.org/10.1088/1126-6708/2002/08/012}{\emph{JHEP} {\bf 08}
  (2002) 012}, [\href{http://arxiv.org/abs/hep-th/0206071}{{\tt
  hep-th/0206071}}].

\bibitem{Drukker:2000ep}
N.~Drukker, D.~J. Gross and A.~A. Tseytlin, \emph{{Green-Schwarz string in
  AdS(5) $\times$ S5: Semiclassical partition function}},
  \href{http://dx.doi.org/10.1088/1126-6708/2000/04/021}{\emph{JHEP} {\bf 04}
  (2000) 021}, [\href{http://arxiv.org/abs/hep-th/0001204}{{\tt
  hep-th/0001204}}].

\bibitem{Beccaria:2022ypy}
M.~Beccaria, G.~P. Korchemsky and A.~A. Tseytlin, \emph{{Strong coupling
  expansions in $\mathbf{\mathcal N=2}$ superconformal theories and the Bessel
  kernel}},  \href{http://arxiv.org/abs/2207.11475}{{\tt 2207.11475}}.

\bibitem{Bobev:2022grf}
N.~Bobev, P.-J. De~Smet and X.~Zhang, \emph{{The planar limit of the
  $\mathcal{N} = 2$ $\mathbf{E}$-theory: numerical calculations and the large
  $\lambda$ expansion}},  \href{http://arxiv.org/abs/2207.12843}{{\tt
  2207.12843}}.

\bibitem{Billo:2019fbi}
M.~Billo, F.~Galvagno and A.~Lerda, \emph{{BPS wilson loops in generic
  conformal $ \mathcal{N} $ = 2 SU(N) SYM theories}},
  \href{http://dx.doi.org/10.1007/JHEP08(2019)108}{\emph{JHEP} {\bf 08} (2019)
  108}, [\href{http://arxiv.org/abs/1906.07085}{{\tt 1906.07085}}].

\bibitem{Beccaria:2020hgy}
M.~Beccaria, M.~Bill\`o, F.~Galvagno, A.~Hasan and A.~Lerda, \emph{{$
  \mathcal{N} $ = 2 Conformal SYM theories at large $ \mathcal{N} $}},
  \href{http://dx.doi.org/10.1007/JHEP09(2020)116}{\emph{JHEP} {\bf 09} (2020)
  116}, [\href{http://arxiv.org/abs/2007.02840}{{\tt 2007.02840}}].

\bibitem{berry-howls}
M.~V. Berry and C.~J. Howls, \emph{{Hyperasymptotics for Integrals with
  Saddles}}, \href{http://dx.doi.org/10.1098/rspa.1991.0119}{\emph{Proc. R.
  Soc. Lond. A} {\bf 434} (1991) 657--675}.

\bibitem{dunne-unsal}
G.~V. Dunne and M.~\"Unsal, \emph{{New Nonperturbative Methods in Quantum Field
  Theory: From Large-N Orbifold Equivalence to Bions and Resurgence}},
  \href{http://dx.doi.org/10.1146/annurev-nucl-102115-044755}{\emph{Ann. Rev.
  Nucl. Part. Sci.} {\bf 66} (2016) 245--272},
  [\href{http://arxiv.org/abs/1601.03414}{{\tt 1601.03414}}].

\bibitem{aniceto-basar-schiappa}
I.~Aniceto, G.~Basar and R.~Schiappa, \emph{{A Primer on Resurgent Transseries
  and Their Asymptotics}},
  \href{http://dx.doi.org/10.1016/j.physrep.2019.02.003}{\emph{Phys. Rept.}
  {\bf 809} (2019) 1--135}, [\href{http://arxiv.org/abs/1802.10441}{{\tt
  1802.10441}}].

\bibitem{Alday:2007mf}
L.~F. Alday and J.~M. Maldacena, \emph{{Comments on operators with large
  spin}}, \href{http://dx.doi.org/10.1088/1126-6708/2007/11/019}{\emph{JHEP}
  {\bf 11} (2007) 019}, [\href{http://arxiv.org/abs/0708.0672}{{\tt
  0708.0672}}].

\bibitem{Basso:2009gh}
B.~Basso and G.~P. Korchemsky, \emph{{Nonperturbative scales in AdS/CFT}},
  \href{http://dx.doi.org/10.1088/1751-8113/42/25/254005}{\emph{J. Phys.} {\bf
  A42} (2009) 254005}, [\href{http://arxiv.org/abs/0901.4945}{{\tt
  0901.4945}}].

\bibitem{Aniceto:2015rua}
I.~Aniceto, \emph{{The Resurgence of the Cusp Anomalous Dimension}},
  \href{http://dx.doi.org/10.1088/1751-8113/49/6/065403}{\emph{J. Phys.} {\bf
  A49} (2016) 065403}, [\href{http://arxiv.org/abs/1506.03388}{{\tt
  1506.03388}}].

\bibitem{Dorigoni:2015dha}
D.~Dorigoni and Y.~Hatsuda, \emph{{Resurgence of the Cusp Anomalous
  Dimension}}, \href{http://dx.doi.org/10.1007/JHEP09(2015)138}{\emph{JHEP}
  {\bf 09} (2015) 138}, [\href{http://arxiv.org/abs/1506.03763}{{\tt
  1506.03763}}].

\bibitem{Drukker:2006ga}
N.~Drukker, \emph{{1/4 BPS circular loops, unstable world-sheet instantons and
  the matrix model}},
  \href{http://dx.doi.org/10.1088/1126-6708/2006/09/004}{\emph{JHEP} {\bf 09}
  (2006) 004}, [\href{http://arxiv.org/abs/hep-th/0605151}{{\tt
  hep-th/0605151}}].

\bibitem{Zarembo:2016bbk}
K.~Zarembo, \emph{{Localization and AdS/CFT Correspondence}},
  \href{http://dx.doi.org/10.1088/1751-8121/aa585b}{\emph{J. Phys.} {\bf A50}
  (2017) 443011}, [\href{http://arxiv.org/abs/1608.02963}{{\tt 1608.02963}}].

\bibitem{Billo:2017glv}
M.~Billo, F.~Fucito, A.~Lerda, J.~F. Morales, {\relax Ya}.~S. Stanev and
  C.~Wen, \emph{{Two-point Correlators in N=2 Gauge Theories}},
  \href{http://dx.doi.org/10.1016/j.nuclphysb.2017.11.003}{\emph{Nucl. Phys.}
  {\bf B926} (2018) 427--466}, [\href{http://arxiv.org/abs/1705.02909}{{\tt
  1705.02909}}].

\bibitem{Cvitanovic:1976am}
P.~Cvitanovic, \emph{{Group Theory for Feynman Diagrams in Non-Abelian Gauge
  Theories}}, \href{http://dx.doi.org/10.1103/PhysRevD.14.1536}{\emph{Phys.
  Rev. D} {\bf 14} (1976) 1536--1553}.

\bibitem{Huang:2016iqf}
J.-H. Huang, \emph{{Group-Theoretic Relations for Amplitudes in Gauge Theories
  with Orthogonal and Symplectic Groups}},
  \href{http://dx.doi.org/10.1103/PhysRevD.95.025015}{\emph{Phys. Rev. D} {\bf
  95} (2017) 025015}, [\href{http://arxiv.org/abs/1612.08868}{{\tt
  1612.08868}}].

\bibitem{Duff:1977ay}
M.~Duff, \emph{{Observations on Conformal Anomalies}},
  \href{http://dx.doi.org/10.1016/0550-3213(77)90410-2}{\emph{Nucl.Phys.} {\bf
  B125} (1977) 334}.

\bibitem{zagier}
D.~Zagier, \emph{{The Mellin transformation and other useful analytic
  techniques}},  in \emph{Quantum Field Theory I: Basics in Mathematics and
  Physics}, pp.~307--323.
\newblock Springer, 2006.

\bibitem{flajolet}
P.~Flajolet, X.~Gourdon and P.~Dumas, \emph{{Mellin transforms and asymptotics:
  Harmonic sums}}, {\emph{Theoretical computer science} {\bf 144} (1995)
  3--58}.

\end{thebibliography}\endgroup
\bibliographystyle{JHEP}
\end{document}